\documentclass[number,10pt,3p,final,twocolumn]{elsarticle}

\usepackage{graphicx}
\graphicspath{{./figures/}}

\usepackage{amssymb}
\usepackage{amsthm}
\usepackage{upgreek}
\usepackage{float}
\usepackage{amsmath}
\usepackage{subfig}
\usepackage{array}
\usepackage{url}
\usepackage{siunitx,booktabs,etoolbox}
\usepackage[para,online,flushleft]{threeparttable}
\usepackage[pdfencoding=auto,psdextra,colorlinks=true,linkcolor=red,citecolor=blue]{hyperref}
\usepackage[colorinlistoftodos]{todonotes}
\usepackage{threeparttable, tablefootnote}
\usepackage{footnote}

\usepackage{lineno}

\biboptions{sort&compress}

\newcommand{\ignore}[1]{}
\journal{Nuclear Instruments and Methods A}

\let\oldequation\equation
\let\oldendequation\endequation

\renewenvironment{equation}
  {\linenomathNonumbers\oldequation}
  {\oldendequation\endlinenomath}

\let\oldalign\align
\let\oldendalign\endalign

\renewenvironment{align}
  {\linenomathNonumbers\oldalign}
  {\oldendalign\endlinenomath}

\begin{document}

\begin{frontmatter}

\author[labelReed]{J.K.~Smith\corref{cor1}}
\ead{smith.jenna.kathleen@gmail.com}
\cortext[cor1]{Corresponding author: 3023 SE Woodstock Blvd., Portland, OR 97202}

\title{Gamma-gamma angular correlation analysis techniques with the GRIFFIN spectrometer}

\author[labelGuelph]{A.D.~MacLean}
\author[labelReed]{W.~Ashfield}
\author[labelTRI]{A.~Chester}
\author[labelTRI]{A.B.~Garnsworthy}
\author[labelGuelph]{C.E.~Svensson}

\address[labelReed]{Department of Physics, Reed College, 3203 SE Woodstock Blvd., Portland, OR, USA, 97202}
\address[labelGuelph]{Department of Physics, University of Guelph, Guelph, ON, Canada, N1G 2W1}
\address[labelTRI]{Physical Sciences Division, TRIUMF, 4004 Wesbrook Mall, Vancouver, BC, Canada, V6T 2A3}




\begin{abstract}
Gamma-gamma angular correlation measurements are a powerful tool for identifying the angular momentum (spin) of excited nuclear states involved in a $\gamma$-ray cascade, and for measuring the multipole orders and mixing ratios of transitions. Though the physical angular correlations are fully calculable from first principles, experimental effects can make the extraction of coefficients and thus conclusions about spins and mixing ratios difficult. In this article we present data analysis techniques developed for the clover detectors of the GRIFFIN spectrometer at TRIUMF-ISAC combined with GEANT4 simulations in order to extract accurate experimental results.
\end{abstract}


\end{frontmatter}


\section{Introduction}

A detailed understanding of nuclear structure benefits from a comparison between theoretical calculations and experimental observations of the properties of excited states in atomic nuclei.
In order to make meaningful comparisons, it is essential to make a positive identification of the angular momentum (spin) for the excited states under study.
Ground state nuclear spins (as well as those of meta-stable states) can be assigned firmly with optical techniques such as laser spectroscopy \cite{Campbell2016} or nuclear magnetic resonance \cite{Neyens2003}. In the case of excited states, experimental techniques for spin assignment utilize selectivity in the particular reaction or decay (i.e. $log(ft)$ values in beta decays) or angular distribution measurements (i.e. angular distribution of transfer reaction products) to constrain or positively assign spins.

In $\gamma$-ray spectroscopy, spins can be identified or constrained based on absolute and relative lifetimes and the angular distribution of the emitted radiation. For a given transition, $\gamma$ rays are emitted in an angular distribution that is dictated by the electromagnetic multipoles of the radiation involved. In an unpolarized sample though, these angular distributions are rendered isotropic by the random orientation of the ensemble of nuclei in the sample. In such a situation, one must rely on measurements of the angle between two $\gamma$ rays emitted sequentially from the same nucleus in order to define the nuclear alignment. These angular correlations have the form:
\begin{equation}
W(\theta) = \sum_{i=0,\textup{even}}^{\infty}{B_{ii} G_{ii}(t) A_{ii} P_{i}(\cos \theta)}
\end{equation}
where $\theta$ is the angle between two successive $\gamma$ rays, $B_{ii}$ specify the initial nuclear orientation, $G_{ii}(t)$ are time-dependent perturbation factors that account for extranuclear interactions which disturb the correlation, and $P_i(\cos \theta)$ are Legendre polynomials. The $A_{ii}$ are a series of coefficients, the values of which are influenced by the spins of the three states involved as well as the multipole order and mixing ratio, $\delta$, for each transition \cite{Rose1967}. Here, we take the situation of an isotropic initial nuclear orientation ($B_{ii}=1$) from excited states populated in $\beta$ decay of a randomly-oriented source, and short excited state lifetimes ($G_{ii}(t)\approx 1$), such that the angular correlation is described as:
\begin{equation}
W(\theta) = \sum_{i=0,\textup{even}}^{\infty}{A_{ii} P_{i}(\cos \theta)}.
\label{eq.W_theory}
\end{equation}
Cascades involving low spins and low multipolarities will be described by only the lowest few terms of this sum; angular momentum considerations will set higher-order coefficients to zero, effectively truncating the series. For these reasons, Equation \ref{eq.W_theory} is often re-written as:
\begin{equation}
W(\theta) = A_{00}[1 + a_2 P_2(\cos \theta)+a_4 P_4(\cos\theta)]
\label{eq:ang-corr}
\end{equation}
where
\begin{equation}
a_{i}=A_{ii}/A_{00}.
\end{equation}
In this article, as only cases involving cascades with an intermediate state spin of $J=2$ are considered, angular momentum considerations render this truncation exact. The same truncation is generally sufficient in the majority of $\beta$-decay experiments, which usually involve the population of low spin states. We note also that for cascades in which the second transition involves a higher multipolarity, the intermediate lifetime will be longer and so the above assumption of $G_{ii}(t)\approx1$ will not hold for such cases. 

It is a simple procedure to fit an experimental distribution with Equation \ref{eq:ang-corr}, but the data gathered by large arrays of detectors are distorted by several experimental factors: uneven distribution of detectors in $\theta$, different detector efficiencies, finite detector size, and the lifetimes of intermediate states. These factors mean that the uncorrected coefficients obtained from the fitted experimental data do not describe the true physical angular correlation of the $\gamma$ rays. These effects typically act to attenuate the asymmetry in the observed angular distributions. Various techniques have been developed to account for these effects in order to extract accurate coefficients from experimental data. In cases where the lifetime of the intermediate state is negligible (the $2^+$ states in $^{60}$Ni, $^{152}$Gd and $^{66}$Zn are sufficiently short-lived, see Table \ref{table.cascades}), the uneven distribution of detectors and different efficiencies can be deconvoluted from the data by counting the crystal pairs at various opening angles and normalizing individual detector signals for the detector-specific, energy-dependent (and sometimes time-dependent) efficiency. This data processing can be difficult and time-consuming, especially for low-statistics peaks or if the efficiency varies over time. In the past, an accounting of the effects of finite detector size has been done by defining attenuation factors $Q_{\ell\ell}$, such that
\begin{equation}
a_\ell = \frac{c_\ell}{Q_{\ell\ell}}
\end{equation}
where $c_{\ell}$ is the coefficient of $P_{\ell}$ extracted from a fit of Equation \ref{eq:ang-corr} to the data. The $Q_{\ell\ell}$ values are specific to the detector size, shape, distance from the source, and the energy of the particular $\gamma$ ray. 
The attenuation coefficients can be calculated in advance but such calculations depend heavily on the particular detector array setup and the energies of the specific $\gamma$ rays involved in the cascade. Examples of such calculations for detectors with simple geometry can be found in Refs. \cite{Krane1972,Krane1973}. While these factors can be calculated analytically for certain detector shapes, the coefficients were often calculated numerically with Monte Carlo simulations \cite{Yates1965,Krane1972,Schmelzenbach2003}. The use of a full simulation in the present work will allow the calculation of differences in these coefficients due to changes in the physical setup of the experiment and incorporate modern cross-section information for the relevant materials.

In this article, we discuss the analysis procedures developed to extract physically relevant angular correlation coefficients from data collected with the high-purity germanium clover detectors used in the Gamma-Ray Infrastructure For Fundamental Investigations of Nuclei (GRIFFIN) \cite{Svensson2013,Rizwan2016}, described briefly in Section \ref{sec:griffin}. These methods are, however, generally applicable to large arrays of $\gamma$-ray detectors. First, in Sections \ref{sec:eventmix}, we present an adaptation of an event-mixing technique to remove energy- and time-dependent efficiency differences between the detectors. In Sections \ref{sec:sim}, \ref{sec:Zfits}, \ref{sec:alg}, and \ref{sec:alp-bet-gam}, we describe a series of methods utilizing simulations to correct for the finite detector sizes with increasing levels of parameterization and approximation. While some methods reduce the achievable precision and accuracy of the measured values they also dramatically reduce the computational cost of each individual measurement. Finally, in Sections \ref{sec:Methods} and \ref{sec:sum} we make comparisons between the methods and summarize the results.

\section{\label{sec:griffin}Experimental details - GRIFFIN}
\begin{table}[t]
\caption{Angles between HPGe crystal pairs in the GRIFFIN geometry with the HPGe detectors at a source-to-detector distance of 11\,cm. Two independent sets of crystal pairs at 86.2 degrees have different geometries but angular differences that are the same to four decimal places. The same is true for the two independent sets of crystal pairs at 93.8 degrees. See text for more details of pair counting.} 
\centering 
\begin{tabular}{cc|cc} 
\hline 
\hline 
 & Num. of &  & Num. of \\
Angle ($^\circ$) & Pairs & Angle ($^\circ$) & Pairs\\ [0.5ex] 
\hline 
0.0 & 64 & 91.5 & 128 \\ 
18.8 & 128 & 93.8 & 48 \\
25.6 & 64 & 93.8 & 64 \\
26.7 & 64 & 97.0 & 64 \\
31.9 & 64 & 101.3 & 64 \\
33.7 & 48 & 103.6 & 96 \\ 
44.4 & 128 & 106.9 & 64 \\
46.8 & 96 & 109.1 & 96 \\
48.6 & 128 & 110.1 & 64 \\
49.8 & 96 & 112.5 & 64 \\ 
53.8 & 48 & 113.4 & 64 \\
60.2 & 96 & 115.0 & 96 \\
62.7 & 48 & 116.9 & 64 \\
63.1 & 64 & 117.3 & 48 \\ 
65.0 & 96 & 119.8 & 96 \\
66.5 & 64 & 126.2 & 48 \\
67.5 & 64 & 130.2 & 96 \\
69.9 & 64 & 131.4 & 128 \\ 
70.9 & 96 & 133.2 & 96 \\
73.1 & 64 & 135.6 & 128 \\
76.4 & 96 & 146.3 & 48 \\
78.7 & 64 & 148.1 & 64 \\ 
83.0 & 64 & 152.3 & 64 \\
86.2 & 64 & 154.4 & 64 \\
86.2 & 48 & 160.2 & 128 \\
88.5 & 128 & 180.0 & 64 \\ [1ex] 
\hline 
\hline 
\end{tabular}
\label{tab:openingAngles} 
\end{table}

Since the analysis techniques described herein are demonstrated through application to data recorded with the GRIFFIN spectrometer, we give here a brief description of the facility and refer the reader to other publications (\cite{Svensson2013,Rizwan2016,Garnsworthy2017}) for further details.

GRIFFIN is an array of 16 High-Purity Germanium (HPGe) clover detectors arranged in a rhombicuboctahedron geometry around the location at which the radioactive beam is implanted. Each clover contains four electrically-independent crystals for a total of 64 individual HPGe crystals in the spectrometer. In typical operation, the TRIUMF-ISAC facility provides beams of radioactive isotopes that are stopped at the center of the array and subsequently decay. The distribution of radioactivity on the tape has a diameter of less than 5\,mm. In addition to the HPGe clover detectors which detect the $\gamma$ rays, ancillary detectors are available which can detect emitted beta, alpha, proton, and neutron radiation. The configuration of the HPGe clovers is also variable: the clovers can be arranged in a close-packed, high-efficiency geometry with the front face of each detector at a source-to-detector distance of 11\,cm, or an optimized peak-to-total geometry with full Compton and background suppression shields and the HPGe detectors at a distance of 14.5\,cm from the source. Some ancillary detectors require the removal of one or more clover detectors. The particular detector configuration thus varies based on the experimental needs.

\begin{table*}
\caption{\label{table.cascades}Details of the $\gamma-\gamma$ cascades used in this work. Subscripts of $i,x,f$ are used to indicate the initial, intermediate, and final states, respectively. Mixing ratios are taken from Refs. \cite{Browne2013,Martin2013,Browne2010}. The $a_2,a_4$ values are calculated from the experimentally measured mixing ratios and known nuclear spins.\ignore{taken from IAEA gamma-ray standards \cite{Nichols2004} or calculated directly ($^{66}$Zn, 0-2-0 and 2-2-0). In the case of the $2^+-2^+-0^+$ cascade in $^{66}$Zn, the coefficients were calculated based on the adopted mixing ratio \cite{Browne2010}.} The mixing ratios for the $x-f$ transitions are all zero as all cascades have a purely $E2$ multipolarity for this $x-f$ transition.}
\centering
\begin{tabular}{lcccccccc}
\hline
\hline
Nucleus & $J_i^\pi$-$J_x^\pi$-$J_f^\pi$ & Mult. & $\tau$ & $E_{\gamma}^{i-x}$ & $E_{\gamma}^{x-f}$ & $\delta^{i-x}$ & $a_2$ & $a_4$\\
 & & $(i-x)$ & (ps) & (keV) & (keV) & & & \\
\hline
$^{60}$Ni & $4^+$-$2^+$-$0^+$ & E2(+M3) & 1.06(3) & 1173.2 & 1332.5 & -0.0025(22) & $0.1005(13)$ & $0.0094(3)$ \\
$^{152}$Gd & $3^-$-$2^+$-$0^+$ & E1(+M2) & 46.2(39) &778.9 & 344.3 & 0.003(6) & -0.0691(47) & 0 \\
\hline
$^{66}$Zn & $1^+$-$2^+$-$0^+$ & M1+E2 & 2.42(4) & 2751.8 & 1039.2 & -0.09(3) & -0.147(35) & -0.0061$^{+0.0034}_{-0.0047}$ \\
 & &  &  &  &  & -0.12(2) & -0.112(23) & -0.0108$^{+0.0033}_{-0.0038}$ \\
$^{66}$Zn & $0^+$-$2^+$-$0^+$ & E2 & 2.42(4) & 1333.1 & 1039.2 & - & 0.357 & 1.143 \\
$^{66}$Zn & $2^+$-$2^+$-$0^+$ & M1+E2 & 2.42(4) &  833.5 & 1039.2 & -1.9(3) & $0.30^{+0.05}_{-0.04}$ & $0.256^{+0.015}_{-0.021}$ \\
 &  &  & &  &  & -1.6(2) & $0.34^{+0.04}_{-0.03}$ & $0.23(2)$ \\
\hline
\hline
\end{tabular}
\end{table*}

The HPGe data used in this work was collected in the close-packed geometry with a full complement of 16 clover detectors. Such a geometry contains 4032 possible crystal-pair combinations, resulting in 51 unique opening angles ranging from 19$^\circ$ to 180$^\circ$. Because the two gamma rays measured are distinguishable, we retain the distinguishability of crystal pairs so that crystal pair $(j,k)$ is distinct from crystal pair $(k,j)$. An additional 64 ``pairs'' of crystals with an angular difference of 0$^\circ$ can be included if one considers the sum peaks that result when both $\gamma$ rays of the cascade interact, and are fully absorbed, in the same crystal. The number of HPGe crystal pairs at each of the unique angular differences is shown in Table \ref{tab:openingAngles}. This set of angular differences is used for all the angular correlations in this work. The average angular difference for $\gamma$ rays detected in pairs of HPGe crystals was examined in a GEANT4 \cite{Agostinelli2003,GEANT4-2016} simulation to explore the sensitivity of this angular difference as a function of the $\gamma$-ray energy which could modify the average interaction location within the crystal volume. 
There is an energy-dependence to the detection position within a crystal due to the reduced $\gamma$-ray efficiency at the center of the crystal where material has been removed for the central core contact. However, as these effects are symmetric they cancel when the average is calculated.
The average angular differences obtained from the simulation match very closely with the geometric calculation between the centers of the front face of each crystal.

High-statistics data from the beta decays of $^{60}$Co, $^{152}$Eu, and $^{66}$Ga were utilized to measure the five cascades in the daughter isotopes detailed in Table \ref{table.cascades}.
The $^{60}$Co and $^{152}$Eu decays were observed from commercially available calibration sources mounted at the beam-implantation location.
A source of $^{66}$Ga ($T_{1/2}$=9.49(3)\,hrs \cite{Browne2010}) was created through the delivery of a radioactive beam from ISAC and observed for several half lives.
The data shown here have not utilized the addback of Compton scattered $\gamma$ rays between crystals, nor have the crystal pairs been grouped in any way beyond the unique angle groupings listed in Table \ref{tab:openingAngles}.

\section{\label{sec:eventmix}Treatment of the experimental data - Event-mixing technique}
An algorithm was developed based on the event-mixing ideas used in Refs. \cite{Drijard1984,Lisa1991,LHote1994,Souza2009} that allows the production of robust experimental angular correlation distributions shortly after the data are collected, with minimal explicit calibration.

\subsection{\label{ssec:eventmix_theory}Experimental angular correlation theory}

The continuous physical angular correlation $W(\theta)$ is distorted by experimental effects and measured as a discrete experimental angular difference distribution, $w(\theta_i;E_a,E_b)$. The relationship between these two can be expressed as the following function:
\begin{align}
\label{eq.wab}
&w(\theta_i;E_a,E_b) = \sum_{j,k} A_{jk}(E_a,E_b) I_{jk}(\theta_i;E_a,E_b)\\
&A_{jk}(E_a,E_b) = \int \varepsilon_j(E_a,t)\varepsilon_k(E_b,t) dt\\
&I_{jk}(\theta_i;E_a,E_b)=\int_{\theta=\theta_i-\Delta\theta}^{\theta_i+\Delta\theta}N_{jk}(\theta;E_a,E_b)W(\theta)d\theta
\end{align}
where $w(\theta_i;E_a,E_b)$ is the discretized angular difference histogram between coincident $\gamma$ rays with energies $E_a$ and $E_b$, $\theta_i$ is one of the unique angles between crystal pairs, $j$ and $k$ are indices that iterate over all pairs of crystals which satisfy $\theta_i=|\theta_j-\theta_k|$, $\varepsilon_j(E_a,t)$ is the energy-dependent, time-dependent efficiency of crystal $j$, $N_{jk}(\theta;E_a,E_b)$ is a weighting distribution that describes the crystal pair response at different angles $\theta$ that are subtended by this crystal pair, $W(\theta)$ is the theoretical angular distribution given in Equation \ref{eq.W_theory}, and $\Delta\theta$ is a limit set such that $N_{jk}(\theta;E_a,E_b)$ is zero outside of the range $\theta_i\pm\Delta\theta$. The time integral in the definition of $A_{jk}$ is performed over the full time of the experiment.

A second experimental distribution, $y(\theta_i;E_a,E_b)$ can be defined that is related to a second theoretical distribution $Y(\theta)$, where $Y(\theta)$ is a different angular correlation, but defined with the same general form as $W(\theta)$ was in Equation \ref{eq.W_theory}:
\begin{align}
\label{eq.yab}
&y(\theta_i;E_a,E_b) = \sum_{j,k} A_{jk}(E_a,E_b)J_{jk}(\theta_i;E_a,E_b)\\
&J_{jk}(\theta_i;E_a,E_b)=\int_{\theta=\theta_i-\Delta\theta}^{\theta_i+\Delta\theta}N_{jk}(\theta;E_a,E_b)Y(\theta)d\theta.
\end{align}
Dividing Equation \ref{eq.wab} by Equation \ref{eq.yab} results in:
\begin{equation}
\frac{w(\theta_i)}{y(\theta_i)} = \frac{\sum_{j,k} \varepsilon_j(E_a,t)\varepsilon_k(E_b,t) I_{jk}(\theta_i;E_a,E_b)}{\sum_{j,k} \varepsilon_j(E_a,t)\varepsilon_k(E_b,t) J_{jk}(\theta_i;E_a,E_b)}
\end{equation}
where the explicit energy dependencies of the $w$ and $y$ distributions have been omitted for conciseness. If it is assumed that the distribution $N_{jk}(\theta;E_a,E_b)$ is identical for all crystal pairs $j,k$ that have an opening angle of $\theta_i$, then the sum and angle integral are entirely independent and separable. The sums and $A_{jk}$ factors within them therefore cancel to leave
\begin{equation}
\frac{w(\theta_i)}{y(\theta_i)} = \frac{\int_{\theta=\theta_i-\Delta\theta}^{\theta_i+\Delta\theta}N_i(\theta;E_a,E_b)W(\theta)d\theta}{\int_{\theta=\theta_i-\Delta\theta}^{\theta_i+\Delta\theta}N_i(\theta;E_a,E_b)Y(\theta)d\theta}.
\end{equation}
If $Y(\theta)$ is isotropic then $Y(\theta)=1$, and,
\begin{equation}
\frac{w(\theta_i)}{y(\theta_i)} = \frac{\int_{\theta=\theta_i-\Delta\theta}^{\theta_i+\Delta\theta}N_i(\theta;E_a,E_b)W(\theta)d\theta}{\int_{\theta=\theta_i-\Delta\theta}^{\theta_i+\Delta\theta}N_i(\theta;E_a,E_b)d\theta}.
\label{eq.5assumptions}
\end{equation}

Finally, in the limiting case that $W(\theta)$ can be approximated as constant over the range of $\theta=\theta_i\pm\Delta\theta$, or if $N(\theta_i,E_a,E_b)$ is symmetric about $\theta_i$ and $W(\theta)$ can be approximated as linear, then this equation simplifies further to:
\begin{equation}
\frac{w(\theta_i)}{y(\theta_i)} = W(\theta_i).
\end{equation}

In summary, the detector-, energy-, and time-dependent efficiencies in $w(\theta_i)$ can be removed by dividing that distribution by a distribution $y(\theta_i;E_a,E_b)$ that uses the same detector configuration. It is essential that $N_{jk}(\theta)$ is identical for all crystal pairs $j,k$ within the sum and that $y(\theta_i;E_a,E_b)$ has the same efficiency as a function of energy and time, uses $\gamma$ rays of the same energies, and is isotropic. Such a distribution construction will be addressed in Section \ref{ssec:em_construct}. 

This leaves primarily the impact of finite detector size. If this impact is small, then the ratio of distributions will closely resemble $W(\theta)$. If this impact is large or the desired precision is particularly high, then this final simplification cannot be assumed and further corrections must be applied to the experimental data either by comparison to simulations or otherwise modify the measured coefficients. Sections \ref{sec:sim}-\ref{sec:alp-bet-gam} present methods for these corrections utilizing simulations.

\subsection{\label{ssec:em_construct}Event-mixing plot construction}

\begin{figure}
	\centering
    \includegraphics[width=0.95\linewidth]{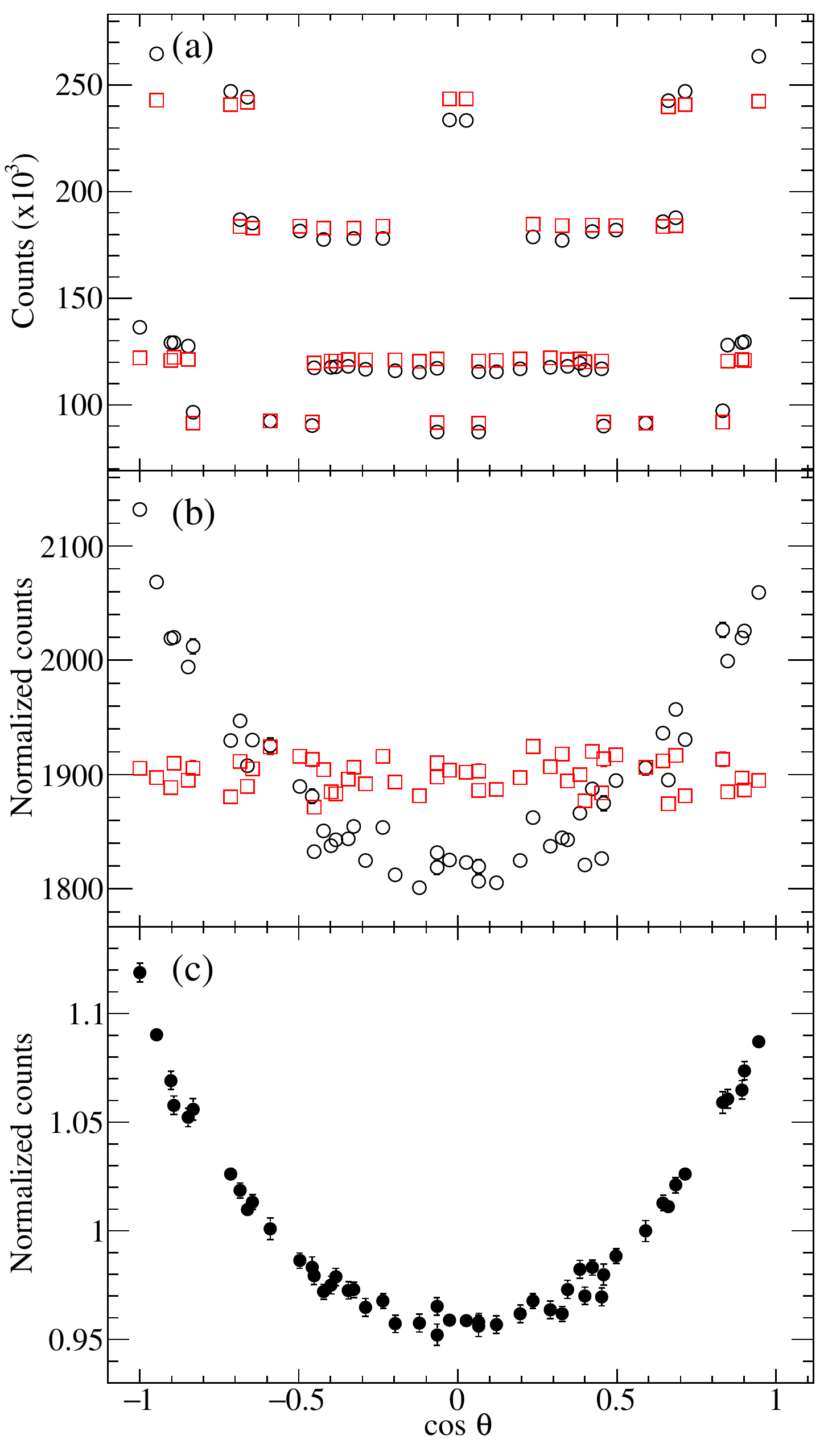}
    \caption{Construction of an angular correlation normalized by event-mixing. (a) Coincidences in the 1332 keV - 1173 keV $4^+\rightarrow 2^+\rightarrow 0^+$ cascade in $^{60}$Ni as a function of angle between detectors (black circles) and event-mixed coincidences in the same cascade (red squares). (b) The two series from the top panel, now divided by the number of crystal pairs at each possible angle. (c) Final angular correlation of the cascade, using the event-mixing technique for normalization.}
    \label{fig:ac1}
\end{figure}

\begin{figure}
	\centering
    \includegraphics[width=0.95\linewidth]{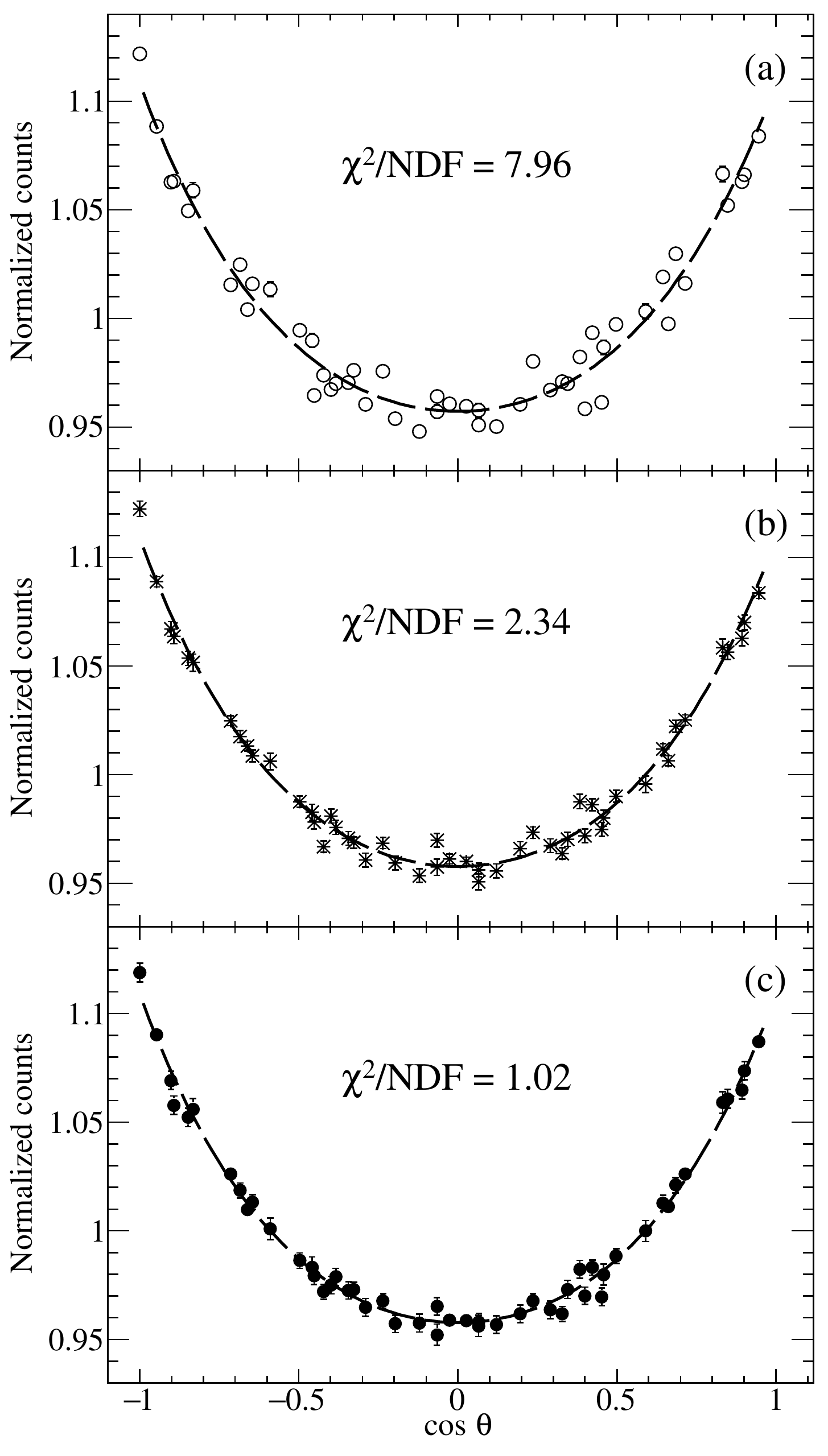}
    \caption{Angular correlations from the 1333 keV - 1173 keV $4^+\rightarrow 2^+\rightarrow 0^+$ cascade in $^{60}$Ni as a function of angle between detectors, normalized by (a) number of crystal pairs at each angle, (b) number of crystal pairs and individual detector efficiencies, and (c) the event-mixing technique. In all panels, the dashed black line shows a scaled simulated distribution. See text for more details. The $\chi^2$/NDF values show the reduced $\chi^2$ for each experimental angular correlation in comparison with the full simulation.}
    \label{fig:ac2}
\end{figure}

In data gathered from large arrays, a plot of coincident counts against angular difference is dominated by the number of crystal pairs at each angle, which in the case of GRIFFIN can vary from 48 to 128. Figure \ref{fig:ac1}a shows the number of 1332-1173 keV coincident counts from the $^{60}$Ni cascade as a function of $\cos(\theta)$ (black circle data points), while Figure \ref{fig:ac1}b shows the same data, but divided by the number of crystal pairs that contribute to each angle (black circle data points). Figure \ref{fig:ac2}a shows the same distribution, but this time in comparison to a simulated $4^+\rightarrow 2^+ \rightarrow 0^+$ correlation from a full GEANT4 simulation (black dashed line) as detailed in the following section. The goodness-of-fit parameter between the full simulation and the data is $\chi^2$/NDF=7.96, where NDF is the number of degrees of freedom.

In order to improve this, one needs to account for the different relative efficiencies of each crystal. Ideally, since the individual angular distributions of the 1332 and 1173\,keV transitions are isotropic, each HPGe crystal should detect the same number of $\gamma$ rays. Any variations in the number of counts seen by each crystal are due to different relative efficiencies. If the counts from each detector are scaled to force the 1332 and 1173\,keV angular distributions to be isotropic, this will correct for the different relative efficiencies. An angular correlation with this correction applied is shown in Figure \ref{fig:ac2}b. Making this correction improves the $\chi^2$/NDF value to 2.34.

The formalism developed in the previous section provides an alternative and more accurate way to make these corrections - by constructing an isotropic distribution from $\gamma$ rays of the same energies and dividing the two angular difference distributions. Experimentally, the challenge is to construct a distribution $y(\theta_i;E_a,E_b)$ with the appropriate characteristics. The $y(\theta_i;E_a,E_b)$ distribution can be created with data collected at the same time as the $w(\theta_i;E_a,E_b)$ distribution. This will satisfy the requirement that both histograms have the same efficiency as a function of $\gamma$-ray energy and time. To also ensure that the distribution is isotropic and uses $\gamma$ rays of the same energies, detected $\gamma$-ray events of the same energies originating from decays of different nuclei (but the same nuclide) can be selected by pairing $\gamma$-ray detections with an unphysically large time difference. The large time difference guarantees that the detections are uncorrelated with each other and thus provides the required isotropic distribution.

Some care must be taken with respect to the selection of gamma rays from different decays and the satisfaction of the time-dependent efficiency criterion. If there are variations in efficiency as a function of time, the events for event-mixing need to have a time difference that is smaller than this time-dependent efficiency fluctuation. As an illustrative example, if the efficiency were to change every thirty seconds for some reason, the average time between $\gamma$ rays that are selected needs to be much less than thirty seconds. This average time, however, also should be much longer than the lifetime of the intermediate state, to ensure that the two $\gamma$ rays are not from the same decay.

The division of the histograms will increase the statistical error in the final distribution. The impact of this can be made negligible by creating a $y(\theta_i;E_a,E_b)$ where each $\gamma$-ray interaction is paired with many other interactions. This increases the statistics and reduces the error in the divisor histogram, avoiding an inflation of the overall error.

Figure \ref{fig:ac1}a shows a raw event-mixed histogram (red squares) in comparison to an associated prompt histogram (black circles). The raw histograms look similar because they are both dominated by the number of crystal pairs that contribute to each angular bin. Dividing out this contribution produces the histograms shown in Figure \ref{fig:ac1}b that are qualitatively different. Here, the isotropic nature of the event-mixed spectrum is much more obvious. The division of these two histograms produces the final angular correlation distribution shown in Figure \ref{fig:ac1}c. The resulting angular correlation compares well with a full simulation ($\chi^2$/NDF of 1.02) as shown in Figure \ref{fig:ac2}c. In this work, the time window for event mixing is between 2\,$\mu$s and 200\,$\mu$s, whereas true coincidence events had time differences of less than 300\,ns.

\section{\label{sec:allsimmethods}Extracting finite detector size effects with simulation}

\begin{figure}
\centering
\includegraphics[width=\columnwidth]{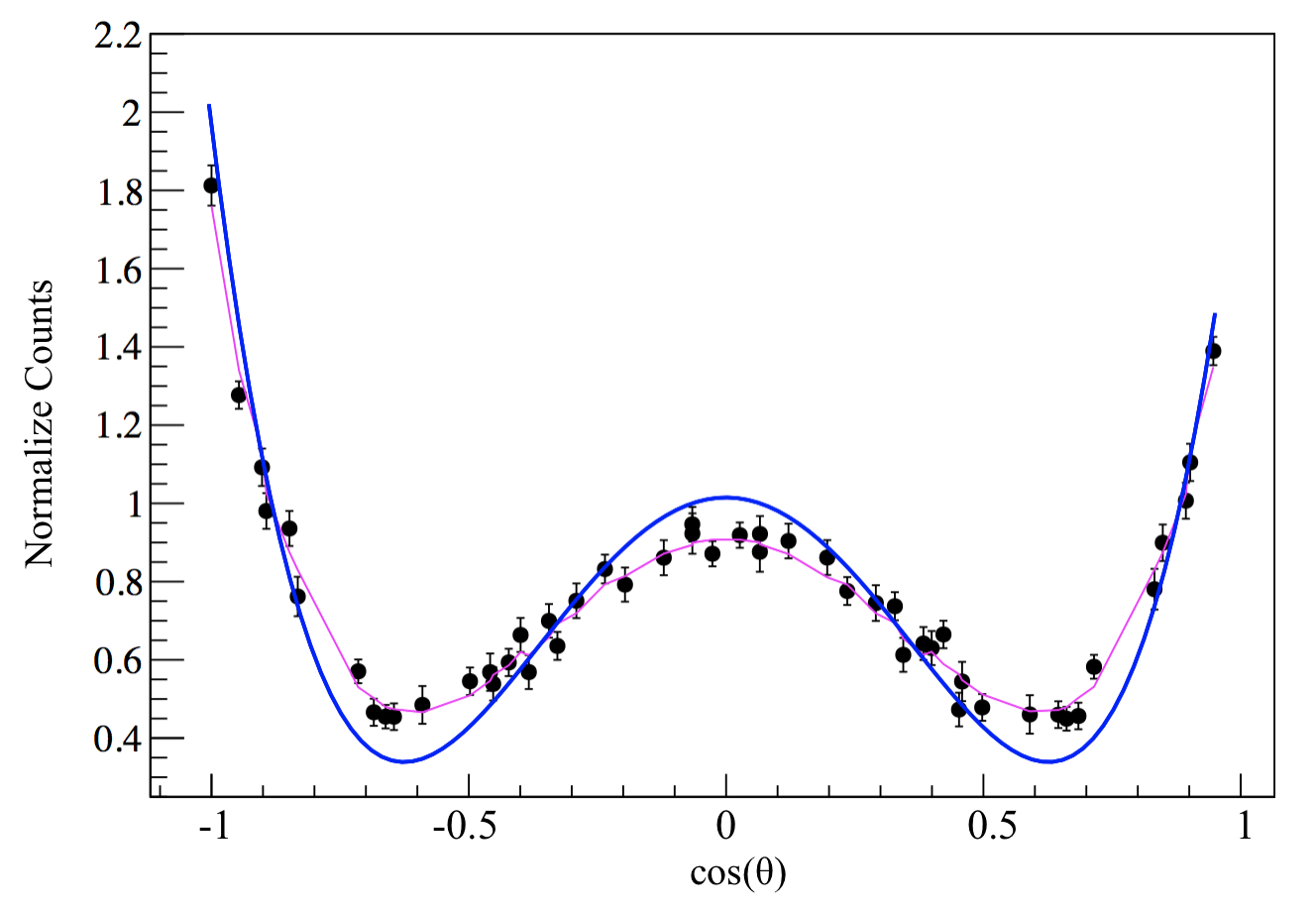}
\caption{\label{fig:Ga020_templatefit}The $\chi^2$/NDF between a full simulation of the $^{66}$Zn $0^+ - 2^+ - 0^+$ cascade (magenta filled line) with $a_2=0.3571$ and $a_4=1.1429$ and data (black points) is 1.01. The blue solid line is the angular correlation expected from theoretically calculated $a_2, a_4$ coefficients without corrections for finite detector size effects.}
\end{figure}

In order to extract the physical angular correlation coefficients in Equation \ref{eq.W_theory}, the slight differences in convolution of the detector response as a function of opening angle ($N_{i}(\theta)$) and the physical angular correlations must still be accounted for.
Even in cases of angular correlations that change slowly over the full opening angular range of a crystal pair, these differences can be appreciable.
Figure \ref{fig:Ga020_templatefit} shows an experimental angular correlation distribution from the $0^+$ -- $2^+$ -- $0^+$ cascade in $^{66}$Zn, which has a strong $P_4$ component and no ambiguity of mixing ratio. This lack of ambiguity ensures that the physical angular correlation must be the blue solid line included in Figure \ref{fig:Ga020_templatefit}, which obviously does not match the experimental data.

To calculate the impact of finite detector size, Compton scattering, and other effects due to $\gamma$-ray interactions with the infrastructure, a Monte Carlo simulation of the full detector setup was created using the GEANT4 framework \cite{GEANT4-2003,Svensson2013,GEANT4-2016}. The GEANT4 radioactive decay and photon evaporation classes were modified to reproduce the appropriate physical angular correlations in user-specified $\gamma$-ray cascades \cite{G4GGAngCorr}. Those correlations were then used as input to the full GRIFFIN Monte Carlo simulation \cite{detectorSIMs} of the $0^+$-- $2^+$-- $0^+$ cascade in $^{66}$Zn. A fit to the high-statistics simulated data provides a `template' which includes all the attenuation effects of the experimental setup. The template (magenta filled line) is compared to the experimental data (black filled points) in Figure \ref{fig:Ga020_templatefit}. Here, the discrete data points of the template are represented by a connected, filled area to distinguish it from the data and still indicate the simulation uncertainty. The statistical uncertainty of the simulation is represented by the width of the colored line, while the statistical uncertainty of the event-mixed angular correlation is indicated by the black error bars on each point. A comparison of the template and the input physical distribution shows the impact of the distortion due to experimental effects. Similar examples are shown in the Appendix to demonstrate the accuracy of the simulations for cascades in $^{152}$Gd and $^{60}$Ni.

With the importance of finite detector size effects and the efficacy of the GRIFFIN Monte Carlo simulation established, the following sections describe four methods for incorporating these effects and extracting physically relevant information from the $\gamma-\gamma$ angular correlation data. In all cases, we show the efficacy of the methods with comparisons to source data of the cascades described in Table \ref{table.cascades}, specifically extracting mixing ratios, $\delta$, of transitions between the initial states (of spins $J_i$) to the intermediate states (all of spin $J_x=2$). For Methods 2-4, we also extract angular correlation coefficients $a_2$ and $a_4$.

\subsection{\label{sec:sim}Method 1: Direct comparison to GEANT4 simulation templates}

One approach is to utilize the Monte Carlo simulation to construct a series of templates --- simulated distributions each with a unique set of spin and mixing ratio inputs --- that can be compared directly to data and used to determine mixing ratios. A goodness-of-fit parameter, $\chi^2$/NDF, can be calculated for the comparison of each template to the data. Following the recommendation of Ref. \cite{Robinson1990}, spins with a $\chi^2$/NDF that fall below a 99\% confidence limit are considered as possible assignments; spins that do not reach this limit are excluded. To extract the mixing ratio and its uncertainty, the $\chi^{2}$ values for each possible spin assignment are approximated by a parabola in the minimum. The minimum value of the fitted parabola determines the best-fit mixing ratio while the $1\sigma$ uncertainties are extracted from the mixing ratios which correspond to $\chi^2_{\textup{min}}+1$ of the parabolic fit.
The spin assignments and mixing ratios extracted using this method are presented in Table \ref{table:templatefitresults}. All compare favorably to the literature values for these cascades, with significant improvements in precision: the mixing ratio for the $^{152}$Gd transition increases the precision by a factor of three and the mixing ratios for the $1^+_1\rightarrow 2^+_1$ and $2^+_2\rightarrow 2^+_1$ transitions in $^{66}$Zn are roughly an order of magnitude more precise.

\begin{table}
\caption{\label{table:templatefitresults} Results from the fitting of simulated templates to data from multiple cascades (Method 1). The mixing ratios, $\delta$, of the transition from an initial spin of $J_i$ to an intermediate spin of $J=2$ were determined by a $\chi^2$ analysis comparing simulated data over a wide range of $\delta$ values.} 
\centering
\begin{tabular}{l|c|ccc}
\hline
\hline
Nucleus  & $J_i$  & $\delta_{fit}$ & $\delta_{lit.}$ & Ref.\\
\hline
$^{60}$Ni  & 4 & -0.003(2) & -0.0025(22) & \cite{Browne2013}\\
$^{152}$Gd & 3 &  0.003(2) & 0.003(6) & \cite{Martin2013}\\
\hline
$^{66}$Zn, $1^+$ & 1 & -0.082(2) & -0.09(3) & \cite{Bhat1998}\\
 &    &  & -0.12(2)  & \cite{Gade2002}\\
$^{66}$Zn, $0^+$ & 0 & - & -  & \\
$^{66}$Zn, $2^+$ & 2 & -2.08(4) & -1.9(3)  & \cite{Bhat1998}\\
   & &  & -1.6(2)  & \cite{Gade2002}\\
\hline
\hline
\end{tabular}
\end{table}

\subsection{\label{sec:Zfits}Method 2: Evolution of the angular distribution coefficients}

\begin{figure}
\centering
\includegraphics[width=\columnwidth]{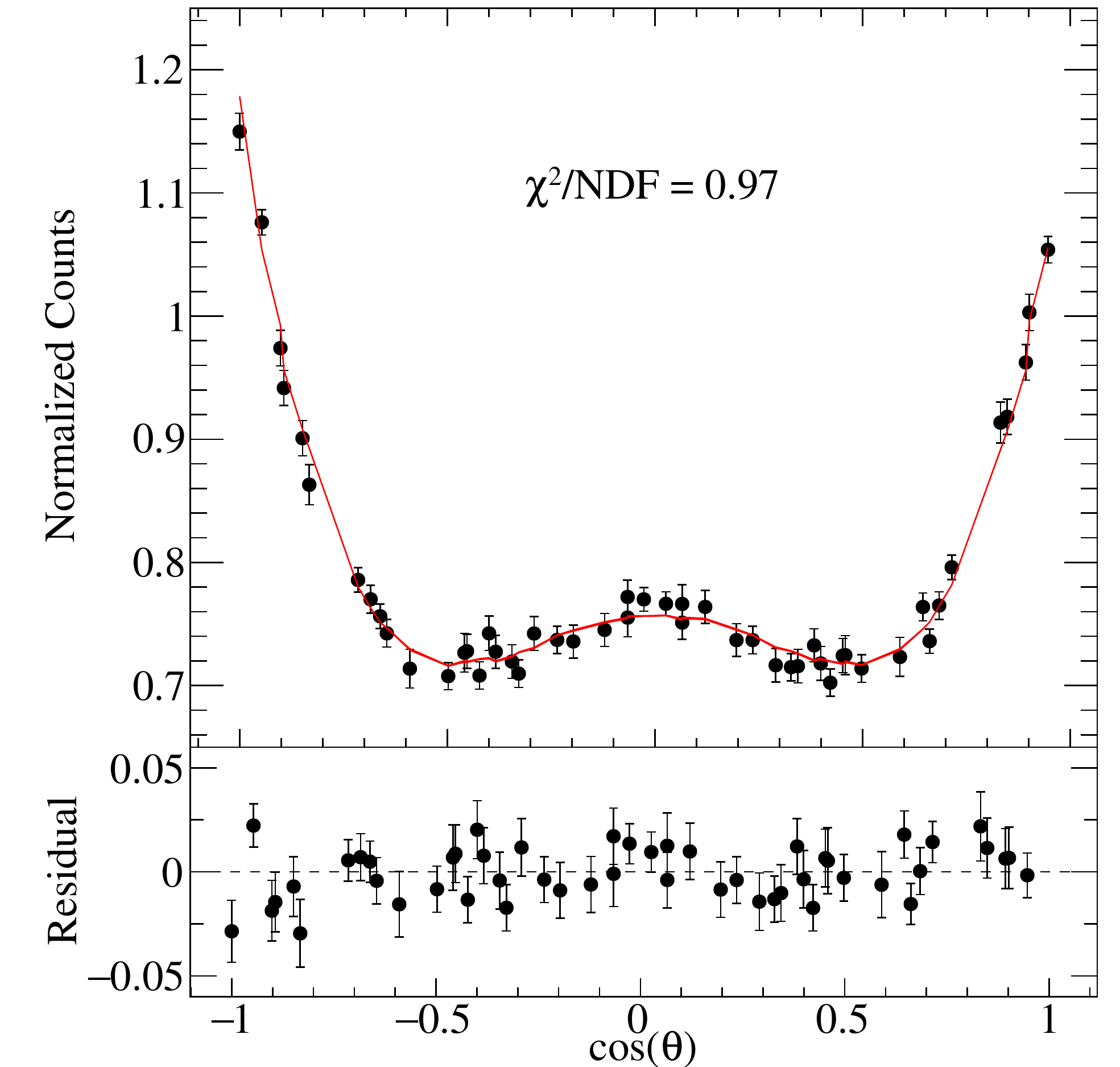}
\caption{\label{fig:Ga220_Zfit}The best Method 2 distribution fit of the $^{66}$Zn $2^+ - 2^+ - 0^+$ cascade (red filled line) and data (black points) has a $\chi^2$/NDF of 0.97 and minimizes with $a_2=0.272(7)$ and $a_4=0.258(10)$. The residual of the fit is shown in the lower panel.}
\end{figure}

The method described in the previous section is effective but time consuming as it requires a simulation to be performed for each combination of spins and mixing ratio that are to be trialled. Alternatively, one can take advantage of the fact that Equation \ref{eq:ang-corr} is simply a linear combination of Legendre polynomials. If the angular correlations of pure Legendre polynomials are simulated, the experimental data can be fitted with a linear combination of those simulated histograms. This is an efficient way to calculate the evolution of the attenuation effects across a range of angular distribution coefficients for a single combination of $\gamma$-ray energies.

While it would be simpler to simulate pure Legendre polynomials, the requirement that the angular distributions always be positive eliminates that option. Instead, the following distributions are defined and simulated individually:
\begin{align}
\label{eq.Z0}
\mathcal{Z}_0(\theta)  &= 1\\
\mathcal{Z}_2(\theta) &= 1 + P_2(\cos\theta) \\
\label{eq.Z4}
\mathcal{Z}_4(\theta) &= 1 + P_4(\cos\theta).
\end{align}
A linear combination of the resultant histograms can be used to construct an angular correlation histogram for any $a_2$, $a_4$ combination:
\begin{align} \label{AC_linear_dis}
\mathcal{Z}_{\textup{sum}}&=x\mathcal{Z}_0 + y\mathcal{Z}_2 + z\mathcal{Z}_4\\
&=A_{00}[(1 - a_2 - a_4)\mathcal{Z}_0 + a_2\mathcal{Z}_2 + a_4\mathcal{Z}_4].
\end{align}
As an example, to construct an angular correlation histogram for a $0^+$--$2^+$--$0^+$ cascade with $a_2=0.357$, $a_4=1.143$, and $A_{00}=1000$, scaling factors of $x=-500$, $y=357$, and $z=1143$ are applied.

One can leverage this linear combination to fit the $\mathcal{Z}$ distributions to experimental data and directly extract $a_2$ and $a_4$ coefficients for comparison to theory. Figure \ref{fig:Ga220_Zfit} shows the best fit of simulated $\mathcal{Z}$ distributions to the $^{66}$Zn $2^+ - 2^+ - 0^+$ angular correlation. The coefficients extracted from this fit are $a_2=0.272(7)$ and $a_4=0.258(10)$ in comparison to previously reported values of $a_2=0.30^{+0.05}_{-0.04}$ and $a_4=0.256^{+0.015}_{-0.021}$ \cite{Bhat1998}.

\begin{figure}
\centering
\includegraphics[width=\columnwidth]{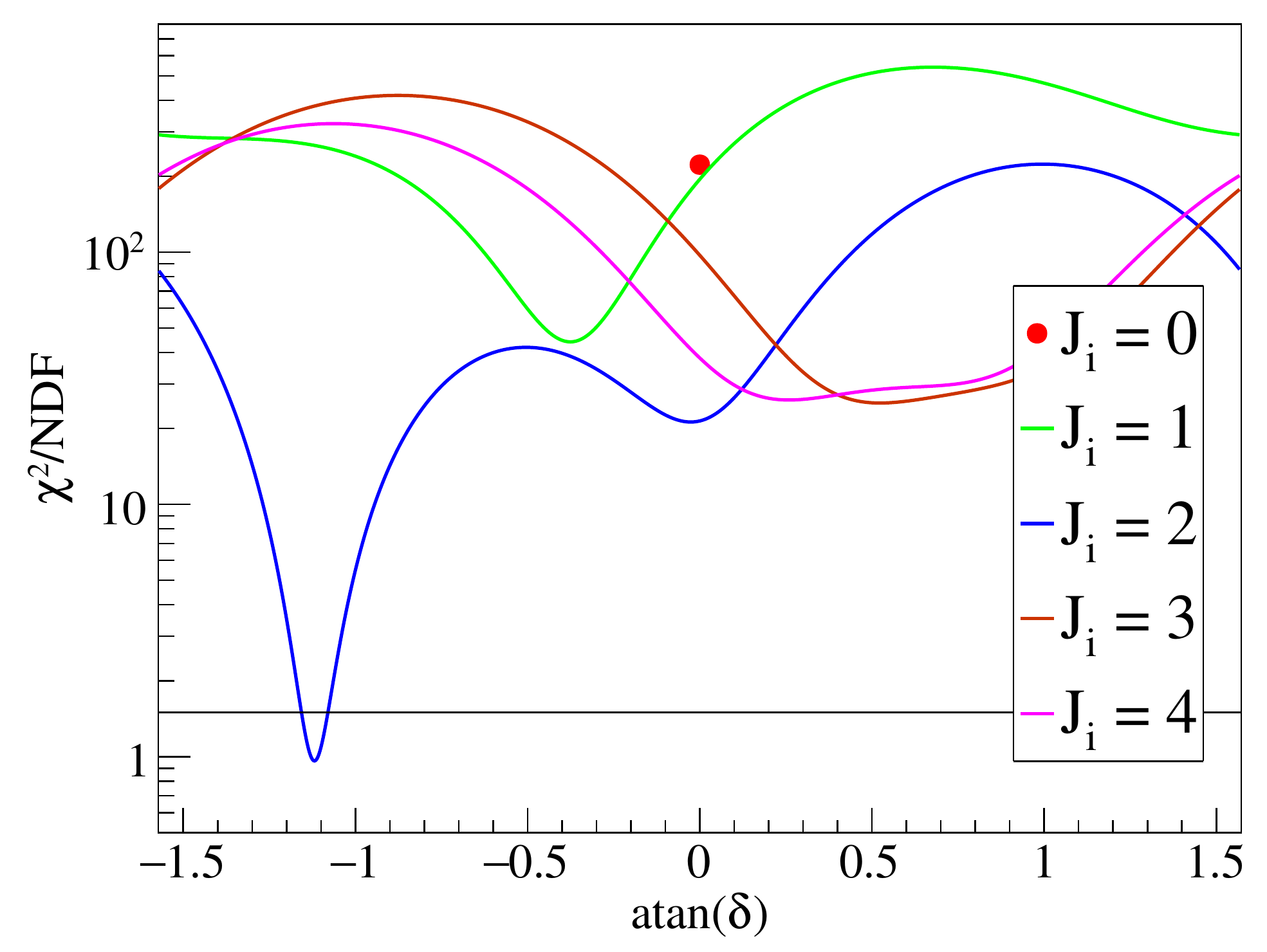}
\caption{\label{fig:Ga220_Zfit_chi2}Using Method 2, a comparison of $\chi^2$/NDF values for potential $J_i=0-4$ and all possible mixing ratios ($\delta$) shows that the best spin-mixing ratio fit to the $^{66}$Zn $2^+ - 2^+ - 0^+$ data using the $\mathcal{Z}$-distribution is made with $J_i=2$ and $\delta=-2.07(4)$. The solid black line indicates the 99\% confidence limit - any spins with goodness-of-fit values below it are considered possible assignments.}
\end{figure}

An alternative approach is to once again follow the methodology recommended by Ref. \cite{Robinson1990}: assume spins and mixing ratios, calculate theoretical $a_2$ and $a_4$ coefficients, construct an equivalent simulated histogram from a linear combination of $\mathcal{Z}$ distribution simulations, and calculate a goodness-of-fit. In a fit with 51 opening angles and two degrees of freedom for the fit (mixing ratio, and overall scaling factor), this corresponds to $\chi^2$/NDF = 1.53. Errors for each minimized mixing ratio are again found from the limits of $\chi^2_{\textup{min}}+1$. Figure \ref{fig:Ga220_Zfit_chi2} shows the $\chi^2$/NDF for such a fit to the data shown in Figure \ref{fig:Ga220_Zfit}, assuming that the intermediate state spin is $J=2$ and the final state spin is $J=0$. The black horizontal line indicates the 99\% confidence interval. The minimum $\chi^2$ is identified with an initial state of $J_i=2$ and $\delta=-2.07(4)$. All other spins are clearly rejected by this analysis. For comparison, the mixing ratio extracted by Method 1 was $-2.08(4)$, but required a much larger number of simulations to explore all possible spin combinations and $\delta$ values, compared to the Method 2 approach of simulating only the three $\mathcal{Z}_0$, $\mathcal{Z}_2$ and $\mathcal{Z}_4$ distributions.

Similar examples to demonstrate the validity of this methodology for other cascades in $^{66}$Zn, $^{152}$Gd, $^{60}$Ni are included in the Appendix. The results from all of these cascades using Method 2 are summarized in Table \ref{table:simfitresults}.

\begin{table*}
\caption{\label{table:simfitresults} Results from the use of Method 2 to fit the data. The $a_2$, $a_4$ values and $J_i$, $\delta$ values were determined independently, as described in the text. See Table \ref{table.cascades} for details of the literature values and \ref{sec:appendix} for level schemes.}
\centering
\begin{tabular}{l|cc|cc|cc|cc}
\hline
\hline
Nucleus & $a_{2,fit}$ & $a_{2,lit.}$ & $a_{4,fit}$ & $a_{4,lit.}$ & $J_{i,fit}$ & $J_{i,lit.}$ & $\delta_{fit}$ & $\delta_{lit.}$ \\
\hline
$^{60}$Ni & 0.100(2) &  0.1005(13) & 0.011(2) & 0.0094(3) & 4 & 4 & -0.003(3) & -0.0025(22) \\
 &  & &  & & 2 &  & 0.193(2) & \\
$^{152}$Gd & -0.068(2) & -0.0691(47) & -0.002(3) & 0.00000 & 3 & 3 & 0.004(2) & 0.003(6)\\
 &  &  &  &  & 4 &   & 3.29(4) & \\
\hline
$^{66}$Zn, $1^+$ & -0.156(4) & -0.147(35) & -0.003(5) & -0.0061$^{+0.0034}_{-0.0047}$ & 1 & 1 & -0.082(3) & -0.09(3) \\
 &  & -0.112(23)  &  & -0.0108$^{+0.0033}_{-0.0038}$ &  & 1 &  & -0.12(2) \\
  &  &  &  &  & 3 & & -0.108(4) & \\
   &  &  &  &  & 4 & & 5.76(14) &  \\
$^{66}$Zn, $0^+$ & 0.33(3) & 0.357 & 1.16(4) & 1.143 & 0 & 0 & - & - \\
$^{66}$Zn, $2^+$ & 0.272(7) & 0.30$^{+0.05}_{-0.04}$  & 0.258(10) & 0.256$^{+0.015}_{-0.021}$ & 2 & 2 & -2.07(4) & -1.9(3) \\
 &  &$0.34^{+0.04}_{-0.03}$  &  & $0.23(2)$  &  &2 &  & -1.6(2) \\
\hline
\hline
\end{tabular}
\end{table*}


\subsection{\label{sec:alg}Method 3: Algebraic approximation of the angular distribution coefficients}

As mentioned earlier, previous work has accounted for the effects of a given experimental setup on the angular correlation via calculated attenuation coefficients, $Q_{\ell \ell}$ \cite{Krane1972,Krane1973}. The physically relevant $a_\ell$ coefficients are obtained by correcting the bare coefficients obtained from a fit of Equation \ref{eq:ang-corr} to the experimental data with the attenuation coefficients. In this method 3, an algebraic approximation is developed and benchmarked in order to algebraically parameterize the attenuation of angular correlation coefficients ($a_2,a_4$) for a particular $\gamma-\gamma$ cascade.

In execution, this method is very similar to method 2 described in the previous section, but the $\mathcal{Z}$-distribution simulations are replaced with attenuated Legendre polynomials that approximate the attenuation of the experimental setup. This substitution allows the derivation of a direct algebraic relationship between bare coefficients, $c_{\ell}$, and the physically relevant $a_{\ell}$ coefficients. This is an essential step towards the final method considered in this work.

\begin{figure}
\centering
\includegraphics[width=\columnwidth]{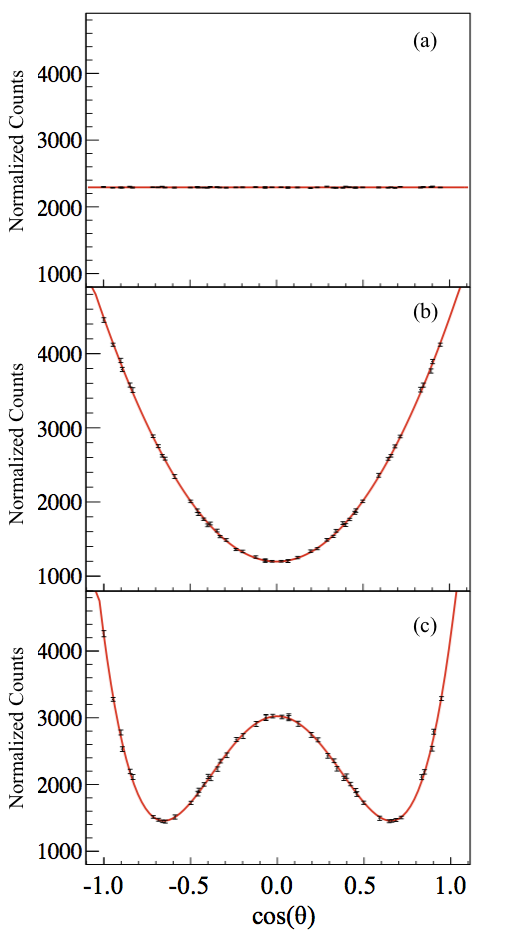}
\caption{\label{fig:Zfuncfits}Set of simulations produced which characterize the $\mathcal{Z}$ distributions described by Equations \ref{eq.L0}-\ref{eq.L4}. (a) The $\mathcal{Z}_0$ distribution and fit. (b) The $\mathcal{Z}_2$ distribution and fit. (c) The $\mathcal{Z}_4$ distribution and fit.}
\end{figure}

The individual $\mathcal{Z}$ distributions used for the fit in Figure \ref{fig:Ga220_Zfit} are shown in Figure \ref{fig:Zfuncfits}. These discrete distributions can be reasonably well approximated by functional forms of the Legendre polynomials:
\begin{align}
\label{eq.L0}
\mathcal{L}_0 &= \alpha\\
\mathcal{L}_2 &= \alpha(1 + \beta P_2) \\
\label{eq.L4}
\mathcal{L}_4 &= \alpha(1 + \gamma P_4)
\end{align}
where $\alpha$ is a common scaling coefficient and $\beta$ and $\gamma$ are coefficients to be fit to the $\mathcal{Z}_2$ and $\mathcal{Z}_4$ distributions, respectively. The form of these equations are the result of an importance truncation based on the magnitude of coefficients fitted using a complete set of Legendre polynomials up to the tenth order. The inclusion of the terms in Equations \ref{eq.L0}-\ref{eq.L4} were found to be necessary and appropriate for a good description of the data. The inclusion of additional terms produced fitted coefficients that were close to zero and returned minimal (if any) improvement in the quality of the fit. This approximation ignores small, bin-by-bin perturbations of the angular correlation in order to allow the parametrization of larger, smoothly-varying features.

An example of this fitting is shown in Figure \ref{fig:Zfuncfits}, with Equations \ref{eq.L0}-\ref{eq.L4} being used to fit the $^{66}$Zn $0^+$ -- $2^+$ -- $0^+$ $\mathcal{Z}_0$ distribution. The $\chi^2$/NDF for the three fits are 1.02, 3.73, and 5.68, indicating a good fit for the $\mathcal{Z}_0$ distribution and increasingly worse fits for the $\mathcal{Z}_{2,4}$ distributions, due principally to geometric attenuation effects not captured by the smooth polynomial. To account for this additional variation, we inflate the uncertainties on the $\beta$ and $\gamma$ coefficients by the square root of the $\chi^2/$NDF for their respective distributions.
In contrast to Methods 1 and 2, which directly compared simulated histograms to data histograms, Method 3 fits the data as a function of $\cos(\theta)$. Each data point represents a geometric arrangement of two crystals that span a finite range of angular differences. Simulations show this range to be approximately $\pm1.3^{\circ}$.  The value of $\alpha$ is dominated by the number of events used in the simulation and the coincidence-detection efficiency. The $\beta$ coefficient is 0.9557(27) and the $\gamma$ coefficient is 0.8498(54), indicating stronger attenuation of the $P_4$ component than the $P_2$ component. This larger attenuation is consistent with the idea that finite detector effects which ``smear'' out a distribution will have more impact on components that change more rapidly as a function of the opening angle. The coefficients from these fits, as well as similar fits to the other four cascades are shown in Table \ref{table:Zfitresults}.

\begin{table}
\caption{\label{table:Zfitresults}Coefficient values for the algebraic approximations (Equations \ref{eq.L0}-\ref{eq.L4}) of the $\mathcal{Z}$ distributions fitted as part of Method 3. The uncertainties in the $\beta$ and $\gamma$ coefficients have been inflated during the fitting procedure as described in the text.}
\begin{center}
\begin{tabular}{l|c|c} 
\hline
\hline
Nucleus & $\beta$ & $\gamma$  \\
\hline
 $^{60}$Ni & 0.9568(27) & 0.8450(52) \\
 $^{152}$Gd& 0.9540(30) & 0.8424(69) \\ 
\hline
$^{66}$Zn, $1^+$ & 0.9567(25) & 0.8542(47)  \\
$^{66}$Zn, $0^+$ & 0.955706(27) &  0.8498(54) \\
$^{66}$Zn, $2^+$ & 0.956265(30) & 0.8472(56) \\
\hline
\hline
\end{tabular}
\end{center}
\end{table}

\begin{table*}
\caption{\label{table:algfitresults}Results from the use of Method 3 to fit the data. The $a_2$, $a_4$ values and $J_i$, $\delta$ values were determined independently, as described in the text. See Table \ref{table.cascades} for details of the literature values and \ref{sec:appendix} for level schemes.}
\centering
\begin{tabular}{l|cc|cc|cc|cc}
\hline
\hline
Nucleus & $a_{2,fit}$ & $a_{2,lit.}$ & $a_{4,fit}$ & $a_{4,lit.}$ & $J_{i,fit}$ & $J_{i,lit.}$ & $\delta_{fit}$ & $\delta_{lit.}$ \\
\hline
$^{60}$Ni & 0.101(2) &  0.1005(13) & 0.012(3) & 0.0094(3) & 4 & 4 & -0.002(4) & -0.0025(22) \\
 &  & &  & & 2 &  & 0.195(3) & \\
$^{152}$Gd & -0.068(3) & -0.0691(47) & 0.003(4) & 0.00000 & 3 & 3 & 0.004(3) & 0.003(6)\\
 &  &  &  &  & 4 &   & 3.32(6) & \\
\hline
$^{66}$Zn, $1^+$ & -0.155(6) & -0.147(35) & -0.001(6) & -0.0061$^{+0.0034}_{-0.0047}$ & 1 & 1 & -0.082(4) & -0.09(3) \\
 &  & -0.112(23)  &  & -0.0108$^{+0.0033}_{-0.0038}$ &  & 1 &  & -0.12(2) \\
  &  &  &  &  & 3 & & -0.108(5) &  \\
  &  &  &  &  & 4 & & 5.78(19) &  \\
$^{66}$Zn, $0^+$ & 0.33(4) & 0.357 & 1.16(4) & 1.143 & 0 & 0 & - & - \\
$^{66}$Zn, $2^+$ & 0.272(12) & 0.30$^{+0.05}_{-0.04}$  & 0.257(12) & 0.256$^{+0.015}_{-0.021}$ & 2 & 2 & -2.07(5) & -1.9(3) \\
 &  & $0.34^{+0.04}_{-0.03}$ &  &$0.23(2)$   &  &2 &  & -1.6(2) \\
\hline
\hline
\end{tabular}
\end{table*}

Using the fitted $\beta$ and $\gamma$ coefficients and their respective uncertainties, an algebraic approximation can be made between the bare angular correlation coefficients $c_\ell$ that result from a direct fit of Equation \ref{eq:ang-corr} to the experimental data and the physically meaningful $a_\ell$ parameters. The $\mathcal{Z}_{\textup{sum}}$ function can be re-expressed as
\begin{equation}
\label{approx1}
\mathcal{Z}_{\textup{sum}} \approx(1 - a_2 - a_4)\mathcal{L}_0 + a_2\mathcal{L}_2 + a_4\mathcal{L}_4.
\end{equation}
The $\mathcal{Z}_{\textup{sum}}$ can also be approximated as
\begin{equation}
\label{eq.Zapprox}
\mathcal{Z}_{\textup{sum}} \approx N[1+c_2 P_2+c_4 P_4]
\end{equation}
where $c_2$ and $c_4$ are bare angular correlation coefficients (a fit of Equation \ref{eq:ang-corr} to the experimental data) and $N$ is an overall scaling factor. In comparing Equations \ref{approx1} and \ref{eq.Zapprox}, and incorporating Equations \ref{eq.L0}-\ref{eq.L4}, the $c_2$ and $c_4$ coefficients can be expressed as functions of the physically relevant $a_2$ and $a_4$ coefficients as:
\begin{align}
\label{eq.c2}
c_2 &= \beta a_2\\
\label{eq.c4}
c_4 &= \gamma a_4
\end{align}

Comparing this to previous methods of simply calculating attenuation coefficients $Q_{\ell \ell}$ for a particular detector geometry, we see that $\beta=Q_{22}$ and $\gamma=Q_{44}$.
The complex geometry of the individual GRIFFIN crystals (co-axial HPGe crystals tapered on two sizes) and their various relative orientations (which is different for each angular group) discourages a global direct calculation, but a comparison to a calculation with a simplified, approximate geometry can increase confidence in this method. A calculation of the attenuation coefficients $Q_{22}$ and $Q_{44}$ following the procedures of Refs. \cite{Krane1972,Krane1973}, and assuming axially-symmetric co-axial HPGe crystals, produces energy-dependent coefficients for coincidence energies ranging from 68-5000\,keV of $Q_{22}=0.923-0.941$ and $Q_{44}=0.761-0.814$. The values for the $\beta$ and $\gamma$ coefficients in Table \ref{table:Zfitresults} are within these ranges but can vary by as much as 5\% when comparing the coefficients calculated for specific energies to those determined in this work.

Fitting Equation \ref{eq.Zapprox} (as a continuous function) to the data allows the extraction of the $c_2$ and $c_4$ parameters which can then be related algebraically to $a_2$ and $a_4$. The error propagation for $a_2$ and $a_4$ must incorporate the errors in the $\beta$ and $\gamma$ coefficients as well as any covariance between $c_2$ and $c_4$ found in the fit of Equation \ref{eq.Zapprox}. Using standard error propagation techniques, the final covariance matrix for $a_2$ and $a_4$ is:
\begin{equation}
\begin{bmatrix}
\beta^{-4}c_{2}^{2}v_{\beta} + \beta^{-2}v_{c_2} & \left(\beta\gamma\right)^{-1}v_{c_2,c_4} \\[1.5em]
 \left(\beta\gamma\right)^{-1}v_{c_4,c_2} & \gamma^{-4}c_{4}^{2}v_{\gamma}+\gamma^{-2}v_{c_{4}}
\end{bmatrix}
\end{equation}
where $v_\beta$, $v_\gamma$, $v_{c_2}$, and $v_{c_4}$ are the variances of $\beta$, $\gamma$, $c_2$, and $c_4$, respectively, and $v_{c_2,c_4}=v_{c_4,c_2}$ is the covariance between $c_2$ and $c_4$ \cite{ISOGuide}.
The coefficients extracted from a fit to the $^{66}$Zn $2^+$ -- $2^+$ -- $0^+$ experimental data are $a_2=0.272(12)$ and $a_4=0.257(12)$. These compare well to the coefficients from the previous Method 2 of $a_2=0.272(7)$ and $a_4=0.258(10)$. Some information is lost in this approximation and the results show a modest reduction in precision. Similar $a_2$ and $a_4$ values extracted from fits to other cascades are shown in Table \ref{table:algfitresults}. 

Similarly to the previous methods, this approach can again be used to extract possible $J_i$ and $\delta$ values. In this case, a particular $J_i$ and $\delta$ value are used to calculate $a_2$ and $a_4$ coefficients, which are then used to calculate $c_2$ and $c_4$ coefficients that specify a particular polynomial that can be fit to the data. At each point, uncertainty from the data is added in quadrature with an uncertainty from the theory, propagated from uncertainties in the $\beta$ and $\gamma$ coefficients. The goodness-of-fit metric $\chi^2$/NDF is then extracted and minimized, producing a figure that is essentially identical to Figure \ref{fig:Ga220_Zfit_chi2}. The minimum $\chi^2$ for the $2^+$ -- $2^+$ -- $0^+$ cascade in $^{66}$Zn is identified with an initial state of $J_i=2$ and $\delta=-2.07(5)$. All other spins are again rejected by this analysis. For comparison, the mixing ratio extracted by Method 1 was $-2.08(4)$ and the mixing ratio extracted by Method 2 was $-2.07(4)$. The results of these fits are shown in Table \ref{table:algfitresults}.

\subsection{\label{sec:alp-bet-gam}Method 4: Parameterization of the \texorpdfstring{$\gamma$}{\textgamma}-ray-energy dependence}
\par In a $\gamma$--$\gamma$ cascade, the energy dependence for each of the $\beta$ and $\gamma$ coefficients is given by the product of two attenuation factors, each describing the energy dependence of the experimental detector setup~\cite{Krane1972} for the energies involved in the cascade. If the evolution of the $\beta$ and $\gamma$ coefficients as a function of the cascade $\gamma$-ray energies can be characterized, then there is no need to perform separate simulations for cascades involving different energies.
In order to explore this behavior, the $\mathcal{Z}$ distributions defined in Eqs.~\ref{eq.Z0}--\ref{eq.Z4} were simulated for a series of cascades involving a range of energy pairs $\left(E_{1},E_{2}\right)$. Values of the $\beta$ and $\gamma$ coefficients, and their associated uncertainties were determined following a global fit of Eqs.~\ref{eq.L0}--\ref{eq.L4} to the $\mathcal{Z}$ distributions and are given in Table~\ref{table:coeffs-pureE2E2}.

\par Sets of $\left(E_{1},E_{2},\beta\right)$ and $\left(E_{1},E_{2},\gamma\right)$ points define two surfaces. The energy dependence of the attenuation coefficients was assumed to be identical for each crystal, and the surfaces were fit using the product of two one-dimensional functions
\begin{align}
\label{eq.surfaceFitf}
f(E_{1},E_{2}) = g(E_{1})\times g(E_{2}),
\end{align}
where the function $g(E)$ describes the energy-dependent $\gamma$-ray attenuation of a single GRIFFIN crystal and is given by
\begin{align}
\label{eq.oneDimEneDependence}
\begin{split}
g(E) = C &+ \frac{A}{1+e^{-\lambda_{s}(E-E_{0})}} \\
	&+B\left(1-e^{-\lambda_{e}(E-E_{0})}\right).
\end{split}
\end{align}
The parameter $B\equiv 0$ for $E<E_{0}$, where $E_{0}$ is the point at which the sigmoidal portion of Eq.~\ref{eq.oneDimEneDependence} reaches half its maximum amplitude. The form of Eq.~\ref{eq.oneDimEneDependence} was chosen to satisfy two requirements. Firstly, it has well-defined limiting behavior when the $\gamma$-ray energy is very small or very large, and secondly, it approximates the previously-calculated energy dependence of a single HPGe crystal in the case of simple axially-symmetric co-axial detector geometries, for example in Ref.~\cite{Krane1972}. 

\begin{figure}
\centering
\includegraphics[width=\linewidth]{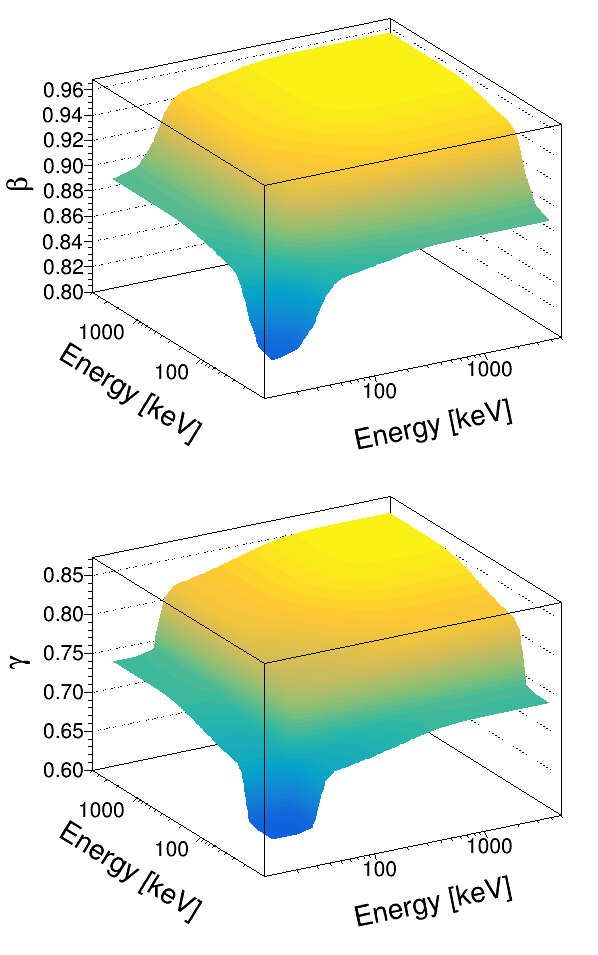}
\caption{\label{fig:Surface}Behaviour of the $\beta$ and $\gamma$ coefficients as a function of the two $\gamma$-ray energies involved in the cascade.}
\end{figure}

\begin{figure}
\centering
\includegraphics[width=\linewidth]{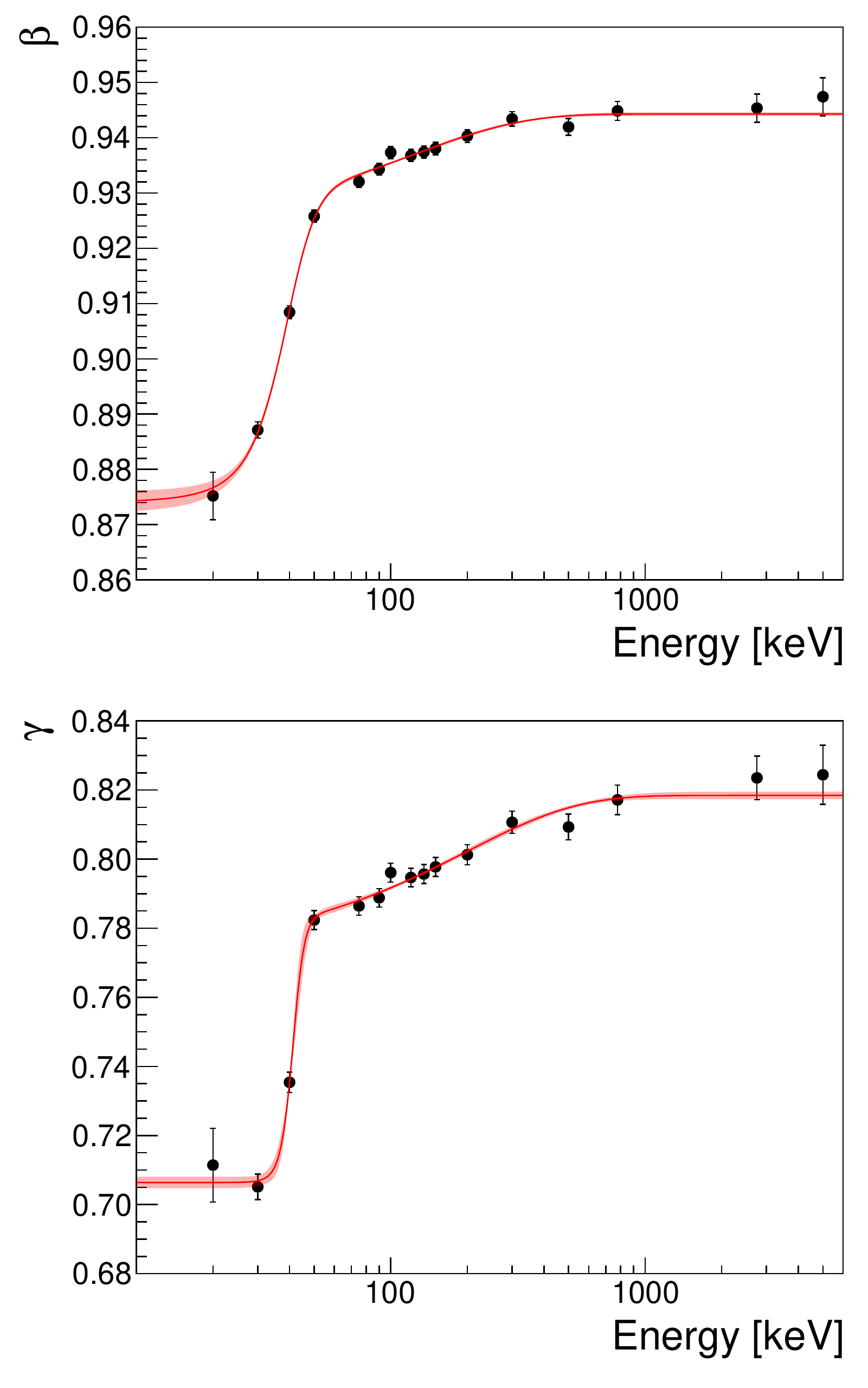}
\caption{\label{fig:Projections}Projection of the $\beta$ and $\gamma$ coefficients gated on the 68\,keV cascade as a function of the other $\gamma$-ray energy in coincidence in the cascade.}
\end{figure}

Best-fit surfaces of the $\beta$ and $\gamma$ coefficients to the data in Table~\ref{table:coeffs-pureE2E2} are shown in Fig.~\ref{fig:Surface}. The uncertainties given in Table~\ref{table:coeffs-pureE2E2} have been inflated so that $\chi^{2}/\text{NDF}=1$ for the best-fit surfaces. These uncertainty inflation factors are 4.49 and 8.62 for the $\beta$ and $\gamma$ surfaces, respectively. Method-4-measured uncertainties on the $\beta$ and $\gamma$ coefficients at the $1\sigma$ level for a given $(E_{1},E_{2})$ pair are determined from the 68.3\% confidence interval of the best-fit surface given by Eq.~\ref{eq.surfaceFitf}, which is calculated using standard techniques available under the ROOT data analysis framework~\cite{ROOT}. One-dimensional projections of simulated $\beta$ and $\gamma$ coefficient values, including only cascades involving the 68\,keV $\gamma$ ray, are shown on Fig.~\ref{fig:Projections}. These figures show projections of the fit to Eq.~\ref{eq.surfaceFitf} along with projections of the corresponding confidence intervals and demonstrate the energy dependence of the $\beta$ and $\gamma$ coefficients. 

Angular correlation coefficients $a_{2}$ and $a_{4}$ as well as mixing ratios $\delta$ for Method 4 are determined using the same error propagation procedure described for Method 3 in Section~\ref{sec:alg}. Examples for the cascades in $^{66}$Zn, $^{152}$Gd, $^{60}$Ni are included in \ref{sec:appendix}. The results generated using Method 4 are presented in Table \ref{table:methodfourresults}.

\begin{table*}
\caption{\label{table:methodfourresults}Results from the use of Method 4 to fit the data. The $a_2$, $a_4$ values and $J_i$, $\delta$ values were determined independently, as described in the text. See Table \ref{table.cascades} for details of the literature values and \ref{sec:appendix} for level schemes.}
\centering
\begin{tabular}{l|cc|cc|cc|cc}
\hline
\hline
Nucleus & $a_{2,fit}$ & $a_{2,lit.}$ & $a_{4,fit}$ & $a_{4,lit.}$ & $J_{i,fit}$ & $J_{i,lit.}$ & $\delta_{fit}$ & $\delta_{lit.}$ \\
\hline
$^{60}$Ni & 0.101(3) &  0.1005(13) & 0.012(3) & 0.0094(3) & 4 & 4 & -0.002(4) & -0.0025(22) \\
 &  & &  & & 2 &  & 0.195(3) & \\
$^{152}$Gd & -0.069(3) & -0.0691(47) & 0.003(4) & 0.00000 & 3 & 3 &  0.004(3)& 0.003(6)\\
 &  &  &  &  & 4 &   & 3.34(6) & \\
\hline
$^{66}$Zn, $1^+$ & -0.155(6) & -0.147(35) & -0.001(6) & -0.0061$^{+0.0034}_{-0.0047}$ & 1 & 1 & -0.083(4) & -0.09(3) \\
 &  & -0.112(23)  &  & -0.0108$^{+0.0033}_{-0.0038}$ &  & 1 &  & -0.12(2) \\
  &  &  &  &  & 3 & & -0.107(5) &  \\
  &  &  &  &  & 4 & & 5.78(19) &  \\
$^{66}$Zn, $0^+$ & 0.33(4) & 0.357 & 1.16(4) & 1.143 & 0 & 0 & - & - \\
$^{66}$Zn, $2^+$ & 0.271(12) & 0.30$^{+0.05}_{-0.04}$ & 0.256(12) & 0.256$^{+0.015}_{-0.021}$ & 2 & 2 & -2.07(5) & -1.9(3) \\
 &  & $0.34^{+0.04}_{-0.03}$ &  & $0.23(2)$  &  &2 &  & -1.6(2) \\
\hline
\hline
\end{tabular}
\end{table*}

\subsection{\label{sec:Methods}Discussion of the methods}
In this article four methods are presented for the correction of the attenuation effects of finite detector size in order to extract accurate $\gamma-\gamma$ angular correlation coefficients. The full 51 unique angles available in the geometry of the spectrometer are utilized without further grouping, binning or folding. In all methods the experimental data is first processed with an event-mixing technique (Section \ref{sec:eventmix}) to correct for any detector-, $\gamma$-ray-energy-, and time-dependent efficiency variations between individual crystals in the array. The angular correlation coefficients and multipole mixing ratio determined using each method for five cascades are compared in Table \ref{tab:comparison}.
\begin{table*}
\begin{minipage}{\textwidth}
\caption{\label{tab:comparison}Comparison of the angular correlation coefficients ($a_2,a_4$) and multipole mixing ratios ($\delta$) for the initial to intermediate transitions determined with the four methods presented in this work.}
\begin{center}
\begin{tabular}{
 @{}
 l
 l
 *{5}{S[table-format=-2.6]}
 @{}
}
\hline
\hline
& & {$^{60}$Ni} & {$^{152}$Gd} & {$^{66}$Zn, $1^+$} & {$^{66}$Zn, $0^+$} & {$^{66}$Zn, $2^+$}\\
\hline
$a_2$ & Literature: &  0.1005(13) & -0.0691(47) & -0.147(35) & 0.357 & {0.30$^{+0.05}_{-0.04}$}\\
& & & & -0.112(23) & & {0.34$^{+0.04}_{-0.03}$}\\
&Method 1:\footnote{$^{1,2}$ The values of the $a_2$ and $a_4$ coefficients are not measured directly in Method 1 but can be calculated by propagating the experimentally measured value and uncertainty for the mixing ratio, $\delta$.}\footnotemark & 0.1002(15) & -0.069(2) & -0.156(3) &  & 0.271(6) \\
&Method 2: & 0.100(2) & -0.068(3) & -0.156(4) & 0.33(3) & 0.272(7)\\
&Method 3:& 0.101(2) & -0.068(3) & -0.155(6) & 0.33(4) & 0.272(12)\\
&Method 4: & 0.101(3) & -0.069(3) & -0.155(6) & 0.33(4) & 0.271(12)\\
&Raw fit: & 0.096(2)  & -0.057(2) & -0.148(4) & 0.31(3) & 0.260(8)\\
\hline
$a_4$ &Literature: & 0.0094(3) & 0 & {-0.0061$^{+0.0034}_{-0.0047}$} & 1.143 & {0.256$^{+0.015}_{-0.021}$}\\ 
& & & & {-0.0108$^{+0.0033}_{-0.0038}$} & & 0.23(2)\\ 
&Method 1:\textcolor{blue}{$^{a}$}\footnotemark{} & 0.0094(3) & 0.000(0) & -0.0051(3) &  & 0.265(2) \\
&Method 2:  & 0.011(2) & 0.002(3) & -0.003(5) & 1.16(4) & 0.258(10)\\ 
&Method 3:  & 0.012(3) & 0.003(4) & -0.001(6) & 1.16(4) & 0.257(12)\\ 
&Method 4:  & 0.012(3) & 0.003(4) & -0.001(6) & 1.16(4) & 0.256(12)\\ 
&Raw fit:  & 0.010(2) & 0.003(3) & -0.001(5) & 0.98(3) & 0.218(10)\\ 
\hline
$\delta$ & Literature:  & -0.0025(22) & 0.003(6) & -0.09(3) & {-} & -1.9(3)\\
  &&  &  & -0.12(2) & {-} & -1.6(2)\\
&Method 1: & -0.0031(24) & 0.0032(23) & -0.0822(24) & {-}  & -2.078(42)\\
&Method 2:  & -0.0028(25) & 0.0036(23) & -0.0819(27) & {-}  & -2.073(39)\\
&Method 3:  & -0.0023(35) & 0.0038(27) & -0.0828(36) & {-}  & -2.072(46)\\
&Method 4:  & -0.0023(35) & 0.0039(27) & -0.0828(36) & {-}  & -2.073(46)\\
&Raw fit:  & -0.0101(32) & 0.0161(25) & -0.0884(24) & {-}  & -2.113(44)  \\
\hline
\hline 
\end{tabular}
\end{center}
\end{minipage}
\end{table*}

The first method (Method 1, Section \ref{sec:sim}) involves comparing the experimental data to a series of simulated templates representing different spin and mixing ratio combinations for that particular $\gamma-\gamma$ cascade. The simulation of the experimental setup captures all the effects that act to attenuate the experimental angular correlation with respect to the theoretical angular correlation.

This is in many respects a brute-force approach and up to forty simulation templates may be needed to assign the spin and identify the best-fit mixing ratio of a transition if the spins in the cascade are not known ahead of time. Typically, a minimum of fifteen templates are required if the spins are already known.

In the second method (Method 2, Section \ref{sec:Zfits}) the evolution of the attenuation effects across all possible values of the angular distribution coefficients ($a_2,a_4$) is reproduced for a particular $\gamma-\gamma$ cascade with a linear combination of three independent simulations.
This linear combination is then fitted to the experimental data. In this method only three simulations are required for each $\gamma-\gamma$ cascade, one for each of the $\mathcal{Z}_{0}$, $\mathcal{Z}_{2}$ and $\mathcal{Z}_{4}$ distributions (Eqs. \ref{eq.Z0}---\ref{eq.Z4}). This dramatically reduces the computational investment required for making each measurement without sacrificing precision. A distinct set of simulations must be generated for each particular combination of $\gamma$-ray energies or for a different experimental setup as the evolution of the effects due to these factors is not yet taken into account.

In the third method (Method 3, Section \ref{sec:alg}) an algebraic approximation is developed such that this evolution is algebraically parameterized across all possible values of the angular distribution coefficients for a particular $\gamma-\gamma$ cascade. In practice this method is very close to the procedure of Method 2 but each $\mathcal{Z}$-distribution simulation is replaced with an algebraic approximation using $\alpha,\beta,\gamma$ parameters. This approximation ignores some of the detailed geometric effects on attenuation and introduces a modest reduction in the precision of the results but is an essential step in the development of Method 4.

A fourth method (Method 4, Section \ref{sec:alp-bet-gam}) is to further parameterize the behavior of the coefficients of the algebraic approximation ($\beta$ and $\gamma$) as a function of the two $\gamma$-ray energies. Once this dependence is known, no further simulations are required for additional angular correlation measurements made with the same experimental setup. This allows simple corrections to be applied to the bare coefficients obtained from fitting the experimental data with Equation \ref{eq:ang-corr} in order to convert them directly to the true unattenuated angular distribution coefficients for comparison with theory.

\begin{figure}
\centering
\includegraphics[width=\columnwidth]{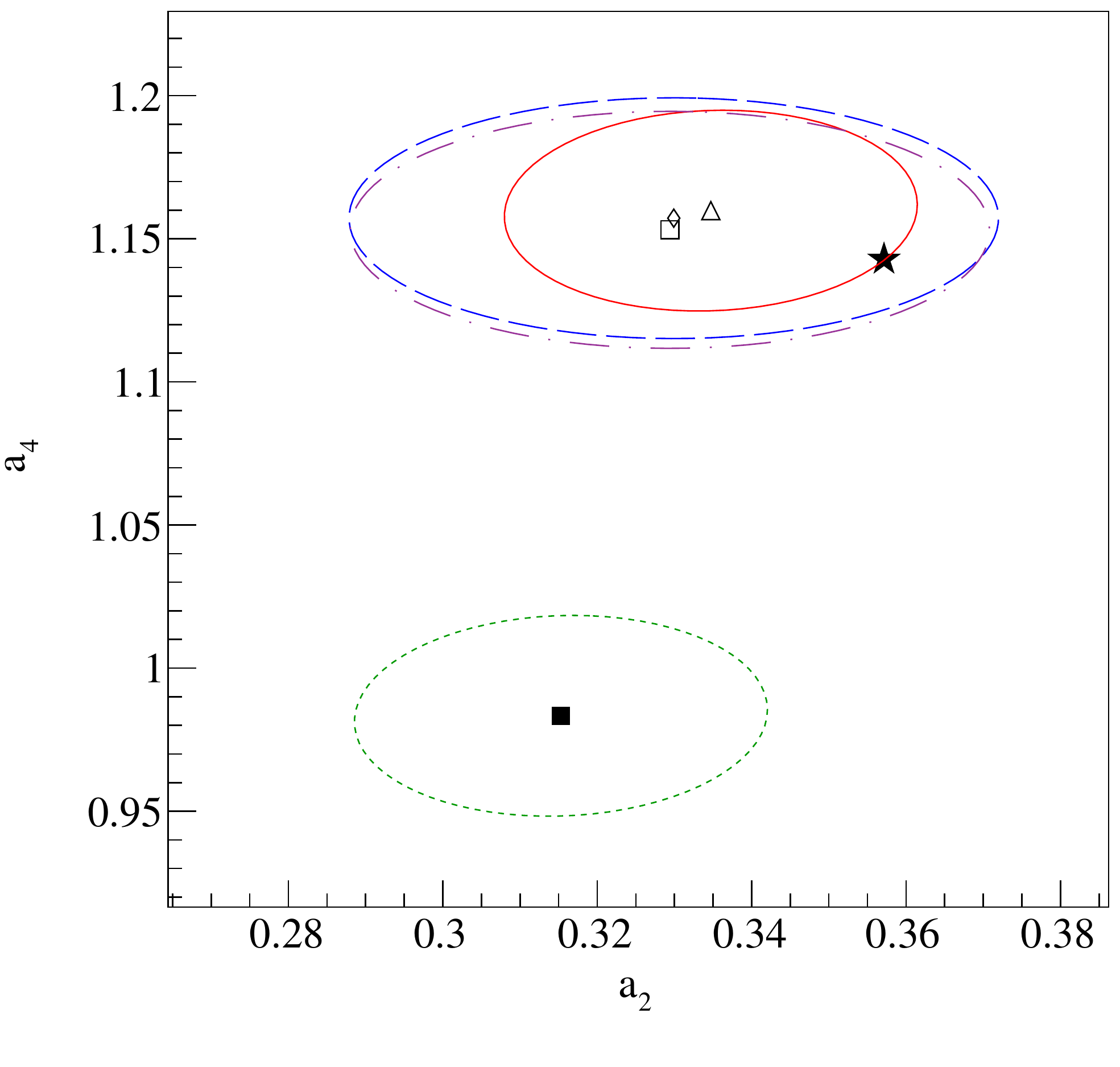}
\caption{\label{fig:Ga020_ellipses}A comparison of minimized $a_2$ and $a_4$ values extracted from fitting the $^{66}$Zn $0^+ - 2^+ - 0^+$ data with methods described in the paper. The expected $a_2$ and $a_4$ values are indicated by the star. Minimized $a_2$, $a_4$ values and $1\sigma$ confidence intervals are shown for a bare fit of Equation \ref{eq:ang-corr} (filled square, green dotted ellipse), Method 2 (open triangle, red solid ellipse), Method 3 (open diamond, blue dashed ellipse), and Method 4 (open square, purple dot-dashed ellipse).}
\end{figure}


Figure \ref{fig:Ga020_ellipses} compares the extraction of $a_2$ and $a_4$ using Methods 2, 3, and 4 with a naive fit of the data and the literature value for the $^{66}$Zn $0^+ - 2^+ - 0^+$ cascade. For this set of spins, both transitions are pure $E2$ multipolarity (both mixing ratios must be zero) and thus angular correlation coefficients are well-defined. The bare fit (filled square, green dotted error ellipse) is significantly different from the theoretical value (filled star). All other methods used here are a significant improvement in the accuracy of the angular correlation coefficients. All methods 2, 3 and 4 (open triangle with the red solid error ellipse, open diamond with the blue dashed error ellipse and open square with the purple dot-dashed error ellipse, respectively) agree within $1\sigma$ with the theoretical value. Similar figures for the $^{152}$Gd, $^{60}$Ni, and other $^{66}$Zn cascades are included in \ref{sec:appendix} and Table \ref{tab:comparison} shows a comparison of $a_2,a_4$ values for all methods.

The results for the mixing ratios shown in Table \ref{tab:comparison} indicate that Methods 1-4 produce results which are consistent with the literature values. 
In contrast, bare fits of the data produce mixing ratios that are inconsistent with the literature values for 3 out of the 4 cascades measured (Table \ref{tab:comparison}). Without accounting for the finite detector size attenuation effects on the angular correlation, the allowed spins may also be susceptible in some cases to having an incorrect assignment. 

The computational effort required for each method is significantly different. As an example, we consider a GRIFFIN experiment to measure five multipole mixing ratios in five $\gamma-\gamma$ cascades where all spins are known. A comparison between the methods is made in Table \ref{tab:computations}. Method 1 requires a very large amount of computations as it is necessary to run at least 15 simulations for each $\gamma-\gamma$ cascade. Methods 2 and 3 both require 3 simulations per $\gamma-\gamma$ cascade. Method 4 requires an initial set of simulations (in this work 72 were used) to be made to map the evolution of the attenuation effects as a function of $\gamma$-ray energy. However, once this is completed for a given experimental setup, it is not necessary to perform further simulations for future measurements with the same setup. Finally, here we have truncated Equation \ref{eq.W_theory} after the $a_4$ coefficient, but cascades with higher intermediate spins will need to retain more coefficients. In general, one will need to retain coefficients up to order $2J_x$ where $J_x$ is the spin of the intermediate state. These additional terms will linearly increase the number of simulations needed for Methods 2 and 4, but not Method 1.

\begin{table}
\caption{\label{tab:computations}Comparison of the computational investment required to measure five multipole mixing ratios in five $\gamma-\gamma$ cascades of known spins. Each individual GEANT4 simulation requires 45~CPU~days to complete. Values with an asterisk are required only to initially characterize the energy-dependence of a particular experimental setup but require no simulations for subsequent measurements.}
\centering
\begin{tabular}{ccc}
\hline
\hline
Method & Number of & Computational time \\
 & simulations & (CPU days) \\
\hline
1 & 75 & 3375 \\
2 & 15 & 675 \\
4 & 72*  & 3240* \\
\hline
\hline
\end{tabular}
\end{table}

\section{\label{sec:sum}Summary}
In this work a series of methods are presented for the extraction of physical coefficients from $\gamma-\gamma$ angular correlation data making use of the full granularity and angular coverage of the GRIFFIN spectrometer. The use of event-mixed histograms allows for the systematic elimination of time- and energy-dependent relative efficiency variations between individual detectors. Using a Monte Carlo simulation within the GEANT4 framework, a set of three simulated $\mathcal{Z}$ distributions can be used to correct for the finite detector size effects on an arbitrary angular correlation. The best-fit values of $a_2$, $a_4$, as well as spins and mixing ratios, can be extracted by fitting a linear combination of these simulated distributions. By ignoring detailed geometric attenuation effect, an algebraic approximation of the $\mathcal{Z}$ distributions can be used to extract the same quantities. Finally, by characterizing the $\beta$ and $\gamma$ parameters for a particular experimental setup but a wide range of energies, future measurements with the same setup require no additional simulations to extract the same quantities, with only slight reductions in accuracy and precision.

Future work will expand the application of these techniques. The event-mixing strategy can be (and has been) applied to other types of measurements at other facilities, though care should first be taken to confirm its functionality in new applications (i.e. different data acquisition triggers, different initial nuclear alignment). Angular correlations can be constructed with data in different ways such as using the ``addback'' of $\gamma$ rays that Compton scatter from one crystal to another in the same clover, having the angular correlation folded about the symmetric axis at $\theta=90^\circ$, or having similar angles grouped together in order to improve the statistics within a single angular bin. Initial tests have shown that the methods discussed in this article will be applicable in all of these situations, with appropriate modification of the simulated distributions. The extension of these methods to utilize Maximum Likelihood approaches for the treatment of low-statistics experimental datasets will also be considered.

Finally, it is worth reiterating that the methods presented in this work are not specific to the GRIFFIN spectrometer but can be applied to any complex detector array for which a detailed GEANT4 simulation has been developed.

\section{Acknowledgements}
The authors would like to thank G.C.~Ball, G.~Hackman, and K.~Starosta for useful discussions. The GRIFFIN infrastructure has been funded jointly by the Canada Foundation for Innovation, TRIUMF and the University of Guelph. TRIUMF receives funding through a contribution agreement through the National Research Council Canada. C.E.S. acknowledges support from the Canada Research Chairs program. This work was supported by the Natural Sciences and Engineering Research Council of Canada.

\appendix

\section{Additional examples}
\label{sec:appendix}

Examples illustrating and comparing the results obtained with the different analysis methods presented in the main text are provided here. Experimental details of the cascades can be found in Table \ref{table.cascades}. Also provided are the results from fitting simulated $\mathcal{Z}$ distributions for a wide range of energies used for the parameterization of Method 4 (Table \ref{table:coeffs-pureE2E2}).

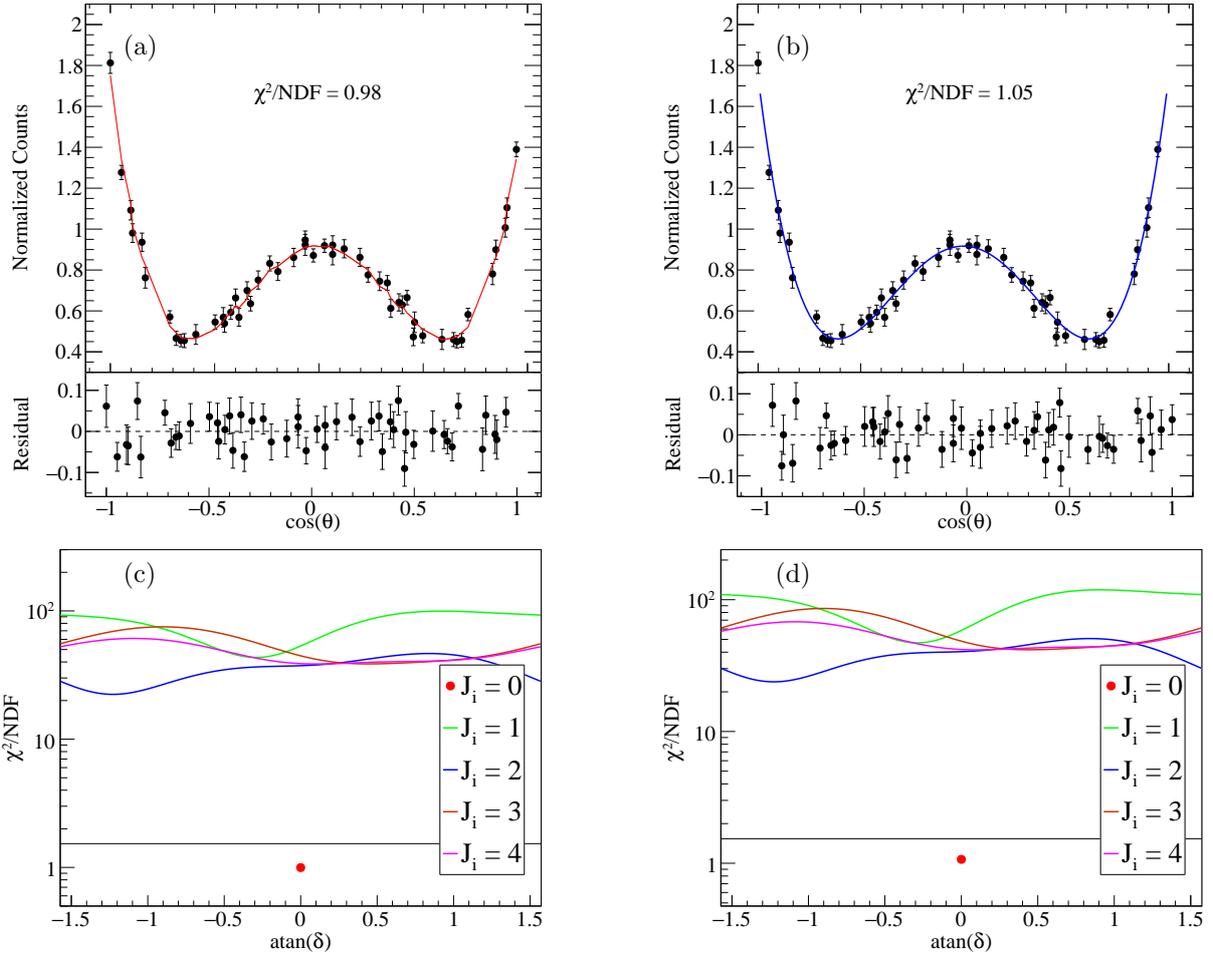
\begin{figure*}
\begin{picture}(100,200)
     \subfloat{%
      \put(0,0){\includegraphics[width=0.45\textwidth]{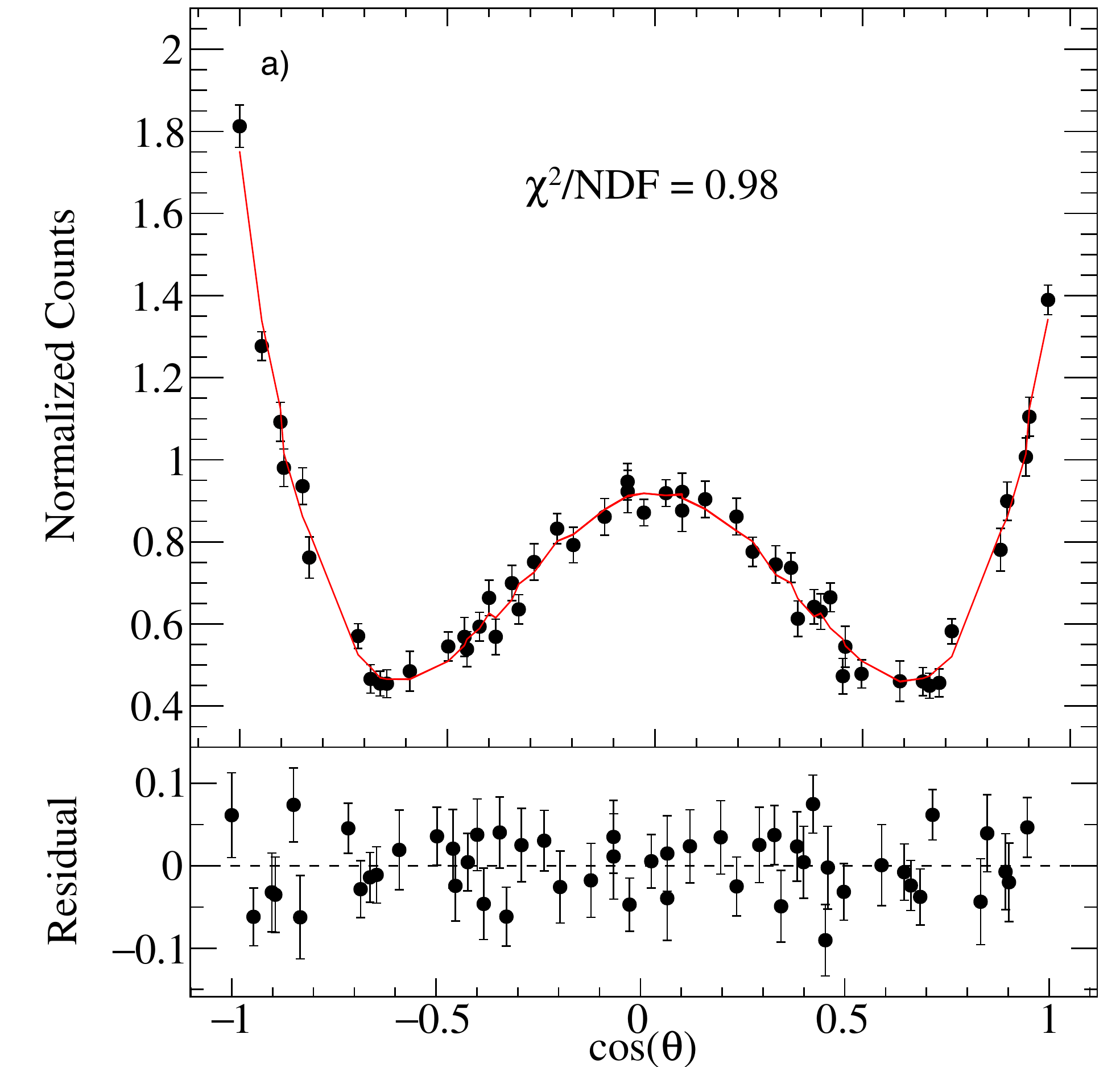}}
\put(50,180){(a)}
}
\end{picture}
     \hfill
\begin{picture}(224,200)
     \subfloat{%
     \put(0,0){\includegraphics[width=0.45\textwidth]{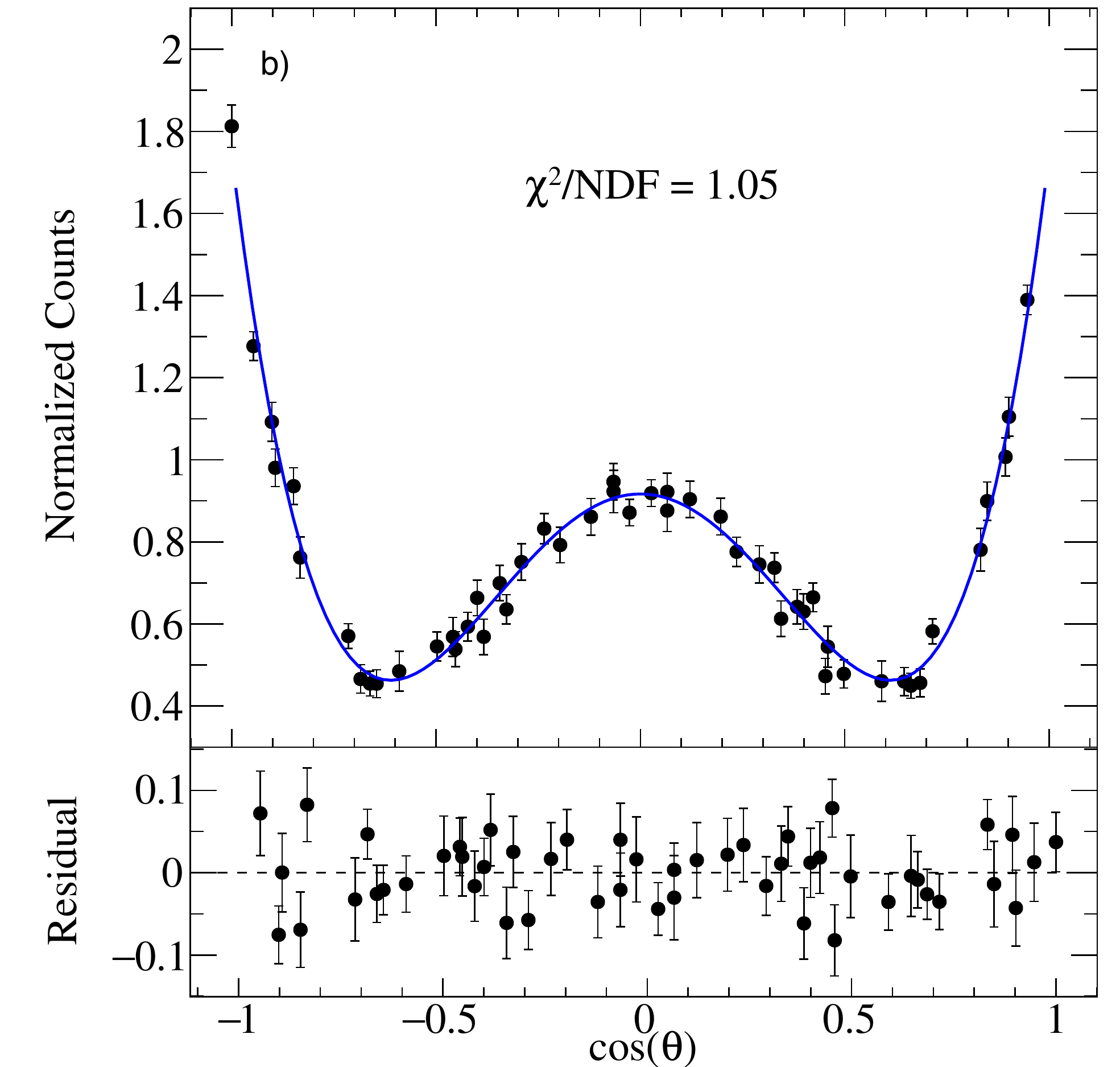}}
     \put(50,180){(b)}
     }
\end{picture}
     \\
 \begin{picture}(100,160)
     \subfloat{%
       \put(0,0){\includegraphics[width=0.45\textwidth]{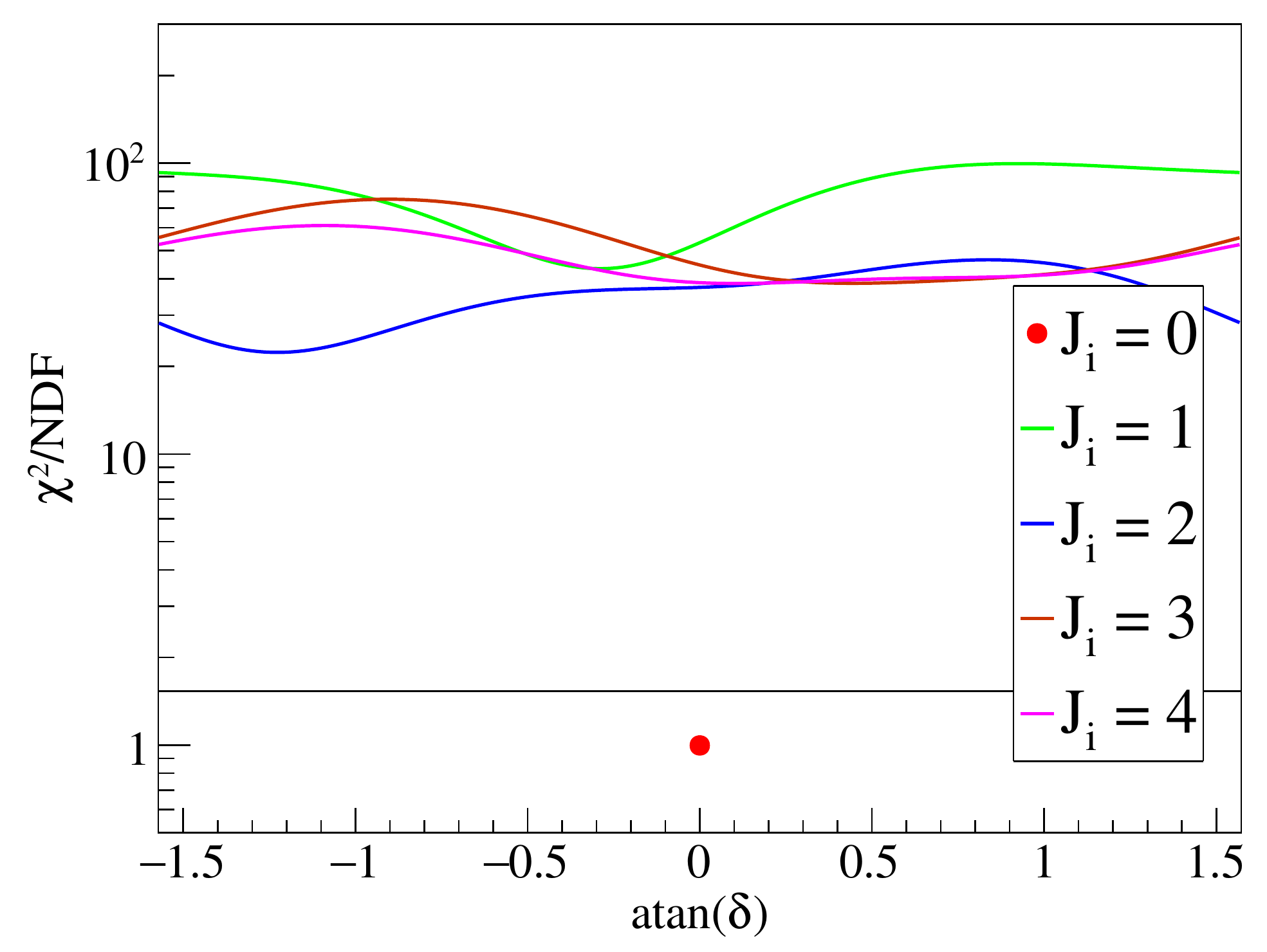}}
       \put(50,142){(c)}
     }
 \end{picture}
     \hfill
\begin{picture}(224,160)
     \subfloat{%
      \put(0,0){ \includegraphics[width=0.45\textwidth]{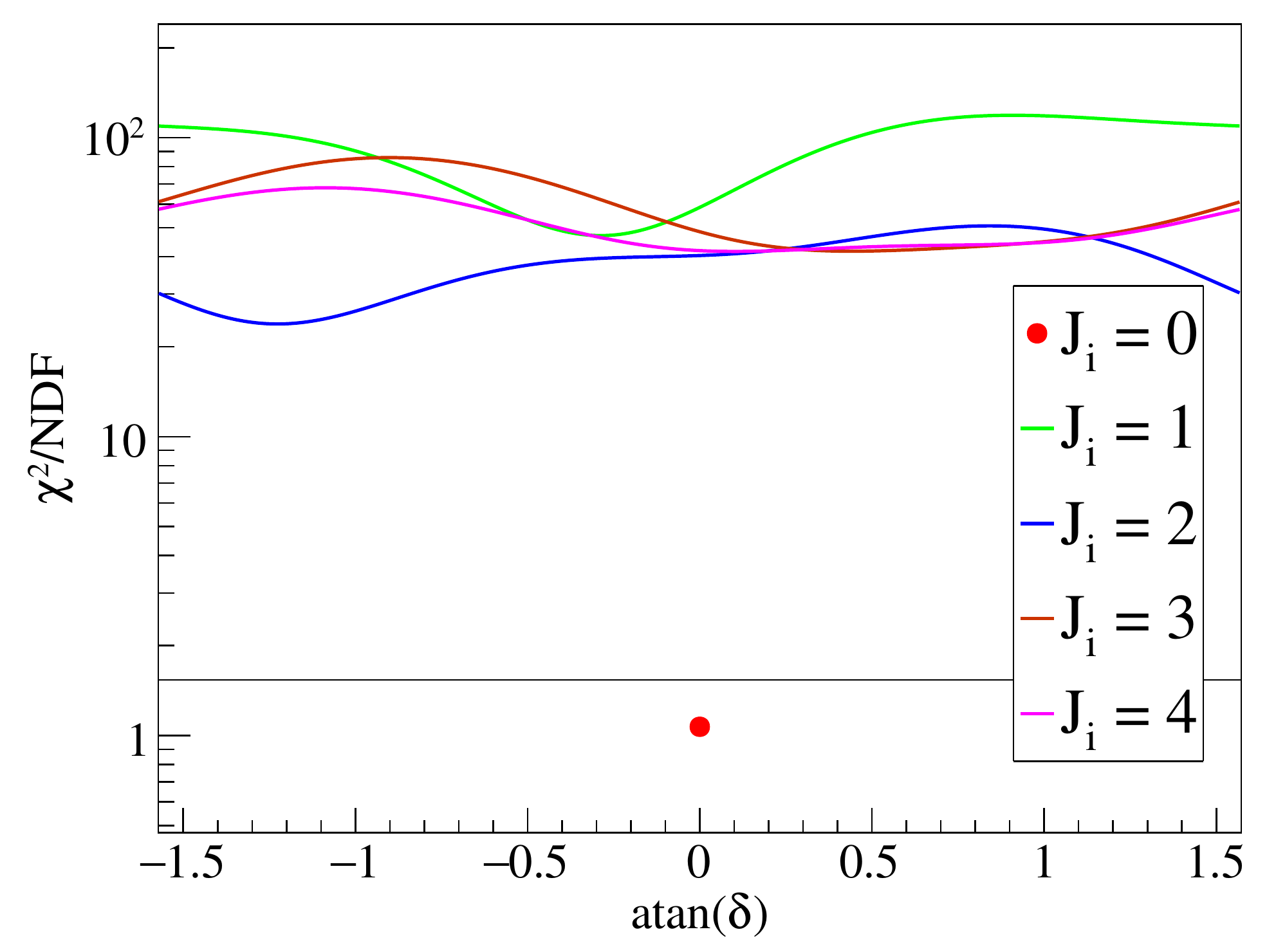}}
      \put(50,142){(d)}
     }
 \end{picture}
     \caption{Plots for the $^{66}$Zn $0^+ - 2^+ - 0^+$ 1333.1 keV - 1039.2 keV cascade. (a) The best Method 2 fit of the $^{66}$Zn $0^+ - 2^+ - 0^+$ cascade (red filled line) and data (black points) has a $\chi^2$/NDF of 0.98 and minimizes with $a_2=0.33(3)$ and $a_4=1.16(4)$. The residual of the fit is shown in the lower panel. (b) The best Method 3 fit of the $^{66}$Zn $0^+ - 2^+ - 0^+$ cascade (blue line) and data (black points) has a $\chi^2$/NDF of 1.05 and minimizes with $a_2=0.33(4)$ and $a_4=1.16(4)$. The residual of the fit is shown in the lower panel. (c) A comparison of $\chi^2$/NDF values for potential $J_i=0-4$ and all possible mixing ratios ($\delta$) shows that the best fit to the $^{66}$Zn $0^+ - 2^+ - 0^+$ data using Method 2 is made with $J=0$. (d) A comparison of $\chi^2$/NDF values for potential $J_i=0-4$ and all possible mixing ratios ($\delta$) shows that the best fit to the $^{66}$Zn $0^+ - 2^+ - 0^+$ data using Method 4 is made with $J=0$.}
     \label{fig:Ga020-4Panel-Fig1}
\end{figure*}

\begin{figure*}
\begin{picture}(100,200)
     \subfloat{%
       \put(0,0){\includegraphics[width=0.45\textwidth]{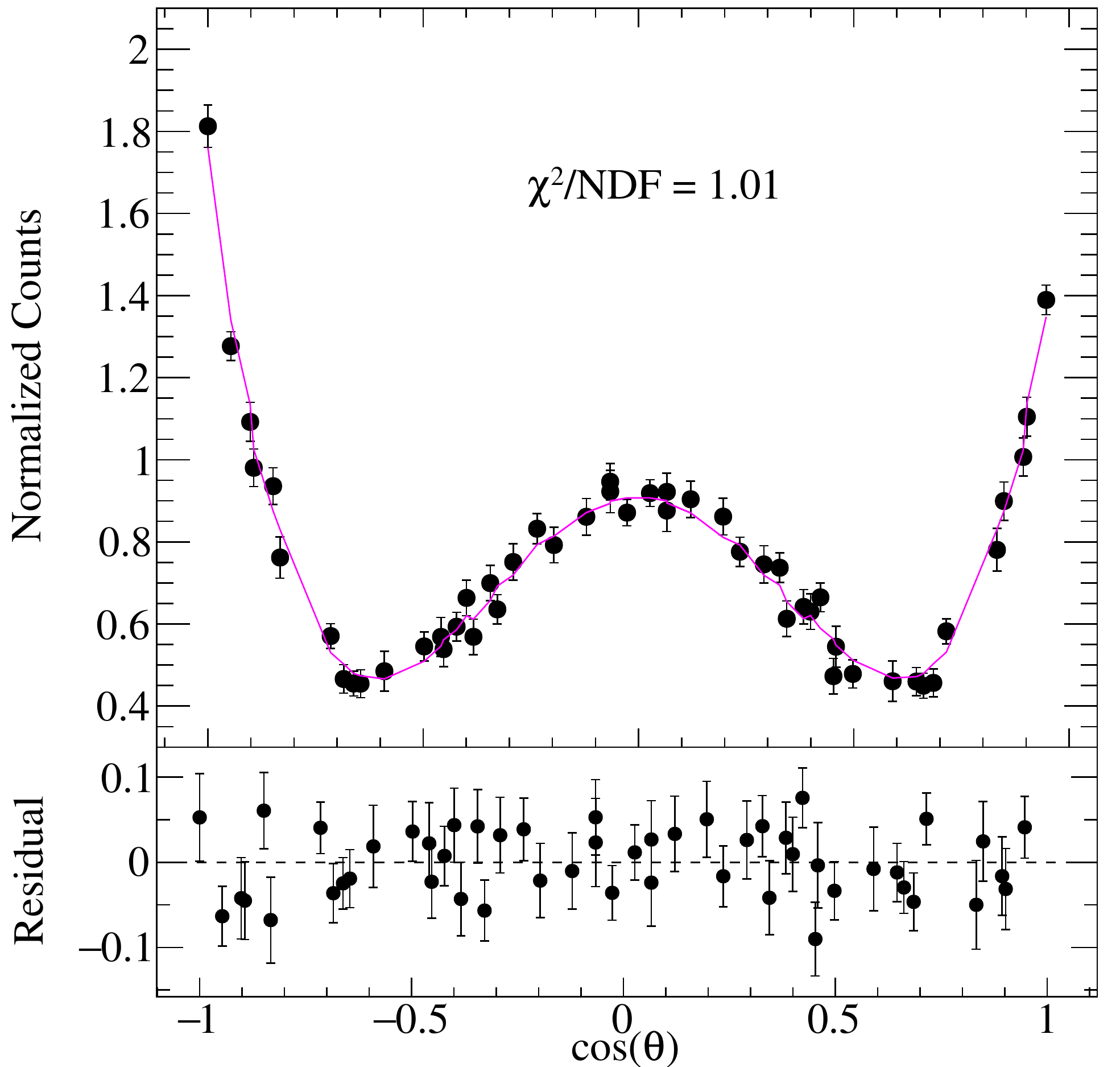}}
     \put(50,180){(a)}  
     }
     \end{picture}
     \hfill
     \begin{picture}(224,200)
     \subfloat{%
       \put(0,0){\includegraphics[width=0.45\textwidth]{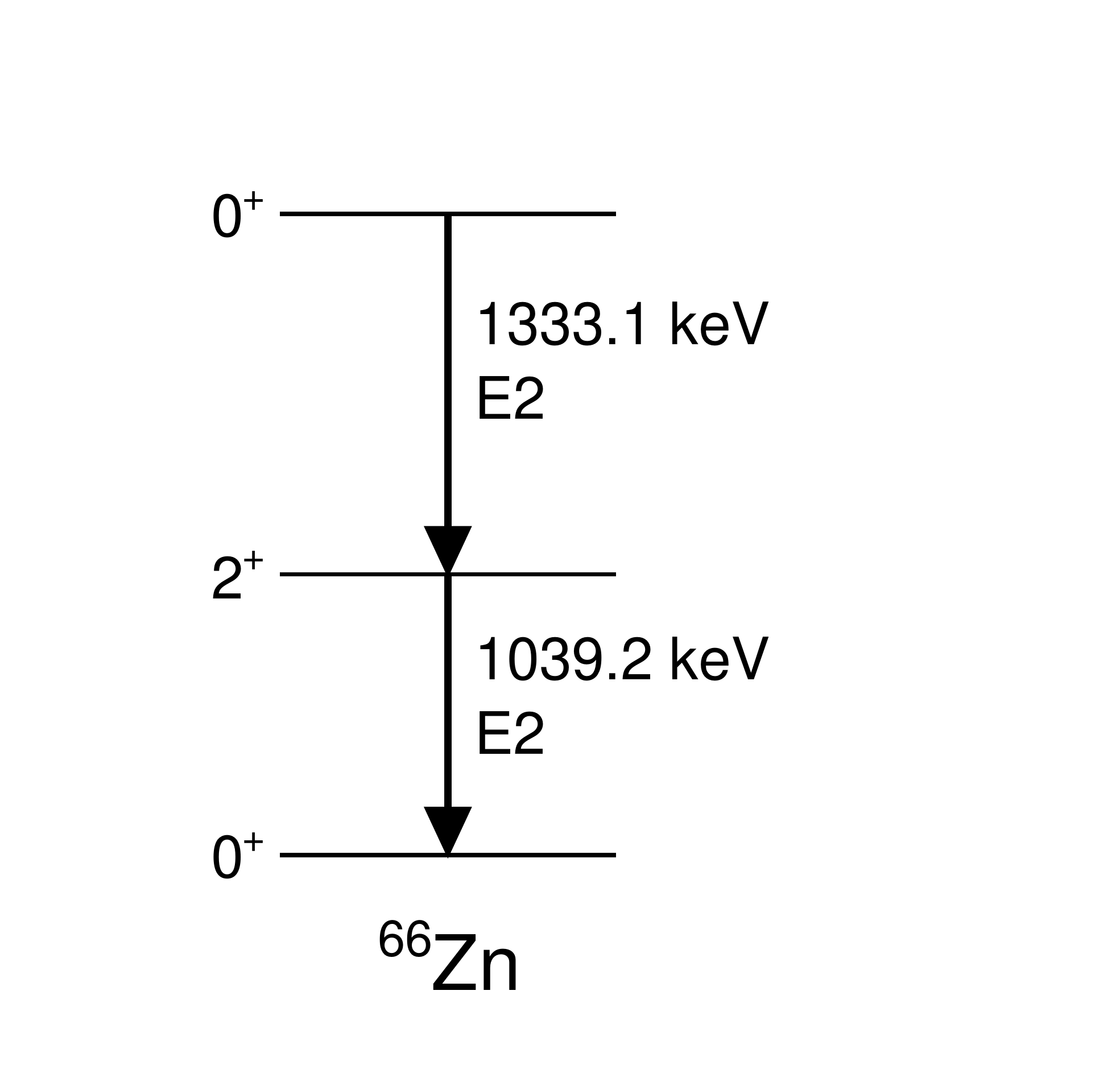}}
       \put(50,180){(b)}
     }
     \end{picture}
     \\
      \begin{picture}(100,200)
     \subfloat{%
     \put(0,0){ \includegraphics[width=0.45\textwidth,height=0.43\textwidth]{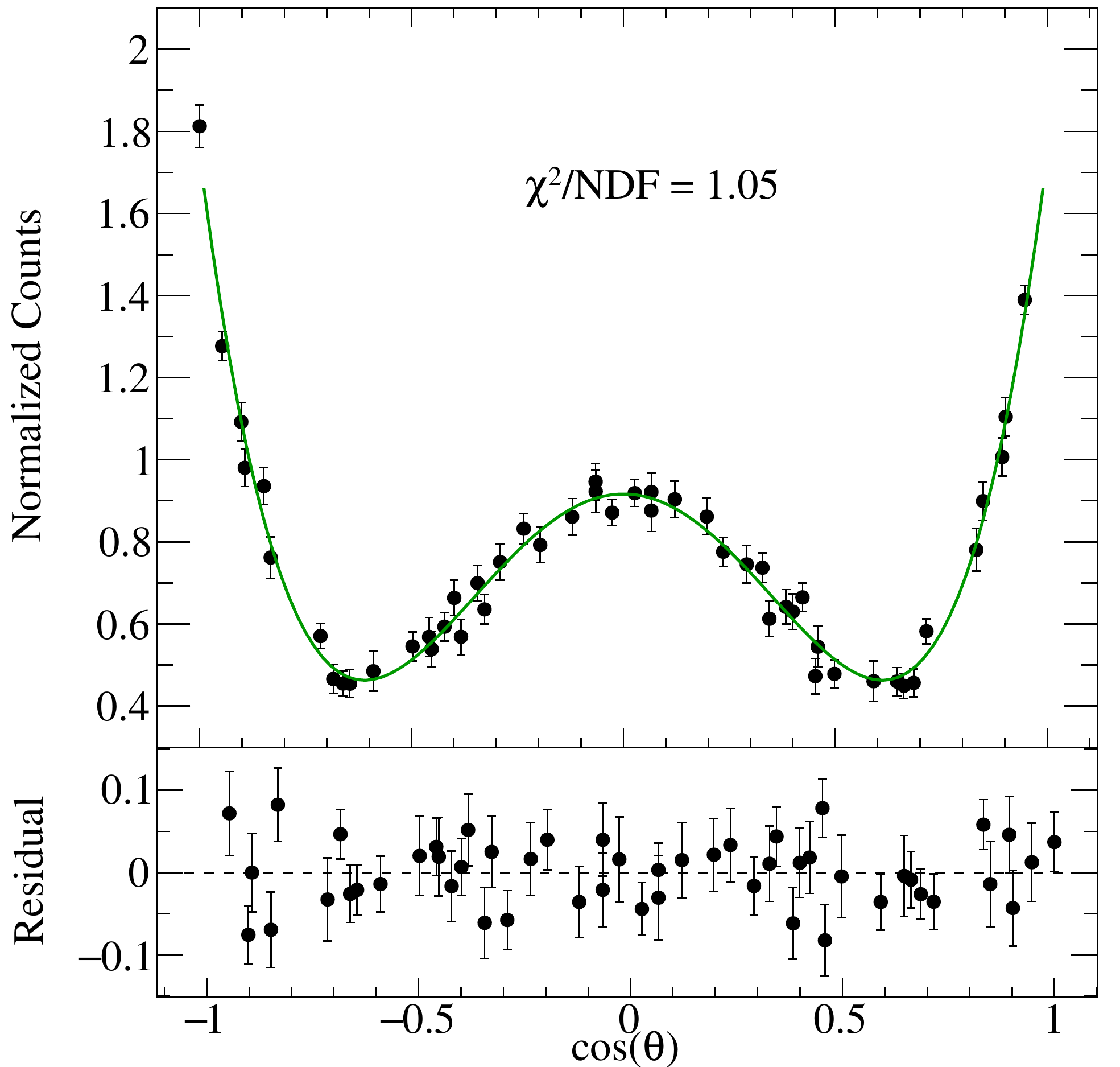}}
     \put(50,172){(c)}
     }
     \end{picture}
     \hfill
    \begin{picture}(224,200)
     \subfloat{%
       \put(0,0){\includegraphics[width=0.45\textwidth]{Ga020_ellipses.pdf}}
       \put(50,172){(d)}
     }
     \end{picture}
     \caption{Plots for the $^{66}$Zn $0^+ - 2^+ - 0^+$ 1333.1 keV - 1039.2 keV cascade. (a) The best Method 1 fit (magenta filled line) to the data (black points) has a $\chi^2$/NDF=1.01. (b) A partial level scheme showing the experimental details of this cascade. (c) A bare fit of Equation \ref{eq:ang-corr} to the data (green line) minimizes with $a_2$=0.31(3) and $a_4$=0.98(3). (d) A comparison of minimized $a_2$ and $a_4$ values extracted from fitting the $^{66}$Zn $0^+ - 2^+ - 0^+$ data with methods described in the paper. The expected $a_2$ and $a_4$ values are indicated by the star. Minimized $a_2$, $a_4$ values and $1\sigma$ confidence intervals are shown for a bare fit of Equation \ref{eq:ang-corr} (filled square, green dotted ellipse), Method 2 (open triangle, red solid ellipse), Method 3 (open diamond, blue dashed ellipse), and Method 4 (open triangle, purple dot-dashed ellipse).}
     \label{fig:Ga020-4Panel-Fig2}
\end{figure*}

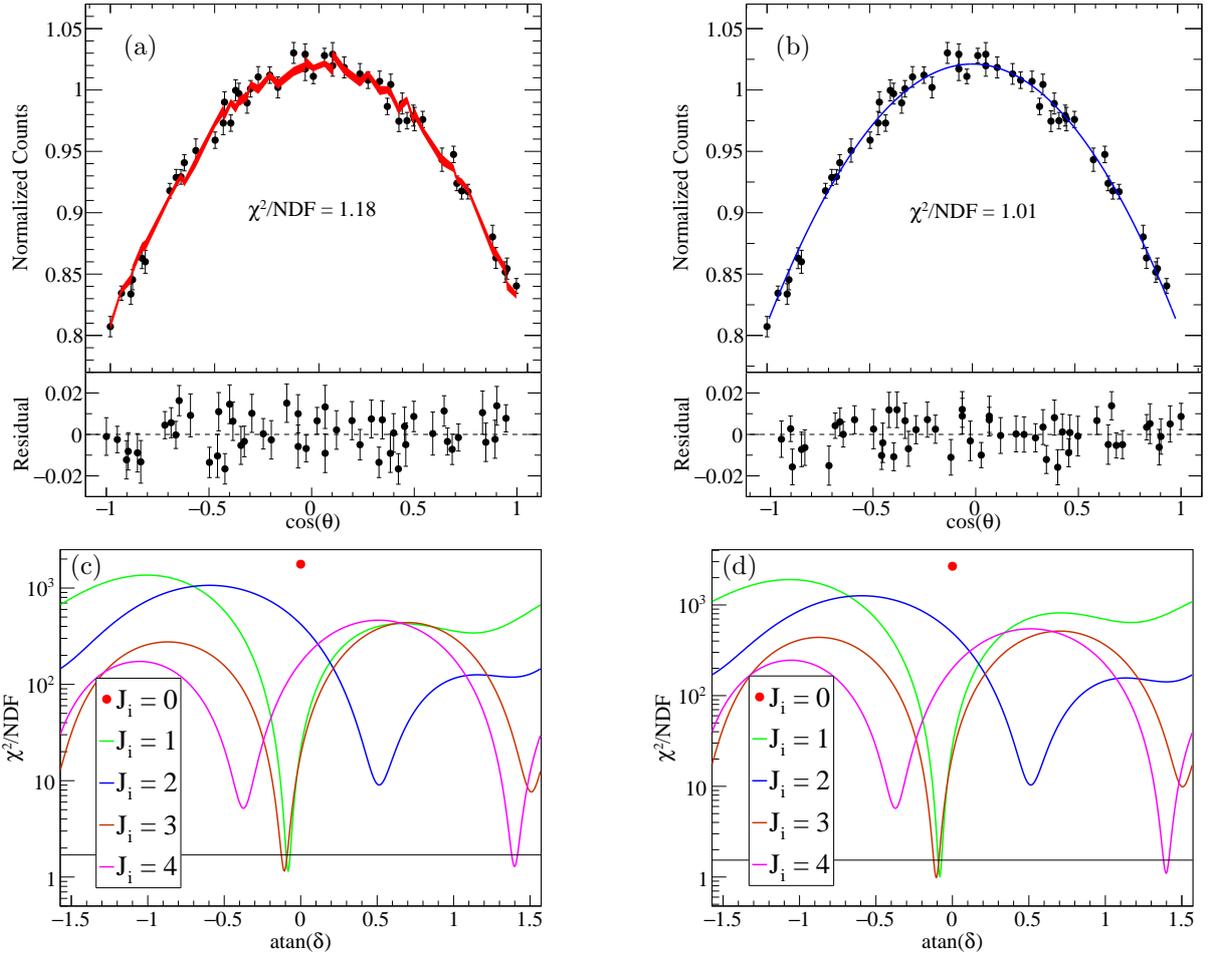
\begin{figure*}
\begin{picture}(100,200)
     \subfloat{%
       \put(0,0){\includegraphics[width=0.45\textwidth]{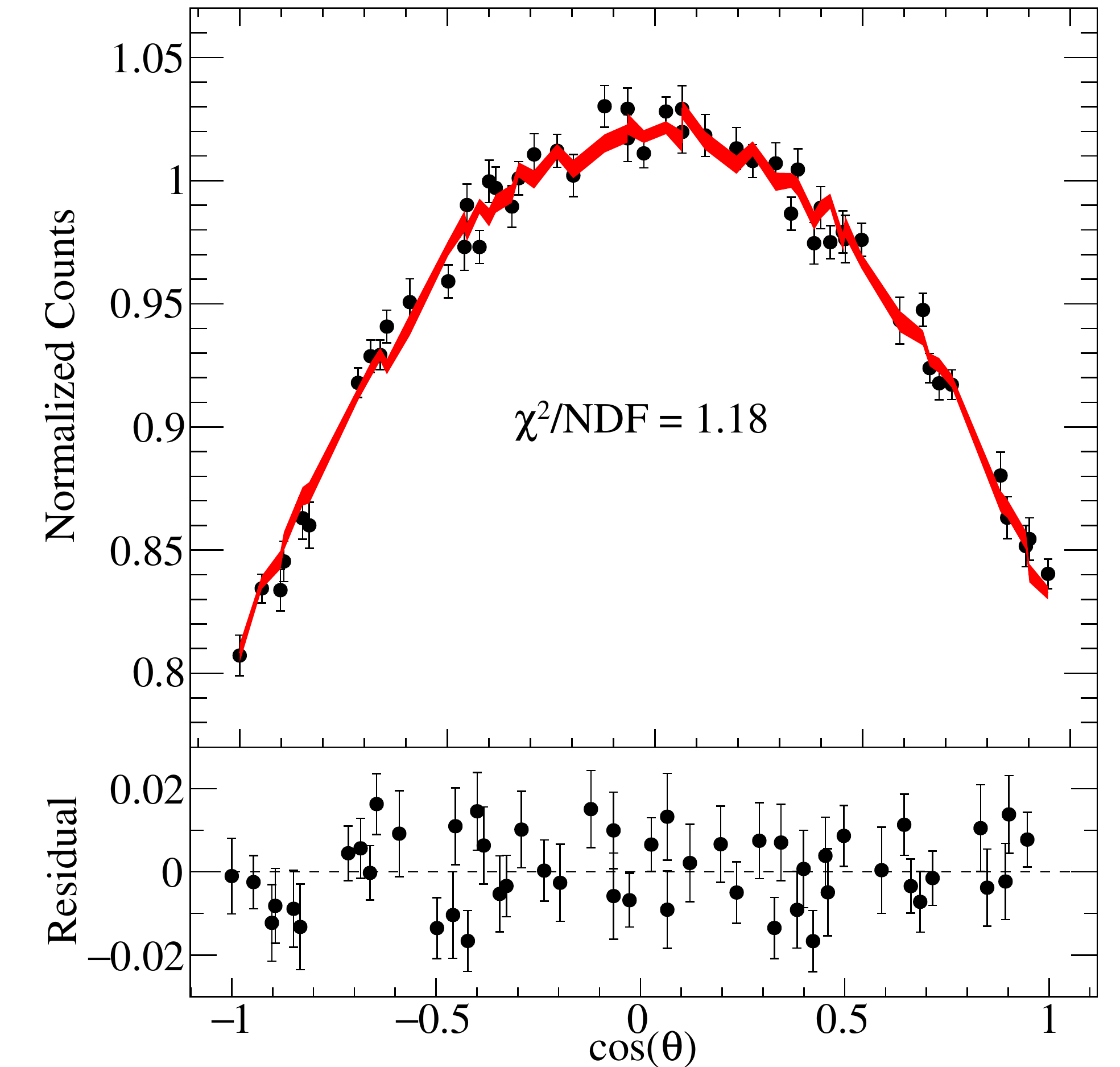}}
    \put(50,180){(a)}
     }
\end{picture}
     \hfill
     \begin{picture}(224,200)
     \subfloat{%
      \put(0,0){ \includegraphics[width=0.45\textwidth]{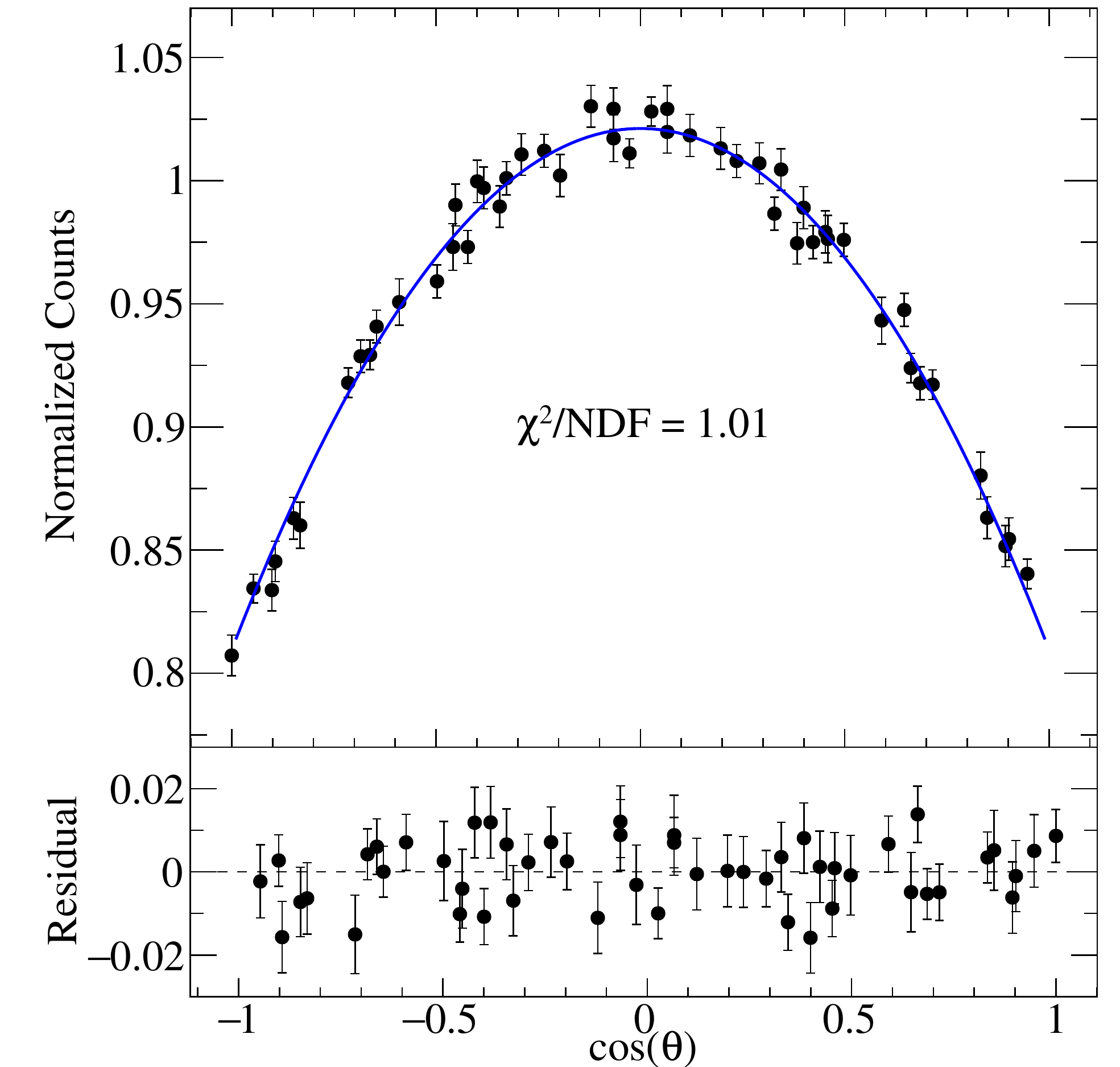}}
      \put(50,180){(b)}
     }
     \end{picture}
     \\
     \begin{picture}(100,160)
     \subfloat{%
       \put(0,0){\includegraphics[width=0.45\textwidth]{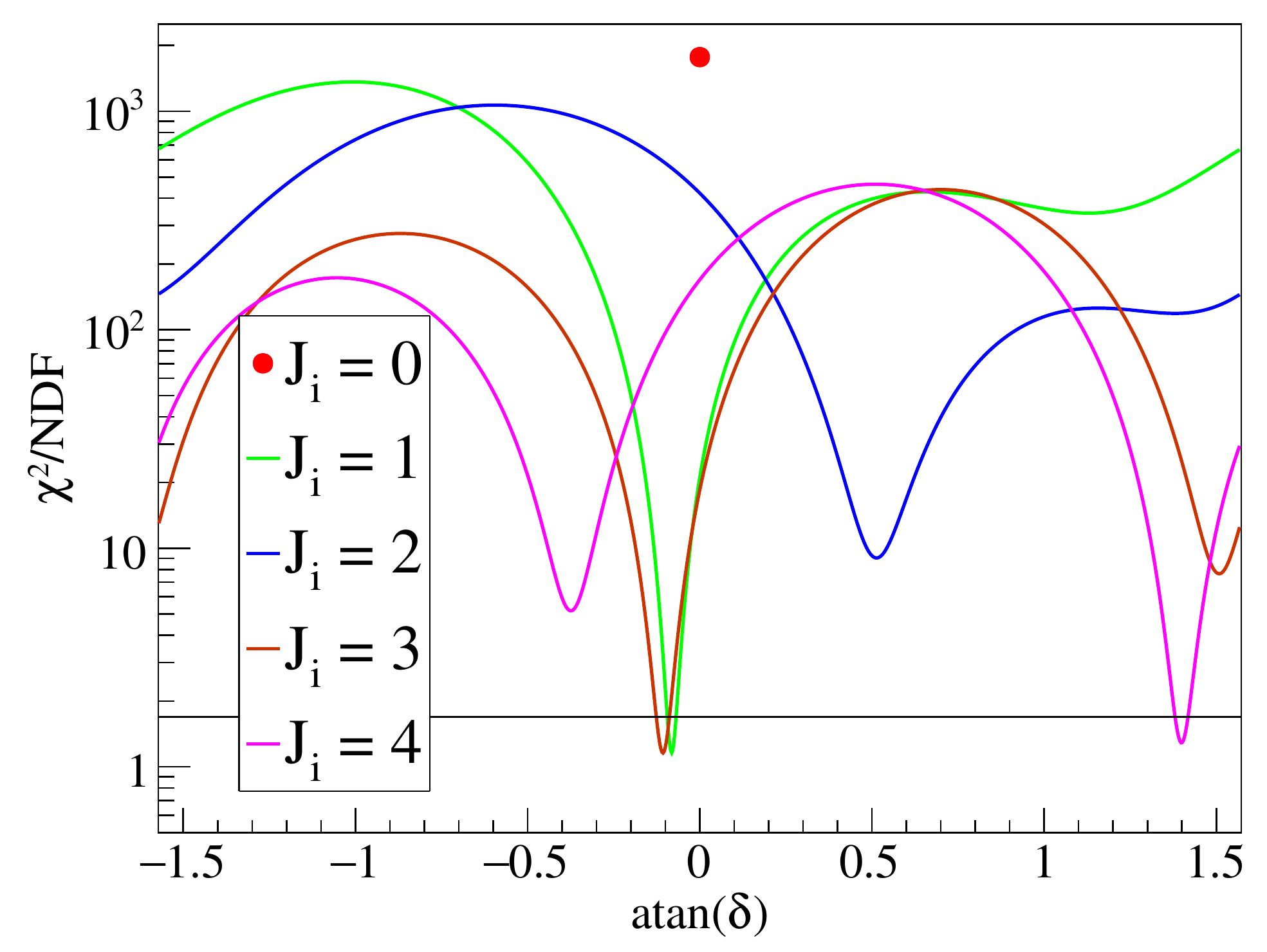}}
       \put(30,145){(c)}
     }
     \end{picture}
     \hfill
     \begin{picture}(224,160)
     \subfloat{%
       \put(0,0){\includegraphics[width=0.45\textwidth]{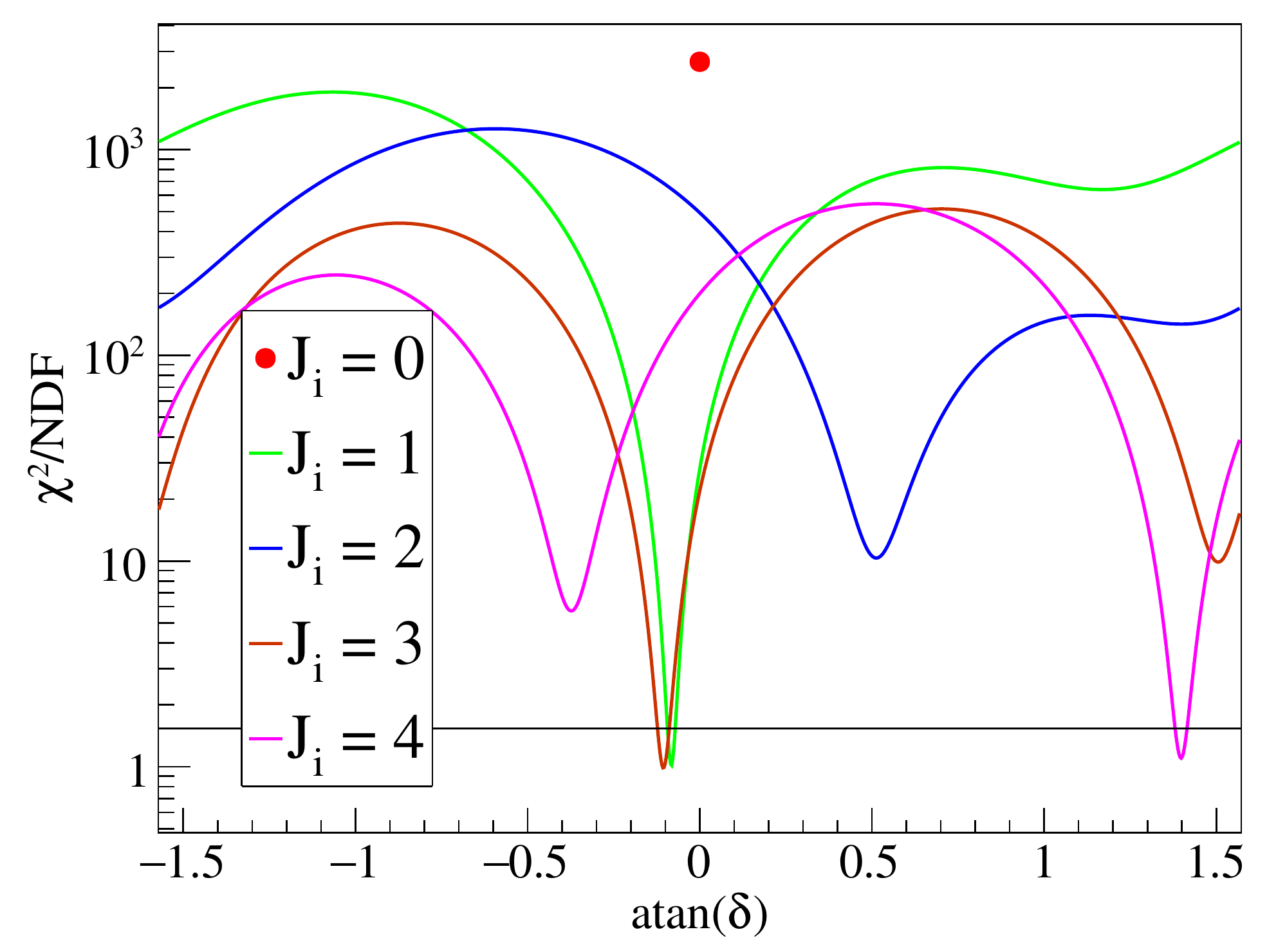}}
       \put(30,145){(d)}
     }
     \end{picture}
     \caption{Plots for the $^{66}$Zn $1^+ - 2^+ - 0^+$ 2751.8 keV - 1039.2 keV cascade. (a) The best Method 2 fit (red filled line) to the data (black points) has a $\chi^2$/NDF of 1.18 and minimizes with $a_2=-0.156(4)$ and $a_4=-0.003(5)$. The residual of the fit is shown in the lower panel. (b) The best Method 3 fit (blue line) to the data (black points) has a $\chi^2$/NDF of 1.01 and minimizes with $a_2=-0.156(5)$ and $a_4=0.002(5)$. The residual of the fit is shown in the lower panel. (c) A comparison of $\chi^2$/NDF values for potential $J_i=0-4$ and all possible mixing ratios ($\delta$) shows that the best fit to the data using Method 2 is made with $J=1,3$ and $4$ with $\delta$ of -0.082(3), -0.108(4) and 5.76(14), respectively. (d) A comparison of $\chi^2$/NDF values for potential $J_i=0-4$ and all possible mixing ratios ($\delta$) shows that the best fit to the data using Method 4 is made with $J=1,3$ and $4$ with $\delta$ of -0.083(4), -0.107(5) and 5.78(19), respectively.}
     \label{fig:Ga120-4Panel-Fig1}
\end{figure*}

\begin{figure*}
\begin{picture}(100,200)
     \subfloat{%
      \put(0,0){\includegraphics[width=0.45\textwidth]{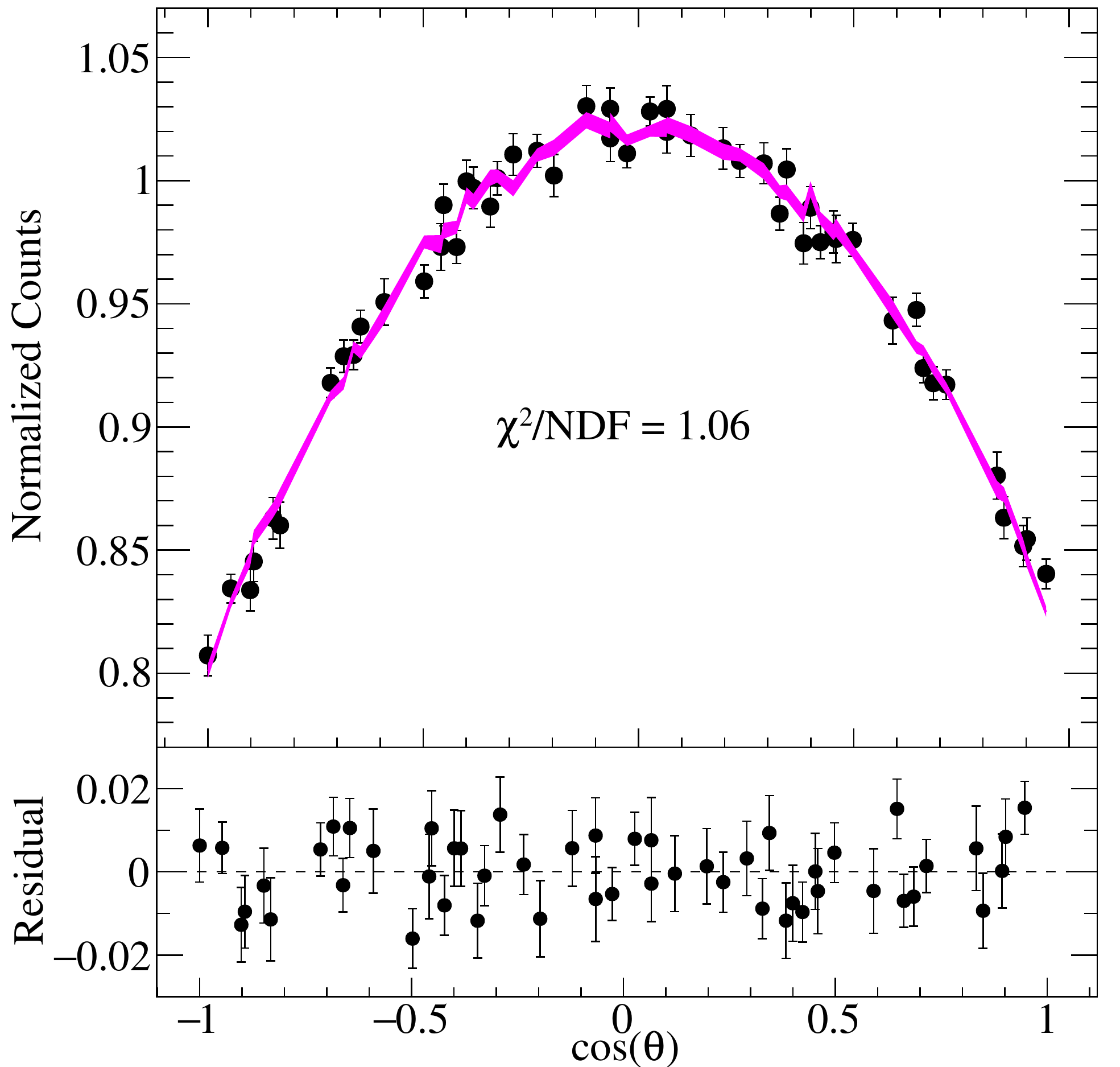}}
      \put(50,180){(a)} 
     }
     \end{picture}
     \hfill
     \begin{picture}(224,200)
     \subfloat{%
       \put(0,0){\includegraphics[width=0.43\textwidth]{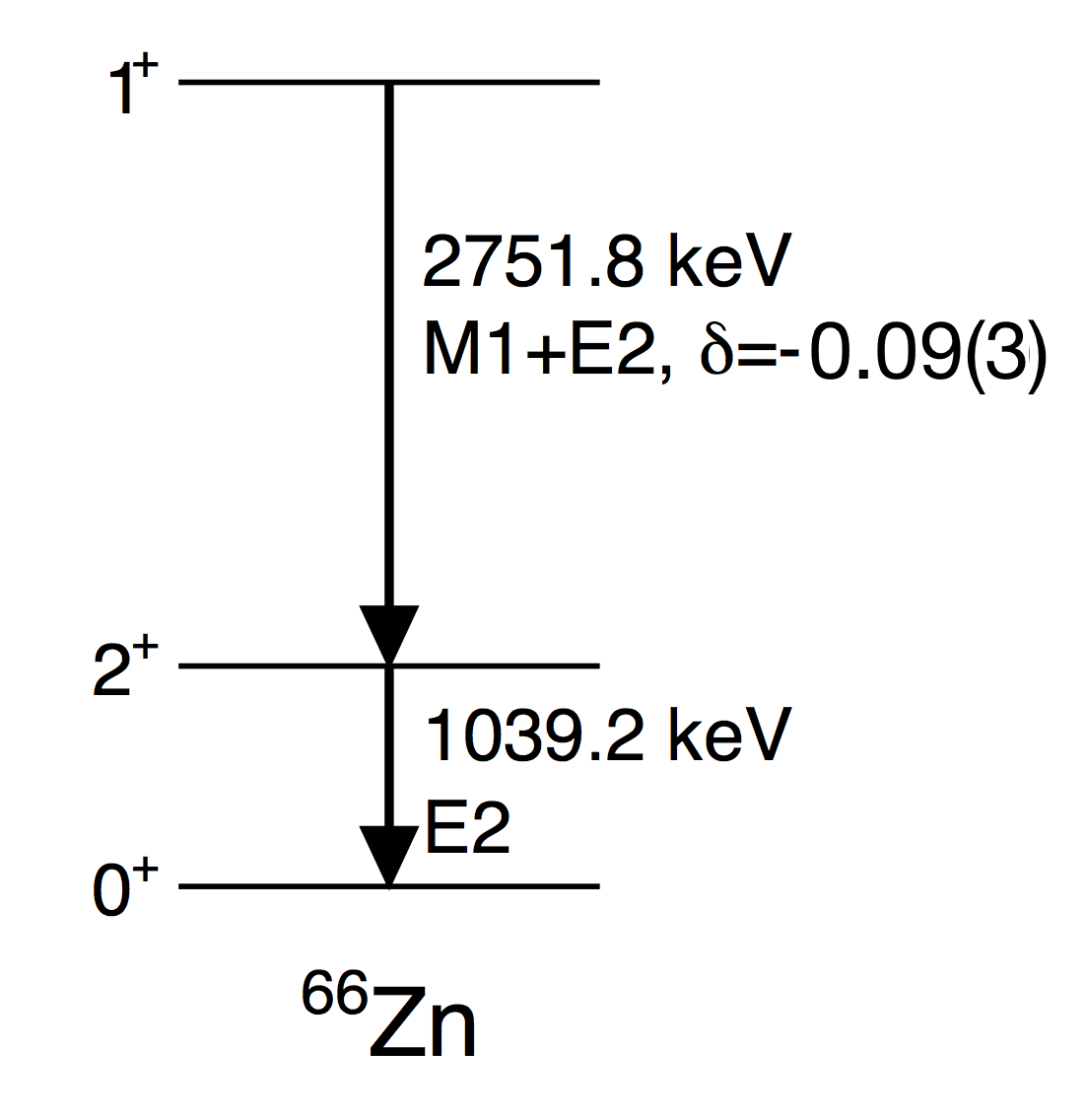}}
     \put(50,180){(b)}
     }
     \end{picture}
     \\
     \begin{picture}(100,200)
     \subfloat{%
       \put(0,0){\includegraphics[width=0.45\textwidth,height=0.39\textwidth]{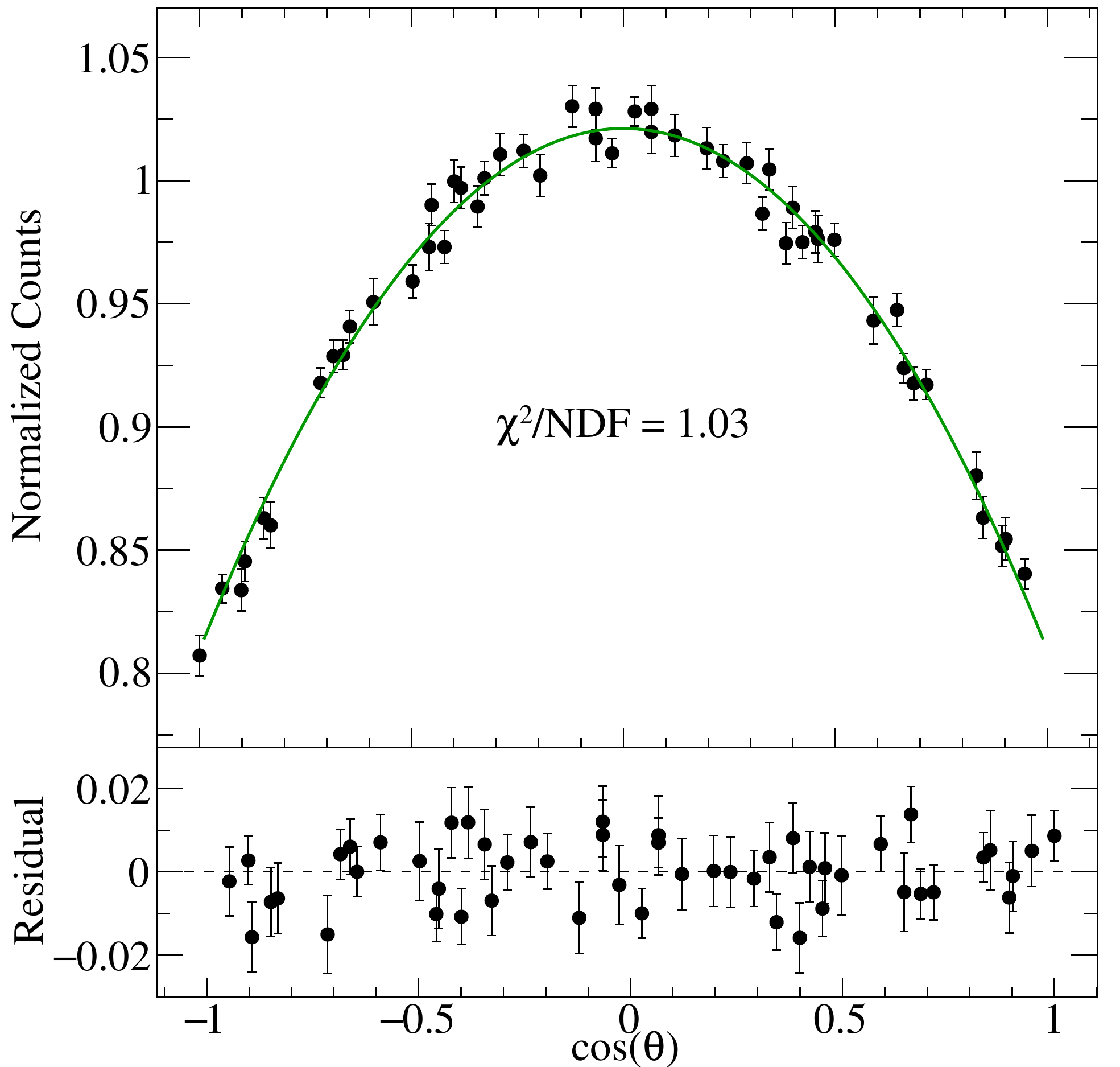}}
     }
     \put(40,160){(c)}
     \end{picture}
     \hfill
     \begin{picture}(224,200)
     \subfloat{%
       \put(0,0){\includegraphics[width=0.45\textwidth]{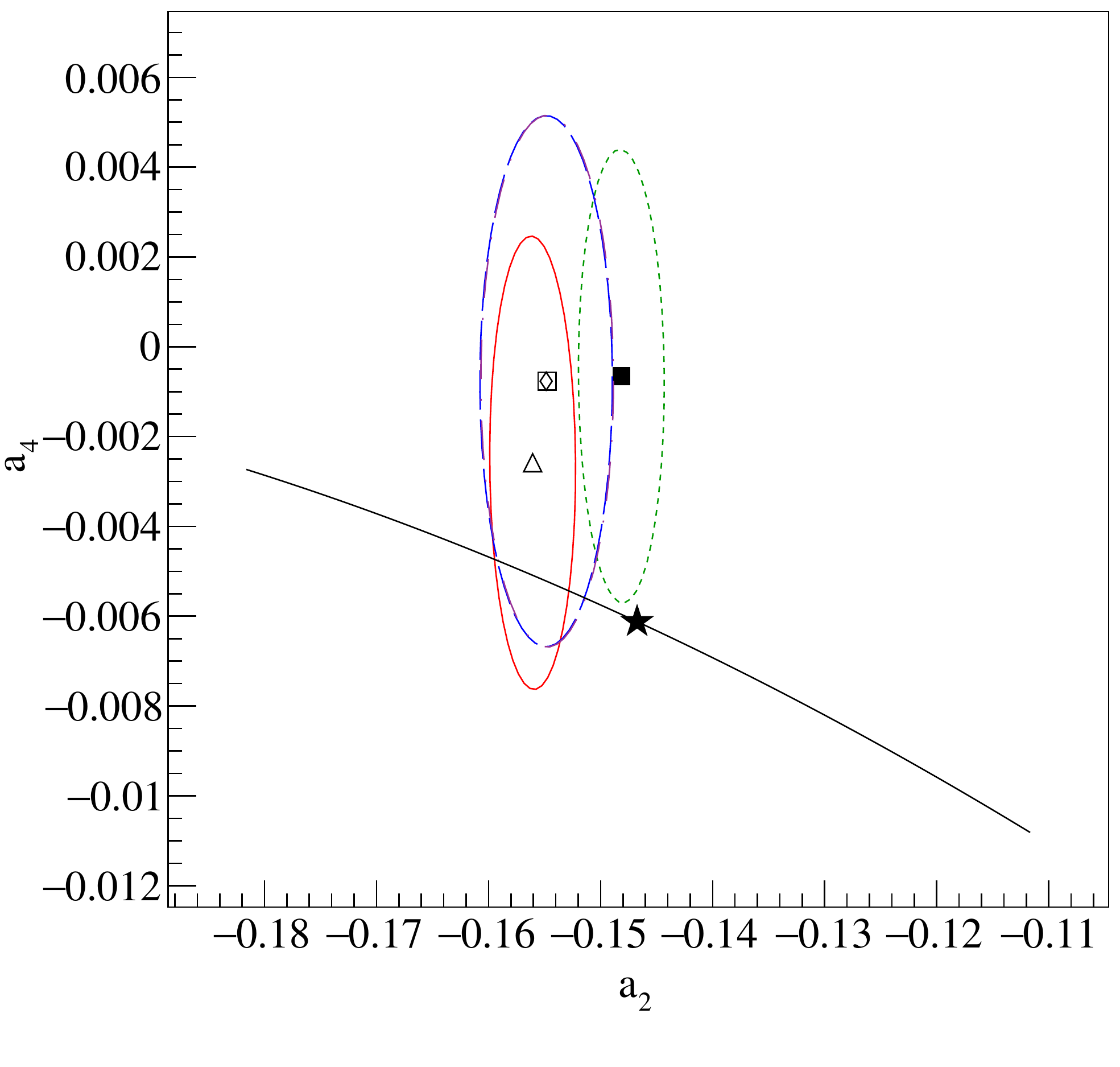}}
       \put(40,180){(d)}
     }
     \end{picture}
     \caption{Plots for the $^{66}$Zn $1^+ - 2^+ - 0^+$ 2751.8 keV - 1039.2 keV cascade. (a) The best Method 1 fit (magenta filled line) to the data (black points) has a $\chi^2$/NDF=1.06. The residual of the fit is shown in the lower panel. (b) A partial level scheme showing the experimental details of this cascade. (c) A bare fit of Equation \ref{eq:ang-corr} to the data (green line) minimizes with $a_2$=-0.148(4) and $a_4$=-0.001(5). (d) A comparison of minimized $a_2$ and $a_4$ values and 1$\sigma$ error fitted to the data with methods described in the paper. The expected $a_2$ and $a_4$ values are indicated by the star, with the black line representing values within the $\delta$ uncertainty. Minimized $a_2$, $a_4$ values and $1\sigma$ confidence intervals are shown for a bare fit of Equation \ref{eq:ang-corr} (filled square, green dotted ellipse), Method 2 (open triangle, red solid ellipse), Method 3 (open diamond, blue dashed ellipse), and Method 4 (open square, purple dot-dashed ellipse).}
     \label{fig:Ga120-4Panel-Fig2}
\end{figure*}

\begin{figure*}
\begin{picture}(100,200)
     \subfloat{%
       \put(0,0){\includegraphics[width=0.45\textwidth]{Ga220_Zfit.pdf}}
     \put(50,180){(a)}
     }
     \end{picture}
     \hfill
     \begin{picture}(224,200)
     \subfloat{%
       \put(0,0){\includegraphics[width=0.45\textwidth]{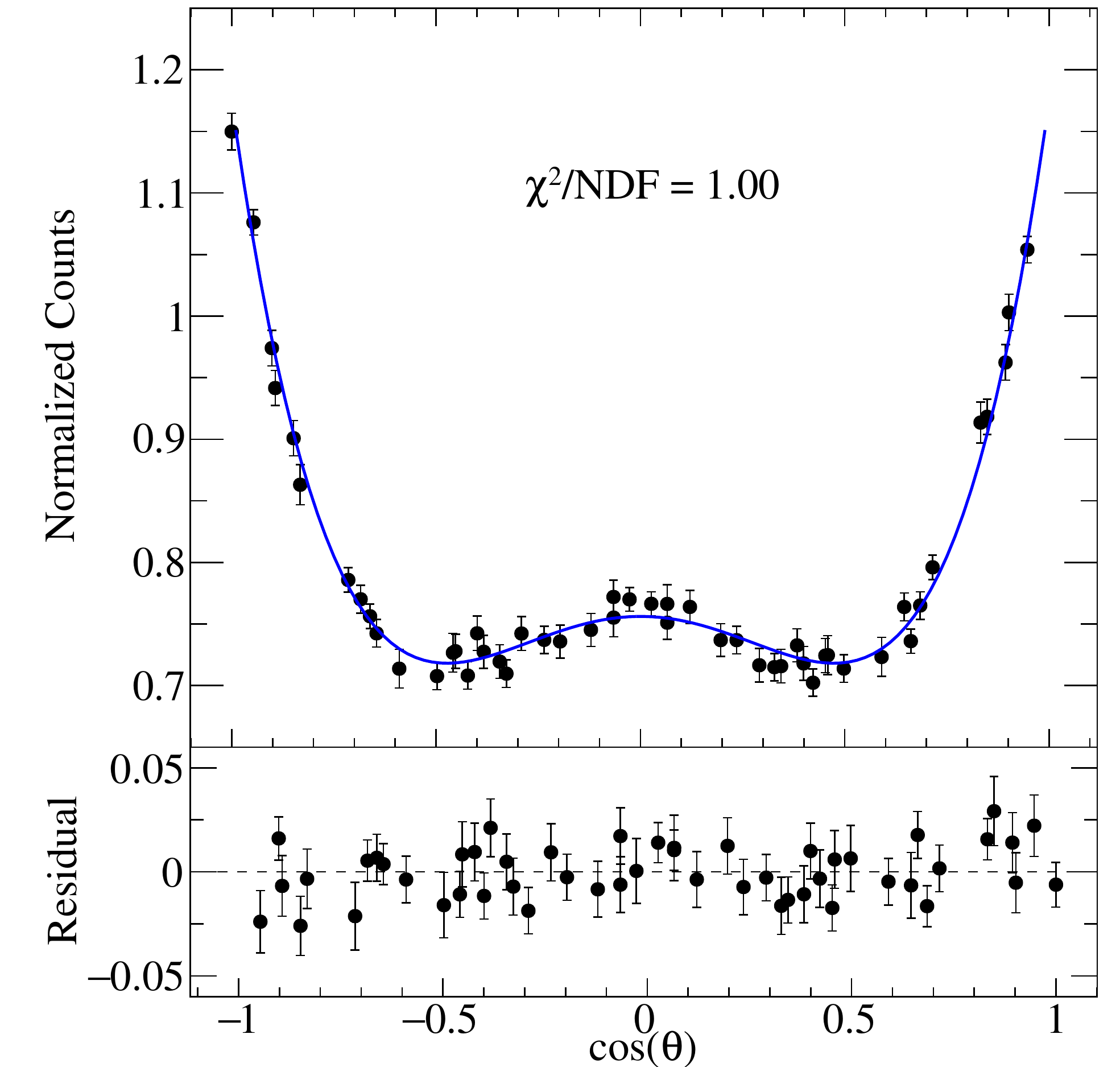}}
       \put(50,180){(b)}
     }
     \end{picture}
     \\
      \begin{picture}(100,160)    
     \subfloat{%
       \put(0,0){\includegraphics[width=0.45\textwidth]{Ga220_Zfit_chi2v2.pdf}}
      \put(30,145){(c)} 
     }
     \end{picture}
     \hfill
 \begin{picture}(224,160)
     \subfloat{%
       \put(0,0){\includegraphics[width=0.45\textwidth]{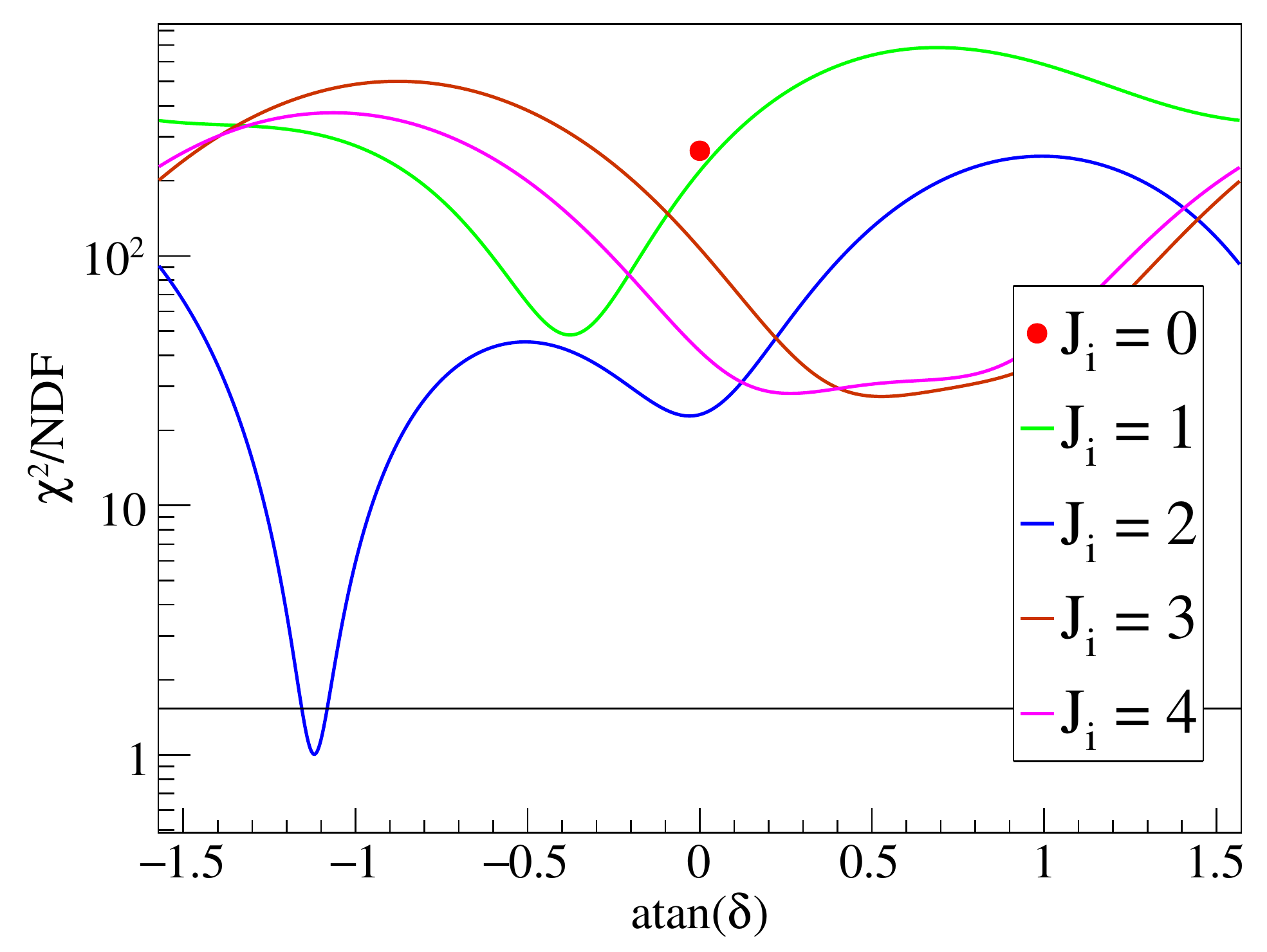}}
       \put(30,145){(d)}
     }
     \end{picture}
     \caption{
     Plots for the $^{66}$Zn $2^+ - 2^+ - 0^+$ 833.5 keV - 1039.2 keV cascade. (a) The best Method 2 fit (red filled line) to the data (black points) has a $\chi^2$/NDF of 0.97 and minimizes with $a_2=0.272(7)$ and $a_4=0.258(10)$. The residual of the fit is shown in the lower panel. (b) The best Method 3 fit (blue line) to the data (black points) has a $\chi^2$/NDF of 1.00 and minimizes with $a_2=0.272(10)$ and $a_4=0.258(10)$. The residual of the fit is shown in the lower panel. (c) A comparison of $\chi^2$/NDF values for potential $J_i=0-4$ and all possible mixing ratios ($\delta$) shows that the best fit to the data using Method 2 is made with $J=2$ with $\delta=-2.07(4)$. (d) A comparison of $\chi^2$/NDF values for potential $J_i=0-4$ and all possible mixing ratios ($\delta$) shows that the best fit to the data using Method 4 is made with $J=2$ with $\delta=-2.07(5)$.
     }
     \label{fig:Ga220-4Panel-Fig1}
\end{figure*}

\begin{figure*}
\begin{picture}(100,200)
     \subfloat{%
\put(0,0){\includegraphics[width=0.45\textwidth]{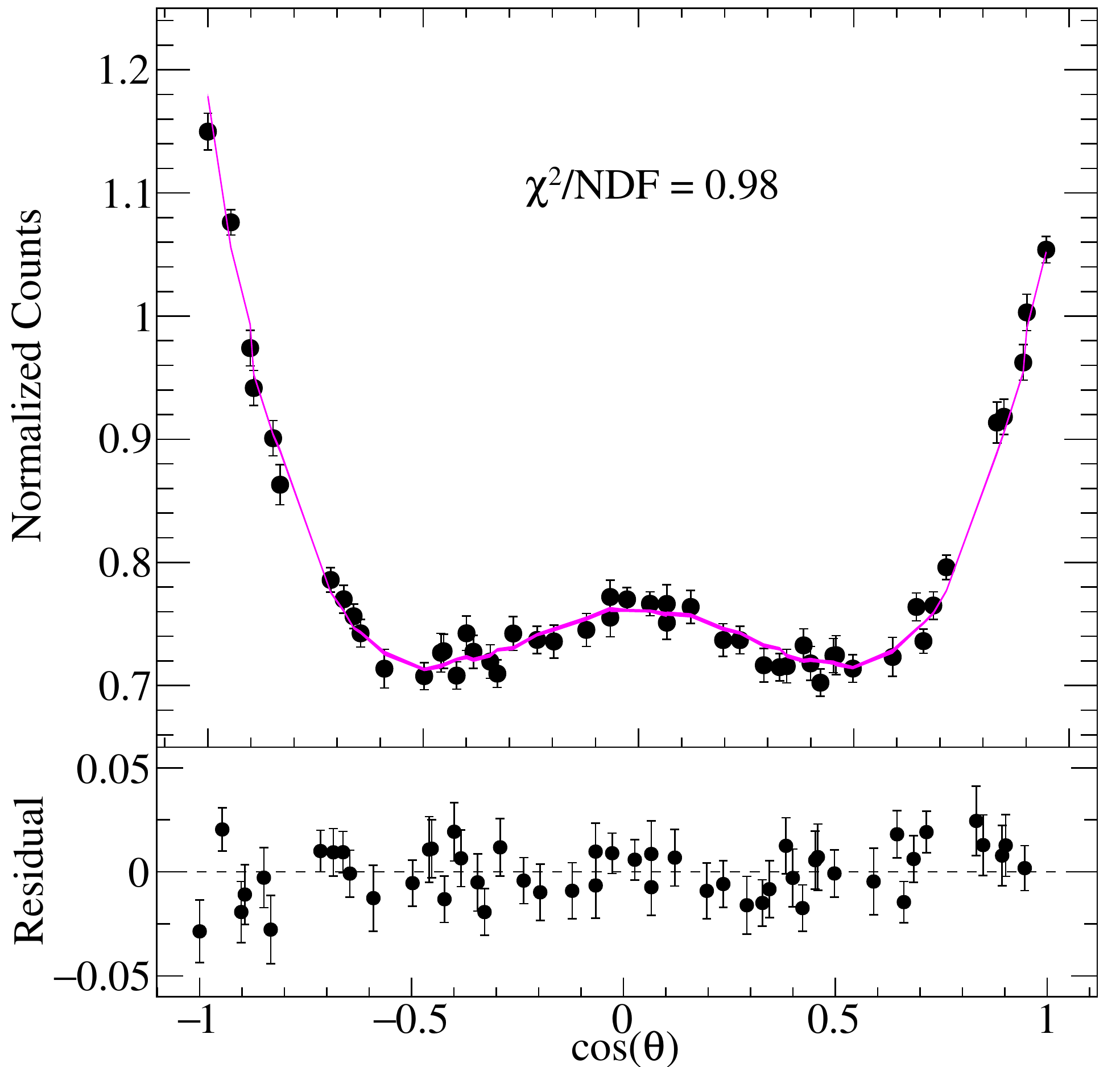}} 
	\put(50,180){(a)}
     }
     \end{picture}
     \hfill 
     \begin{picture}(224,200)
     \subfloat{%
       \put(0,0){\includegraphics[width=0.4\textwidth]{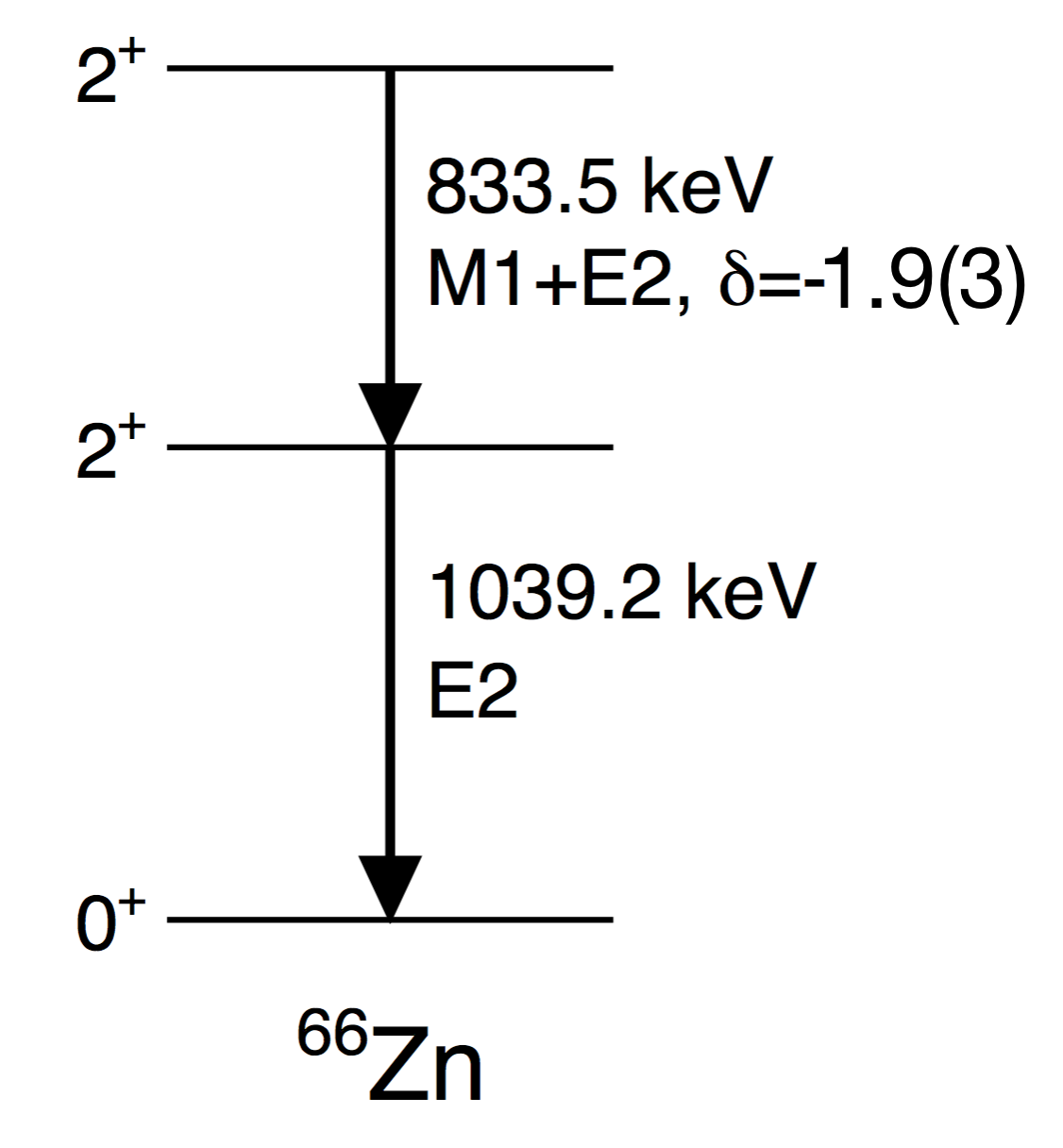}}
       \put(50,180){(b)}
     }
     \end{picture}
     \\
     \begin{picture}(100,200)
     \subfloat{%
     \put(0,0){\includegraphics[width=0.45\textwidth,height=0.42\textwidth]{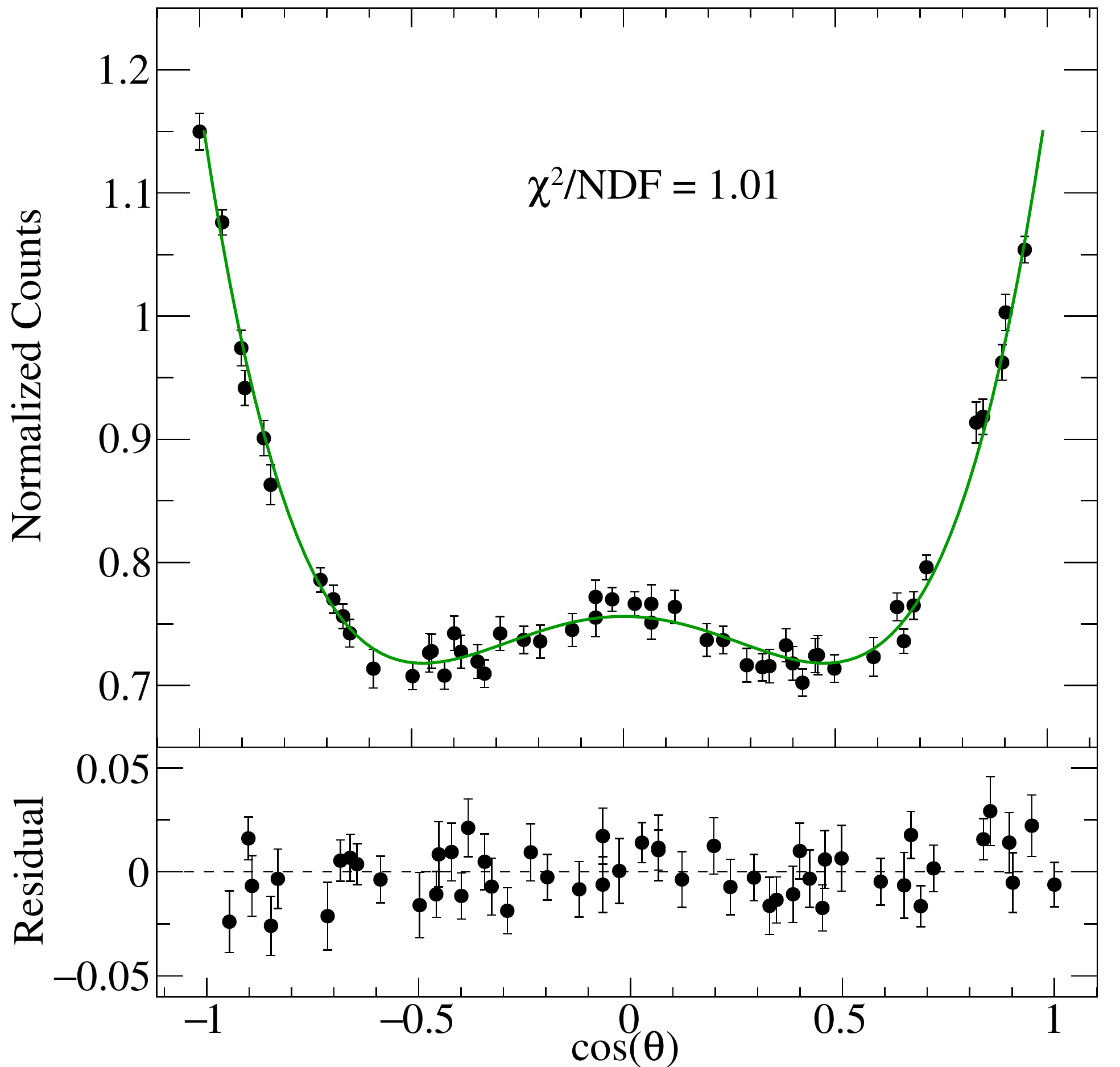}}
     \put(45,180){(c)} 
     }
     \end{picture}
     \hfill
     \begin{picture}(224,200)
     \subfloat{%
       \put(0,0){\includegraphics[width=0.45\textwidth]{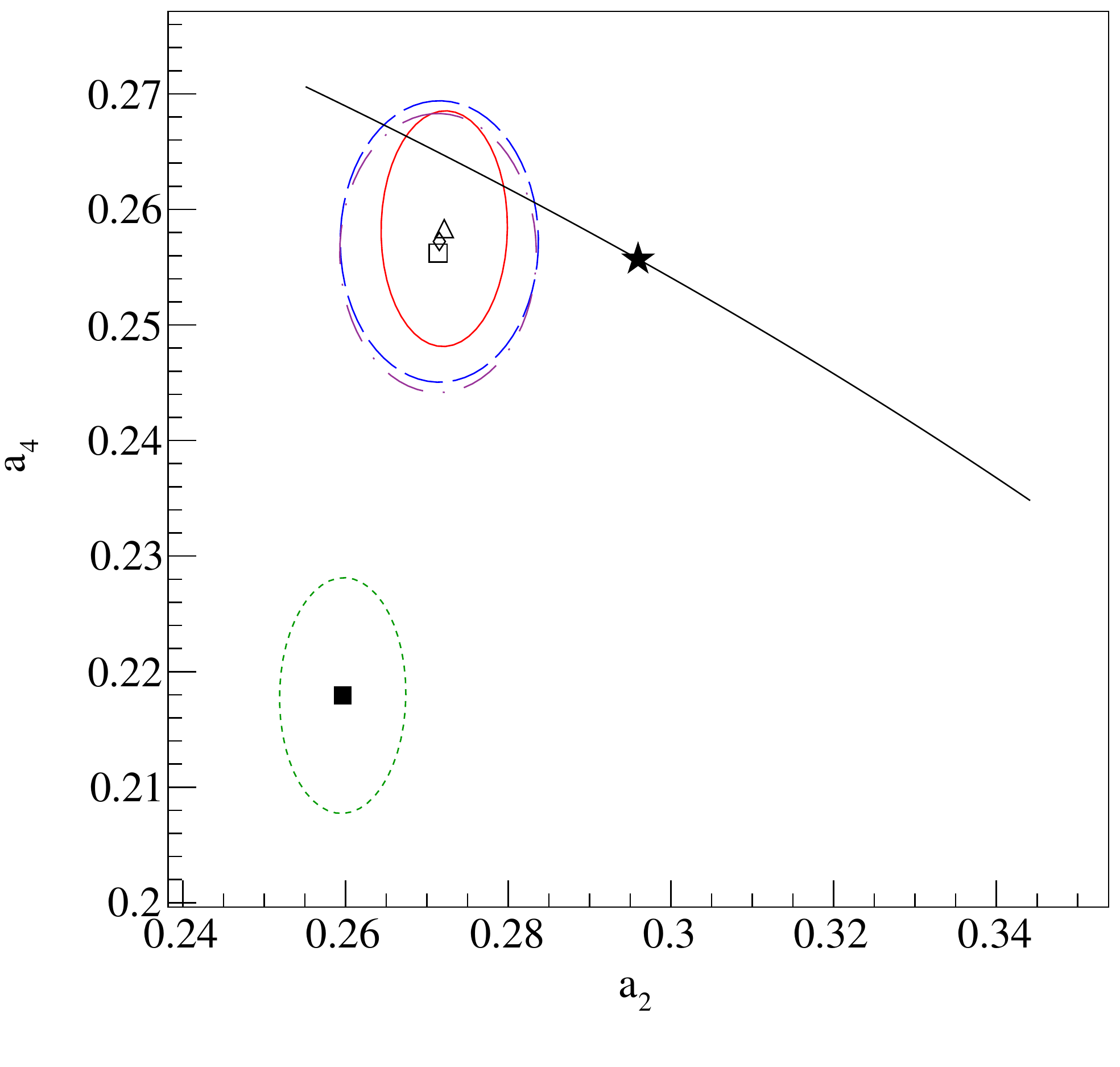}}
       \put(40,185){(d)}
     }
     \end{picture}
     \caption{Plots for the $^{66}$Zn $2^+ - 2^+ - 0^+$ 833.5 keV - 1039.2 keV cascade. (a) The best Method 1 fit (magenta filled line) to the data (black points) has a $\chi^2$/NDF=0.98. The residual of the fit is shown in the lower panel. (b) A partial level scheme showing the experimental details of this cascade. (c) A bare fit of Equation \ref{eq:ang-corr} to the data (green line) minimizes with $a_2$=0.260(8) and $a_4$=0.218(10). (d) A comparison of minimized $a_2$ and $a_4$ values and 1$\sigma$ error fitted to the data with methods described in the paper. The expected $a_2$ and $a_4$ values are indicated by the star, with the black line representing values within the $\delta$ uncertainty. Minimized $a_2$, $a_4$ values and $1\sigma$ confidence intervals are shown for a bare fit of Equation \ref{eq:ang-corr} (filled square, green dotted ellipse), Method 2 (open triangle, red solid ellipse), Method 3 (open diamond, blue dashed ellipse), and Method 4 (open square, purple dot-dashed ellipse).}
     \label{fig:Ga220-4Panel-Fig2}
\end{figure*}

\begin{figure*}
\begin{picture}(100,200)
     \subfloat{%
       \put(0,0){\includegraphics[width=0.45\textwidth]{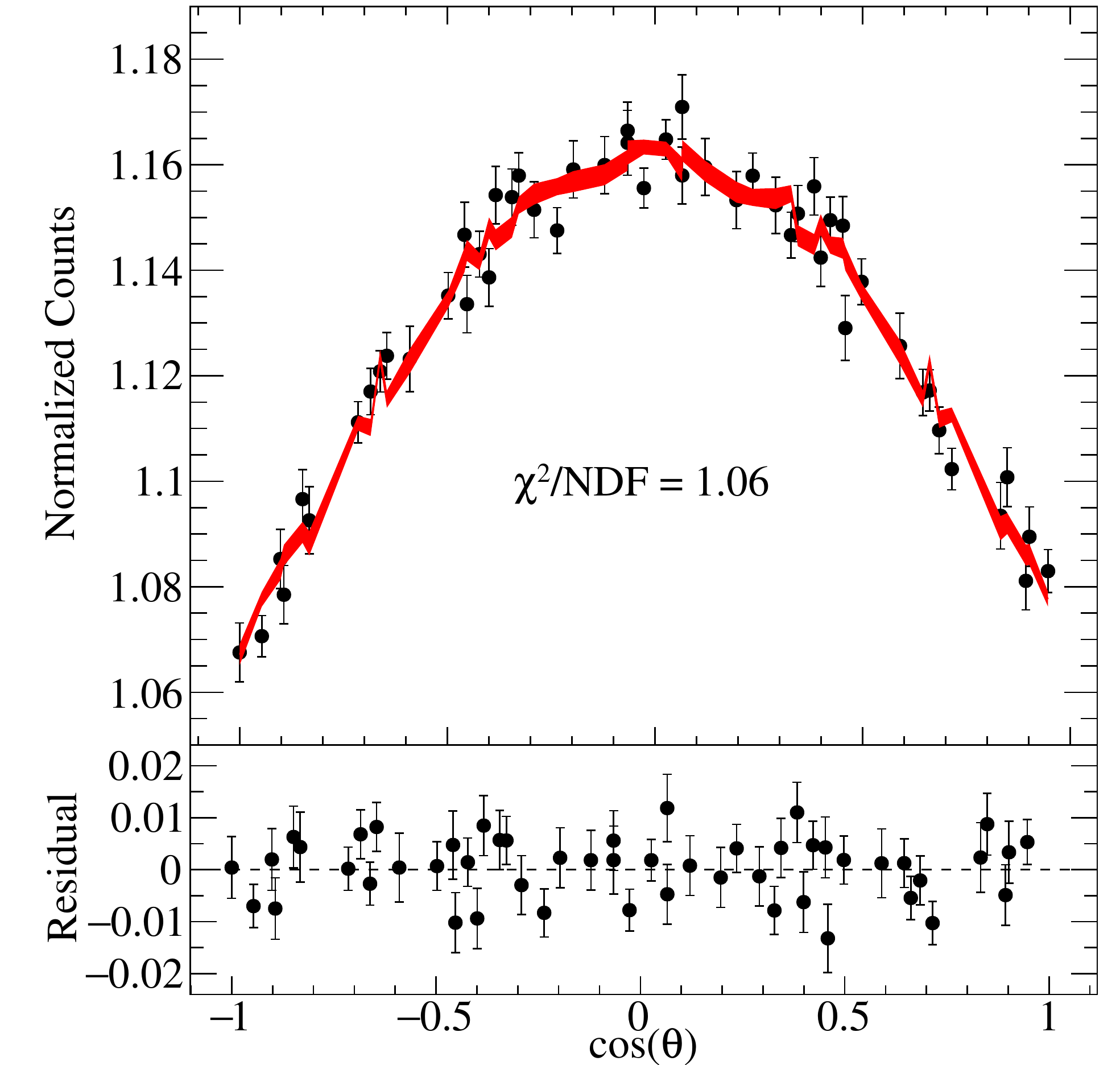}}
     \put(50,180){(a)}
     }
     \end{picture}
     \hfill
    \begin{picture}(224,200)
     \subfloat{%
       \put(0,0){\includegraphics[width=0.45\textwidth]{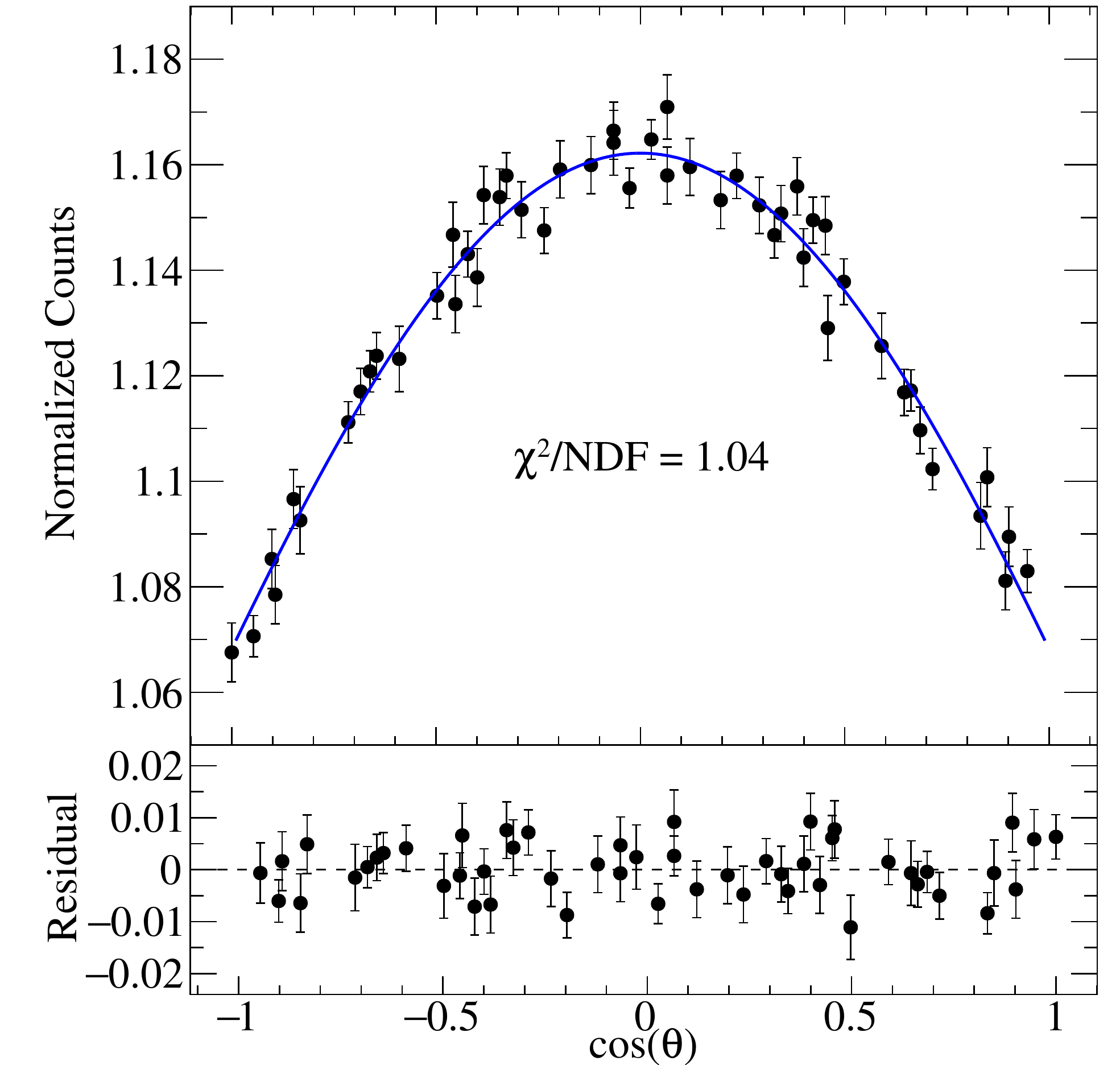}}
       \put(50,180){(b)}
     }
     \end{picture}
     \\
     \begin{picture}(100,160) 
     \subfloat{%
       \put(0,0){\includegraphics[width=0.45\textwidth]{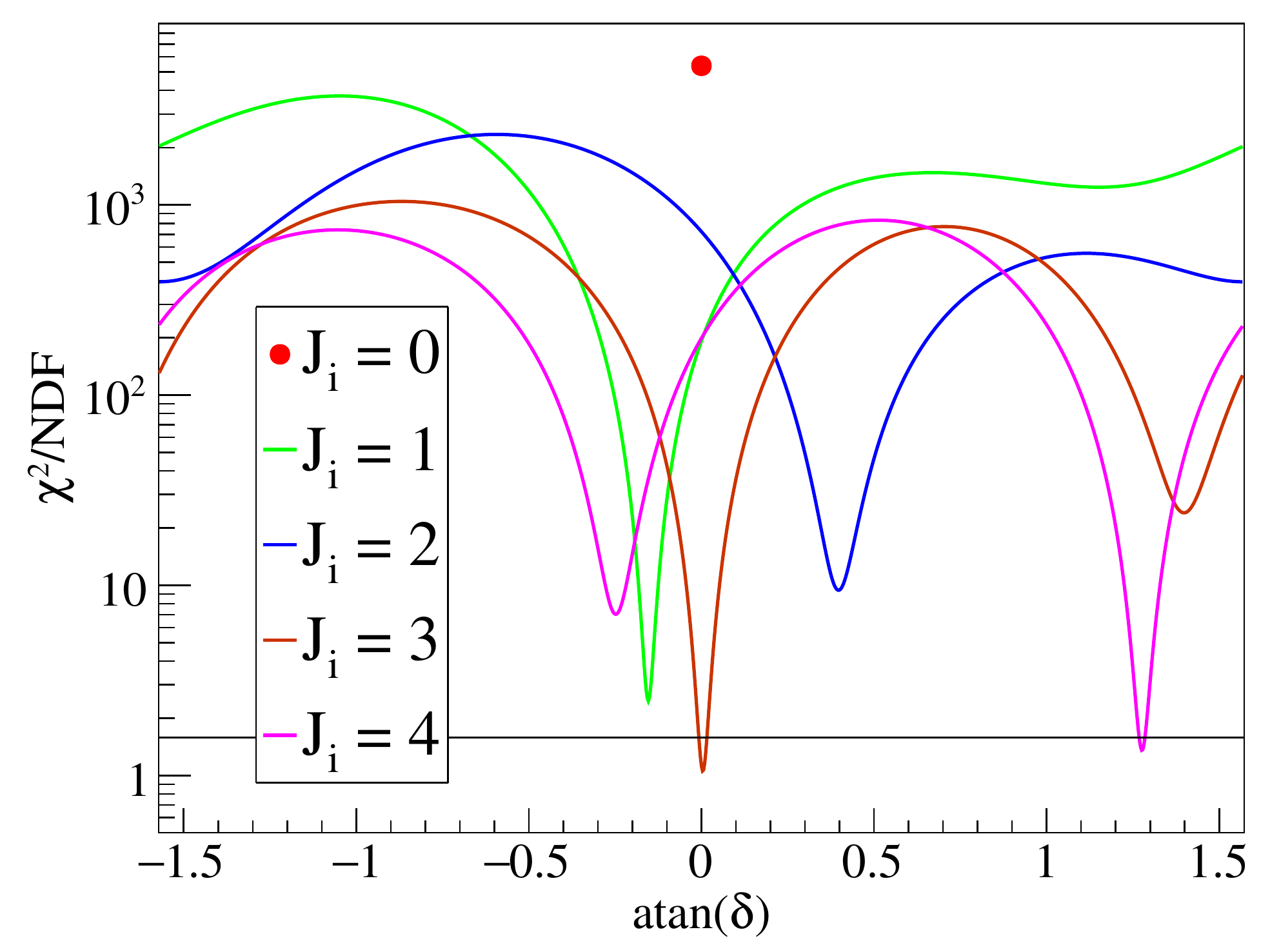}}
       \put(30,145){(c)}
     }
     \end{picture}
     \hfill
     \begin{picture}(224,160)
     \subfloat{%
       \put(0,0){\includegraphics[width=0.45\textwidth]{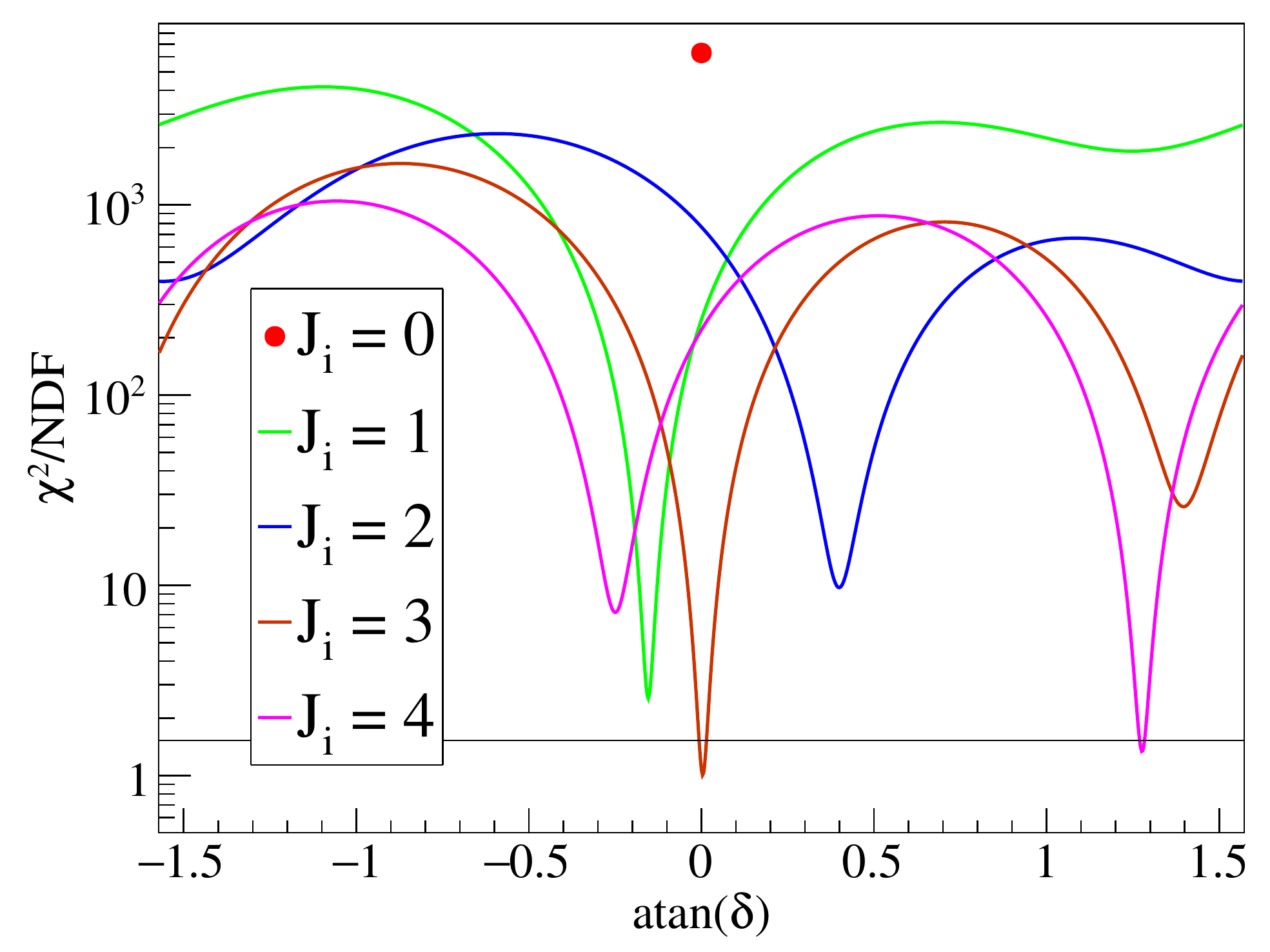}}
     \put(30,145){(d)}
     }
     \end{picture}
     \caption{
     Plots for the $^{152}$Gd $3^- - 2^+ - 0^+$ 778.9 keV - 344.3 keV cascade. (a) The best Method 2 fit (red filled line) to the data (black points) has a $\chi^2$/NDF of 1.06 and minimizes with $a_2=-0.068(3)$ and $a_4=-0.002(3)$. The residual of the fit is shown in the lower panel. (b) The best Method 3 fit (blue line) to the data (black points) has a $\chi^2$/NDF of 1.04 and minimizes with $a_2=-0.068(3)$ and $a_4=0.003(4)$. The residual of the fit is shown in the lower panel. (c) A comparison of $\chi^2$/NDF values for potential $J_i=0-4$ and all possible mixing ratios ($\delta$) shows that the best fit to the data using Method 2 is made with $J=1$ with $\delta=0.004(2)$. (d) A comparison of $\chi^2$/NDF values for potential $J_i=0-4$ and all possible mixing ratios ($\delta$) shows that the best fit to the data using Method 4 is made with $J=3$ with $\delta=0.004(3)$.
     }
     \label{fig:Eu152-4Panel-Fig1}
\end{figure*}
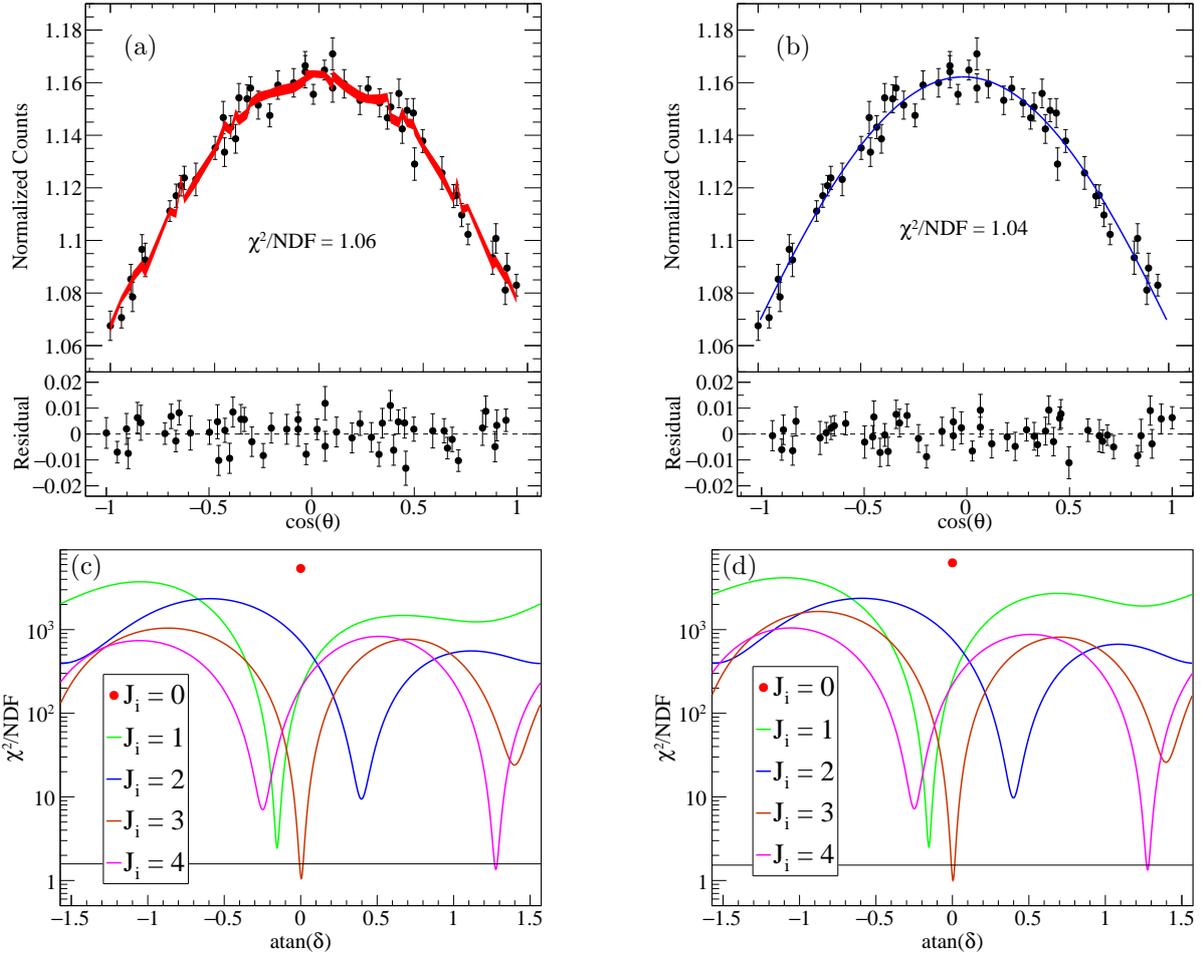

\begin{figure*}
\begin{picture}(100,200)
     \subfloat{%
       \put(0,0){\includegraphics[width=0.45\textwidth]{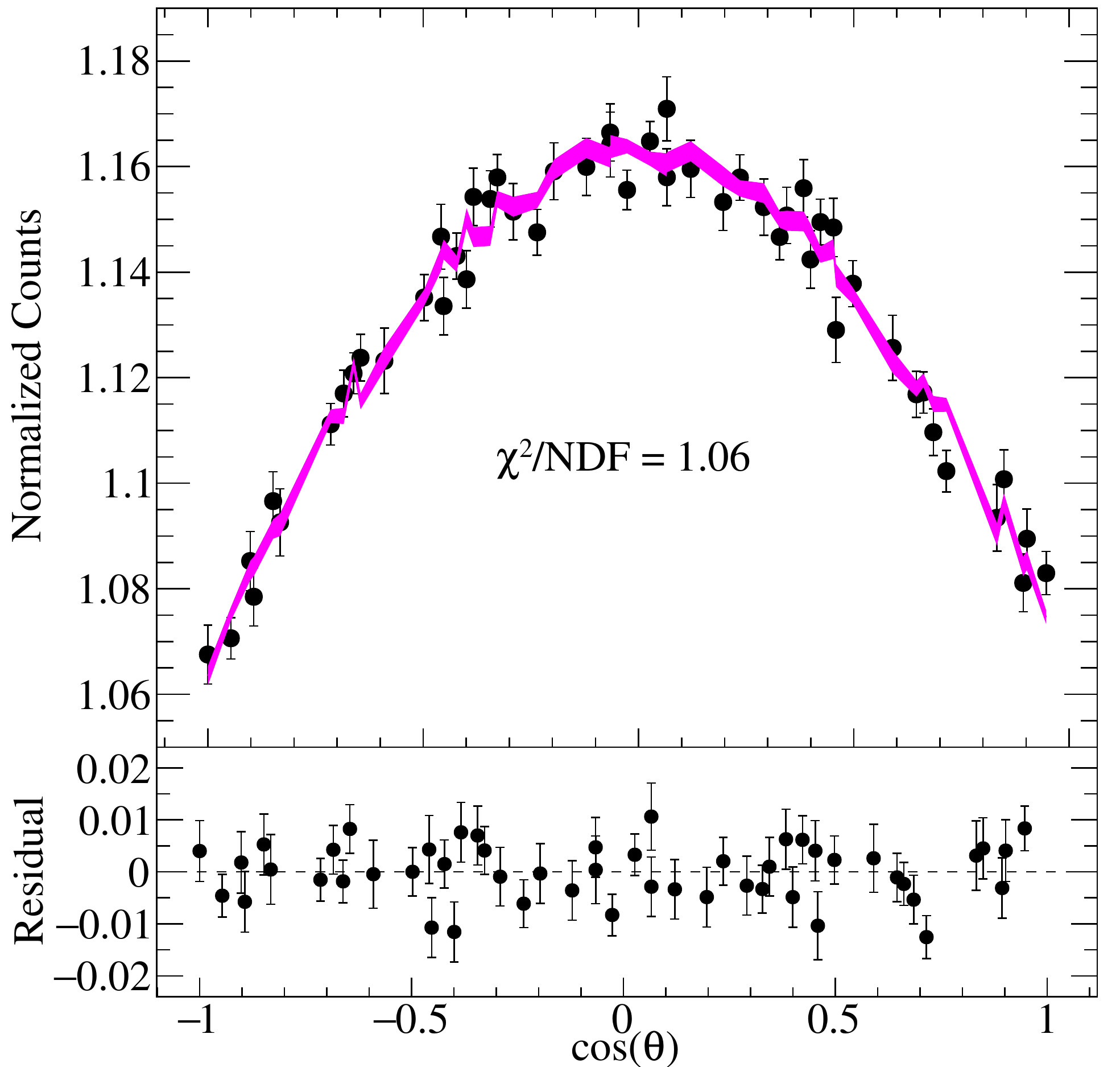}}
       \put(50,180){(a)}
     }
     \end{picture}
     \hfill
     \begin{picture}(224,200)
     \subfloat{%
       \put(0,0){\includegraphics[width=0.45\textwidth]{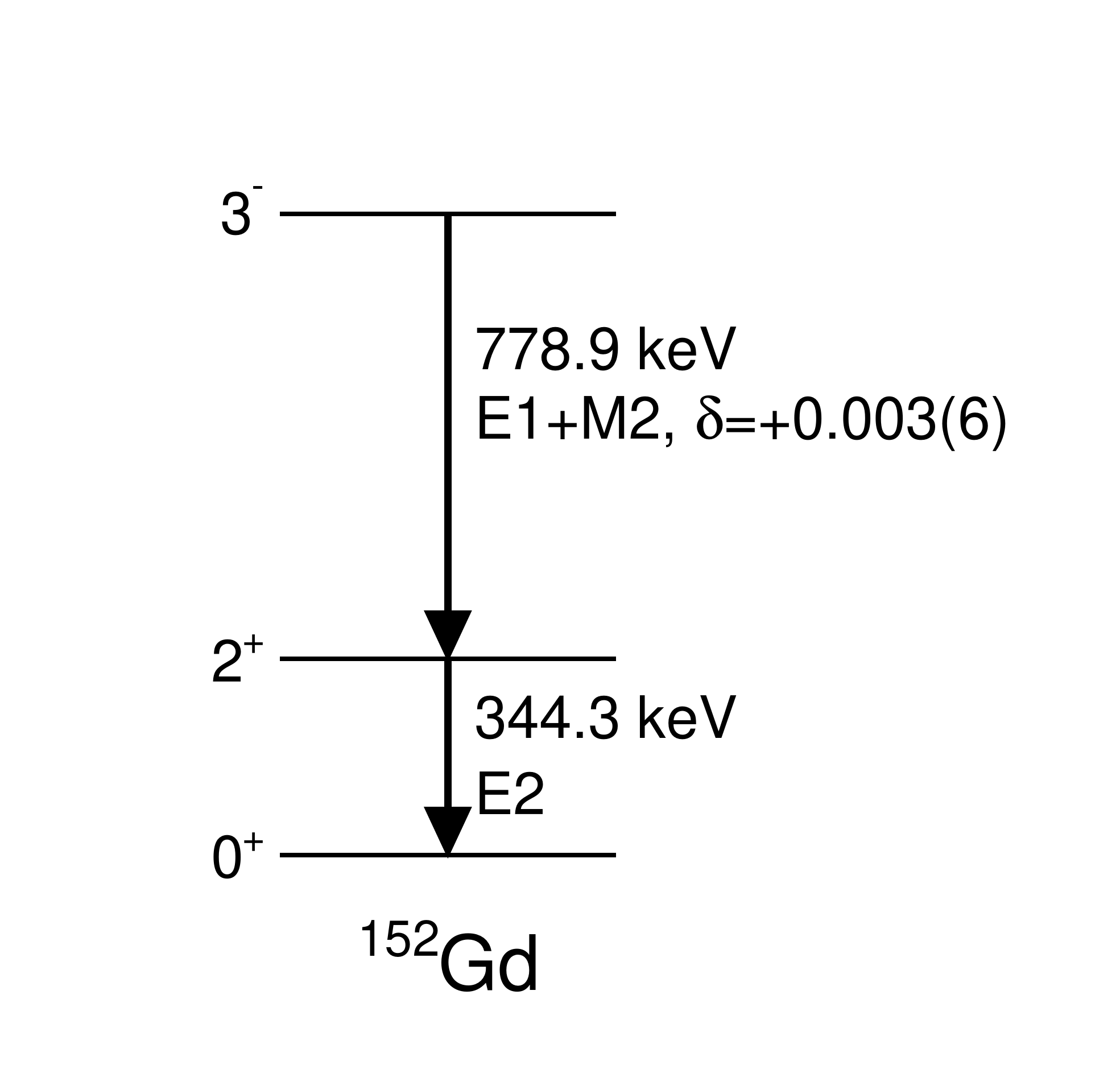}}
       \put(50,180){(b)}
     }
     \end{picture}
     \\
     \begin{picture}(100,200)
     \subfloat{%
       \put(0,0){\includegraphics[width=0.45\textwidth,height=0.39\textwidth]{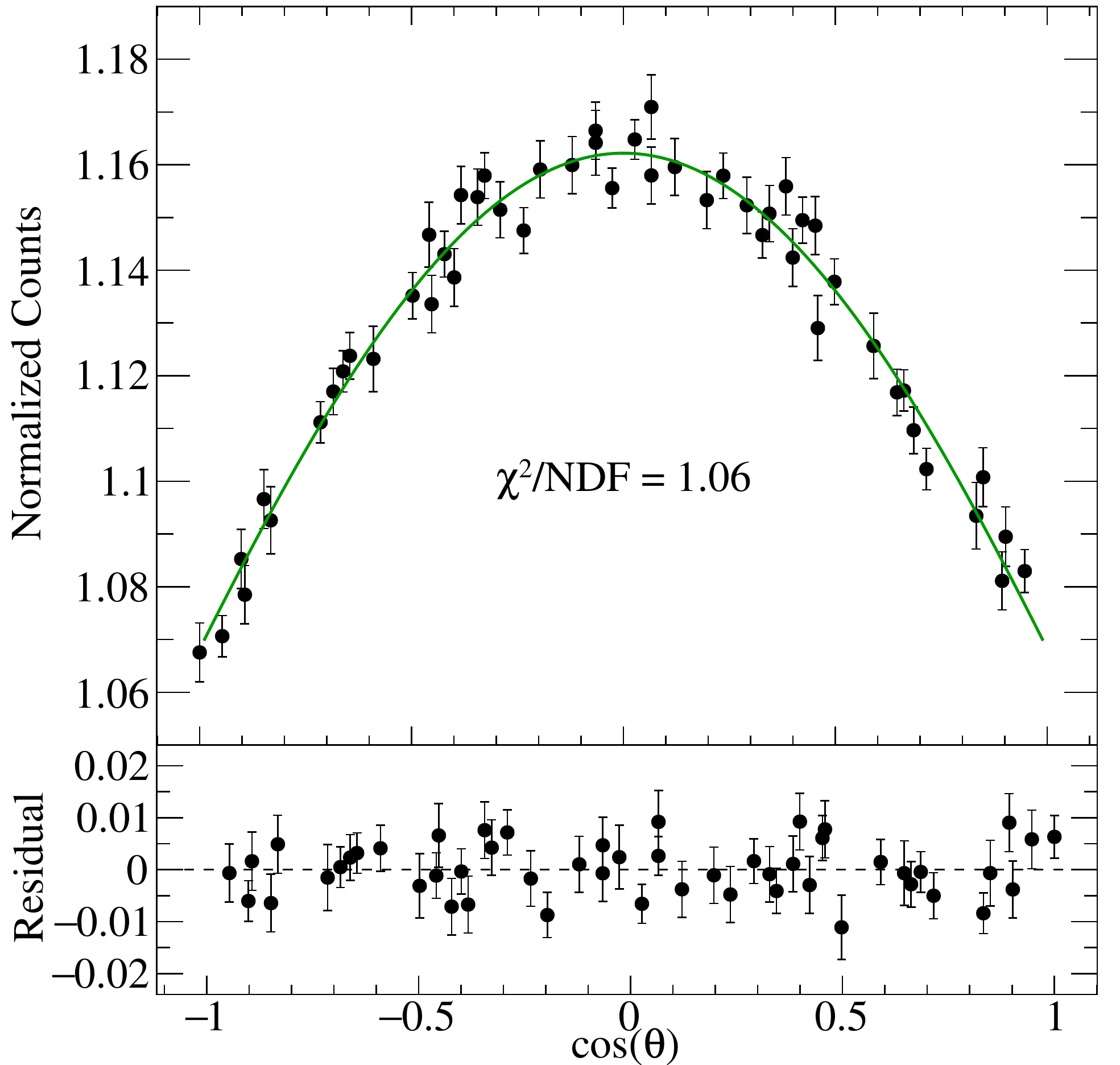}}
       \put(45,170){(c)}
     }
     \end{picture}
     \hfill
     \begin{picture}(224,200)
     \subfloat{%
       \put(0,0){\includegraphics[width=0.45\textwidth]{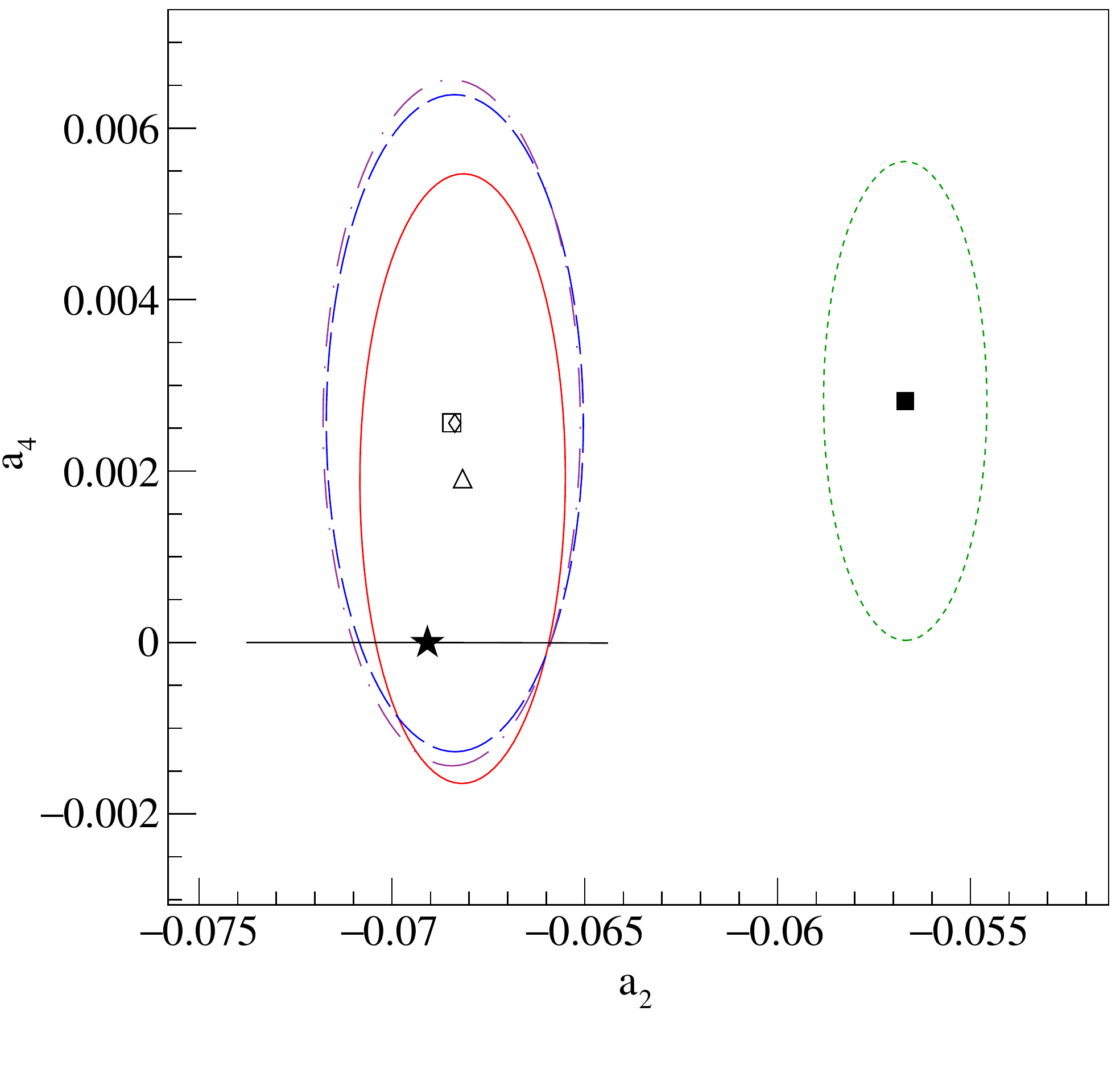}}
       \put(40,185){(d)}
     }
     \end{picture}
     \caption{Plots for the $^{152}$Gd $3^- - 2^+ - 0^+$ 778.9 keV - 344.3 keV cascade. (a) The best Method 1 fit (magenta filled line) to the data (black points) has a $\chi^2$/NDF=1.06. The residual of the fit is shown in the lower panel. (b) A partial level scheme showing the experimental details of this cascade. (c) A bare fit of Equation \ref{eq:ang-corr} to the data (green line) minimizes with $a_2$=-0.057(2) and $a_4$=0.003(3). (d) A comparison of minimized $a_2$ and $a_4$ values and 1$\sigma$ error fitted to the data with methods described in the paper. The expected $a_2$ and $a_4$ values are indicated by the star, with the black line representing values within the $\delta$ uncertainty. Minimized $a_2$, $a_4$ values and $1\sigma$ confidence intervals are shown for a bare fit of Equation \ref{eq:ang-corr} (filled square, green dotted ellipse), Method 2 (open triangle, red solid ellipse), Method 3 (open diamond, blue dashed ellipse), and Method 4 (open square, purple dot-dashed ellipse).}
     \label{fig:Eu152-4Panel-Fig2}
\end{figure*}

\begin{figure*}
\begin{picture}(100,200)
     \subfloat{%
       \put(0,0){\includegraphics[width=0.45\textwidth]{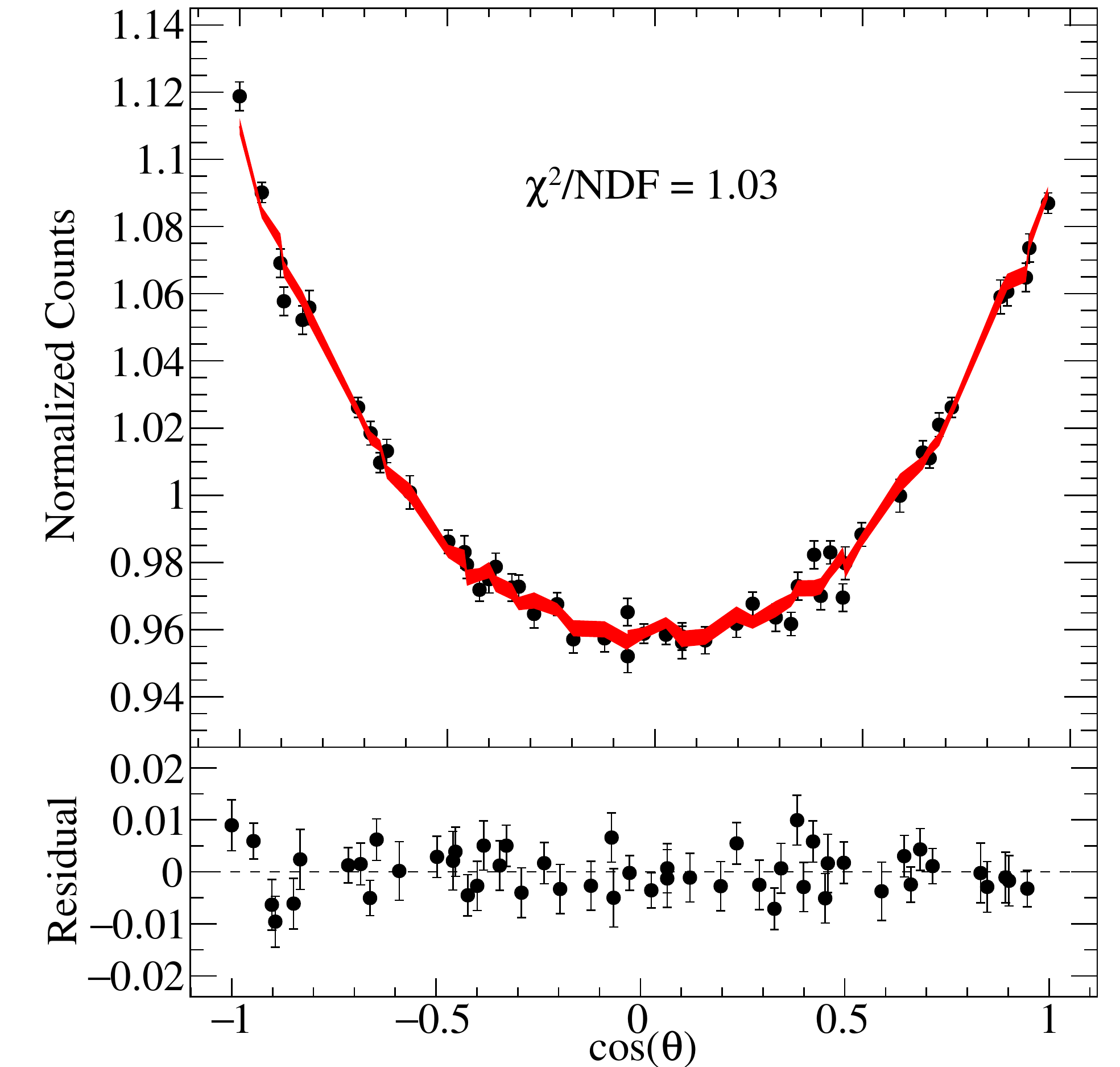}}
	\put(50,180){(a)}
	 }
     \end{picture}
     \hfill
     \begin{picture}(224,200)
     \subfloat{%
       \put(0,0){\includegraphics[width=0.45\textwidth]{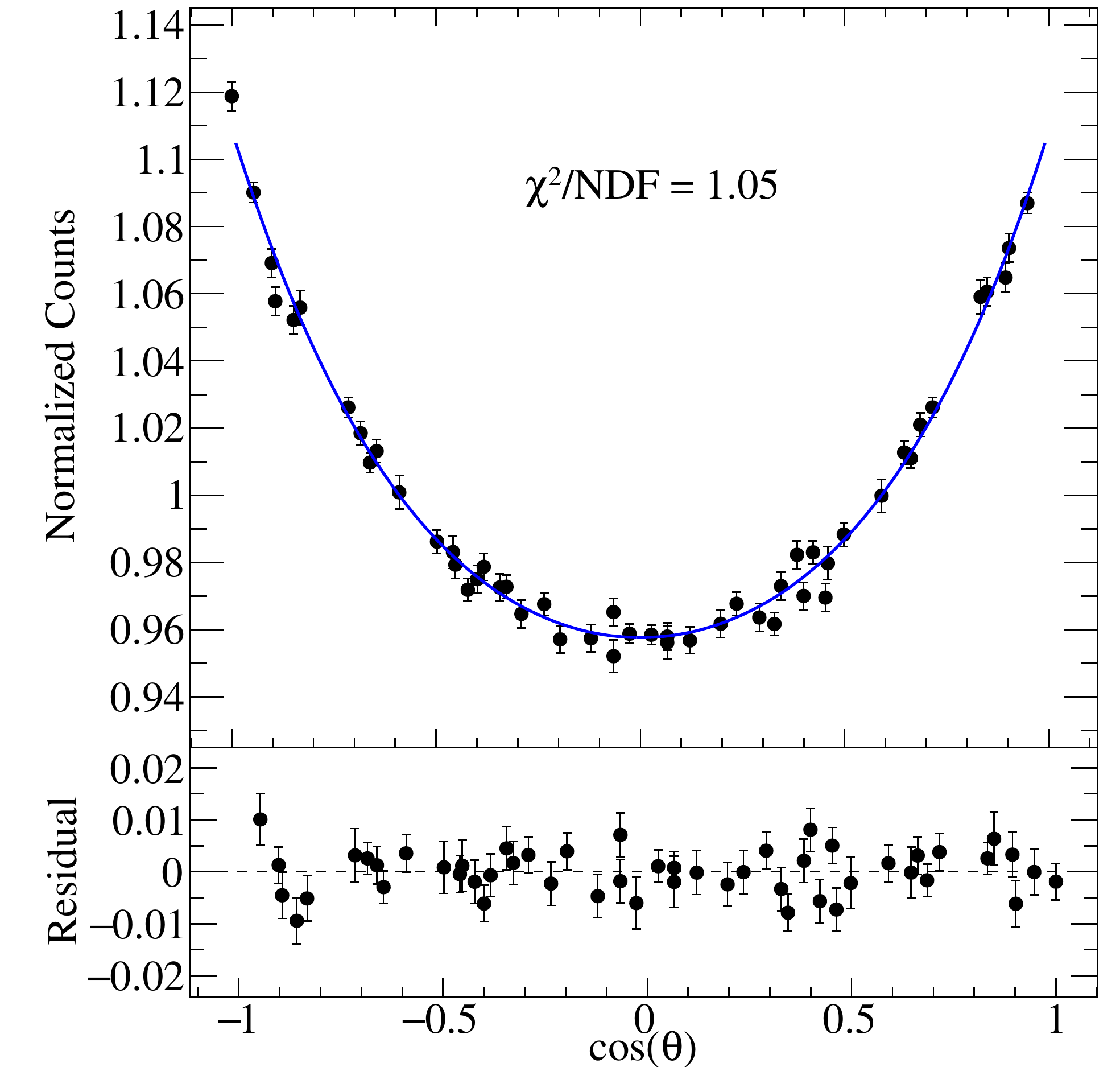}}
       \put(50,180){(b)}
     }
     \end{picture}
     \\
     \begin{picture}(100,160)
     \subfloat{%
       \put(0,0){\includegraphics[width=0.45\textwidth]{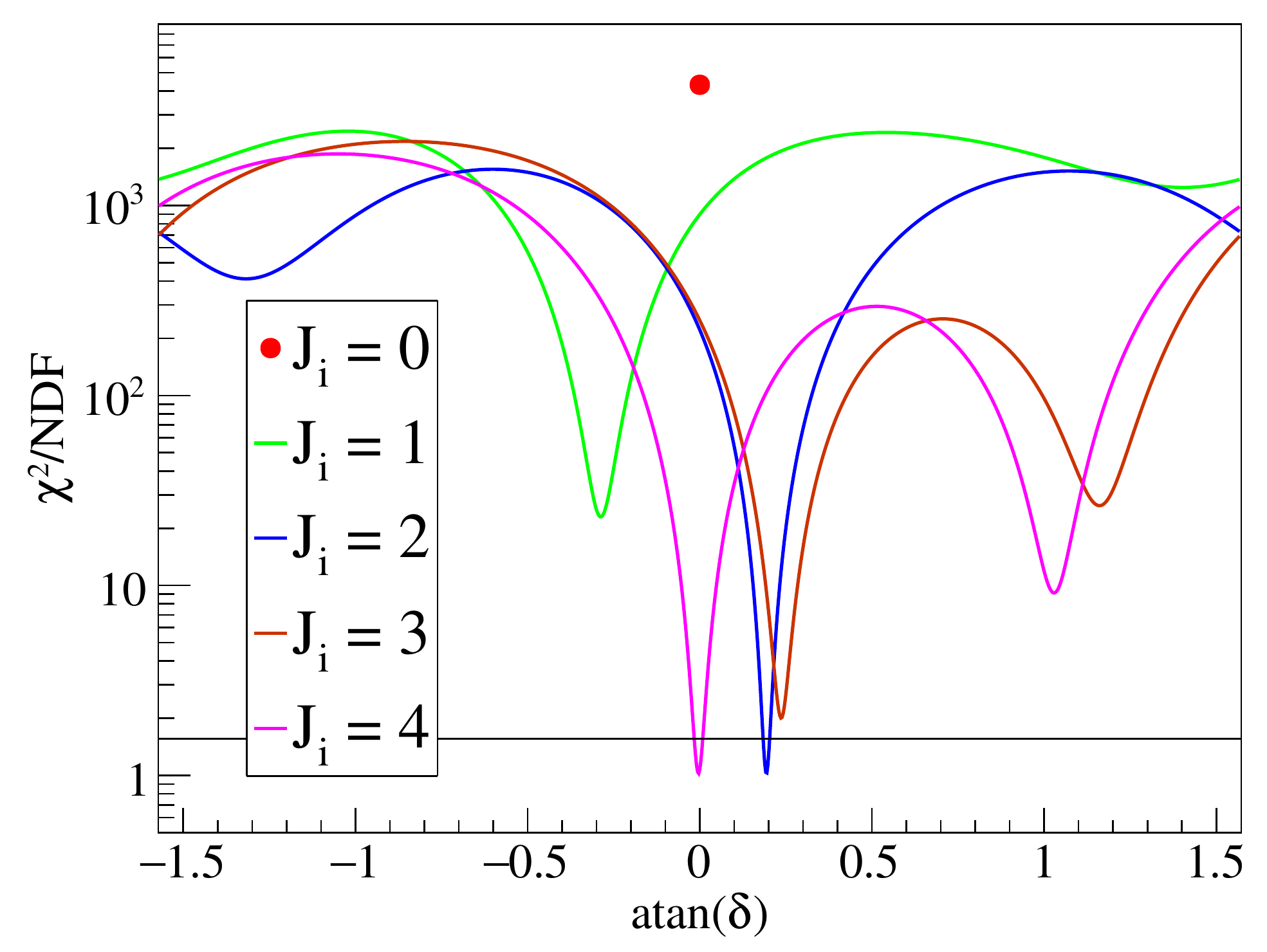}}
      \put(30,145){(c)} 
     }
     \end{picture}
     \hfill
     \begin{picture}(224,160)
     \subfloat{%
       \put(0,0){\includegraphics[width=0.45\textwidth]{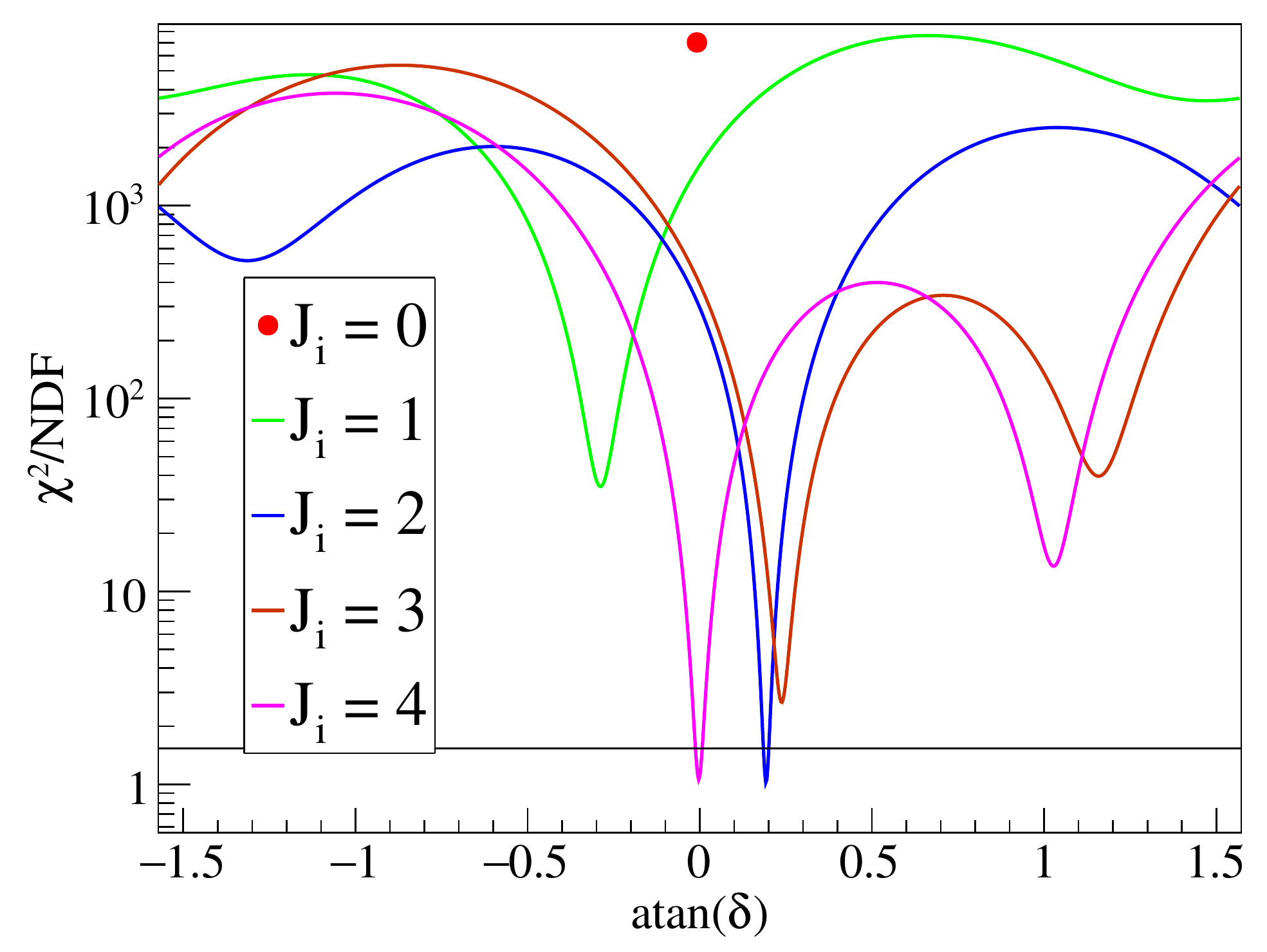}}
     \put(30,145){(d)}  
     }
     \end{picture}
     \caption{
     Plots for the $^{60}$Ni $4^+ - 2^+ - 0^+$ 1332 keV - 1173 keV cascade. (a) The best Method 2 fit (red filled line) to the data (black points) has a $\chi^2$/NDF of 1.03 and minimizes with $a_2=0.100(2)$ and $a_4=0.011(2)$. The residual of the fit is shown in the lower panel. (b) The best Method 3 fit (blue line) of the data (black points) has a $\chi^2$/NDF of 1.05 and minimizes with $a_2=0.100(2)$ and $a_4=0.011(2)$. The residual of the fit is shown in the lower panel. (c) A comparison of $\chi^2$/NDF values for potential $J_i=0-4$ and all possible mixing ratios ($\delta$) shows that the best fit to the data using Method 2 is made with $J=2,4$ with $\delta=0.193(2)$ and -0.003(3), respectively. (d) A comparison of $\chi^2$/NDF values for potential $J_i=0-4$ and all possible mixing ratios ($\delta$) shows that the best fit to the data using Method 4 is made with $J=2,4$ with $\delta=0.195(3)$ and -0.002(4), respectively.
     }
     \label{fig:Co60-4Panel-Fig1}
\end{figure*}

\begin{figure*}
\begin{picture}(100,200)
     \subfloat{%
      \put(0,0){\includegraphics[width=0.45\textwidth]{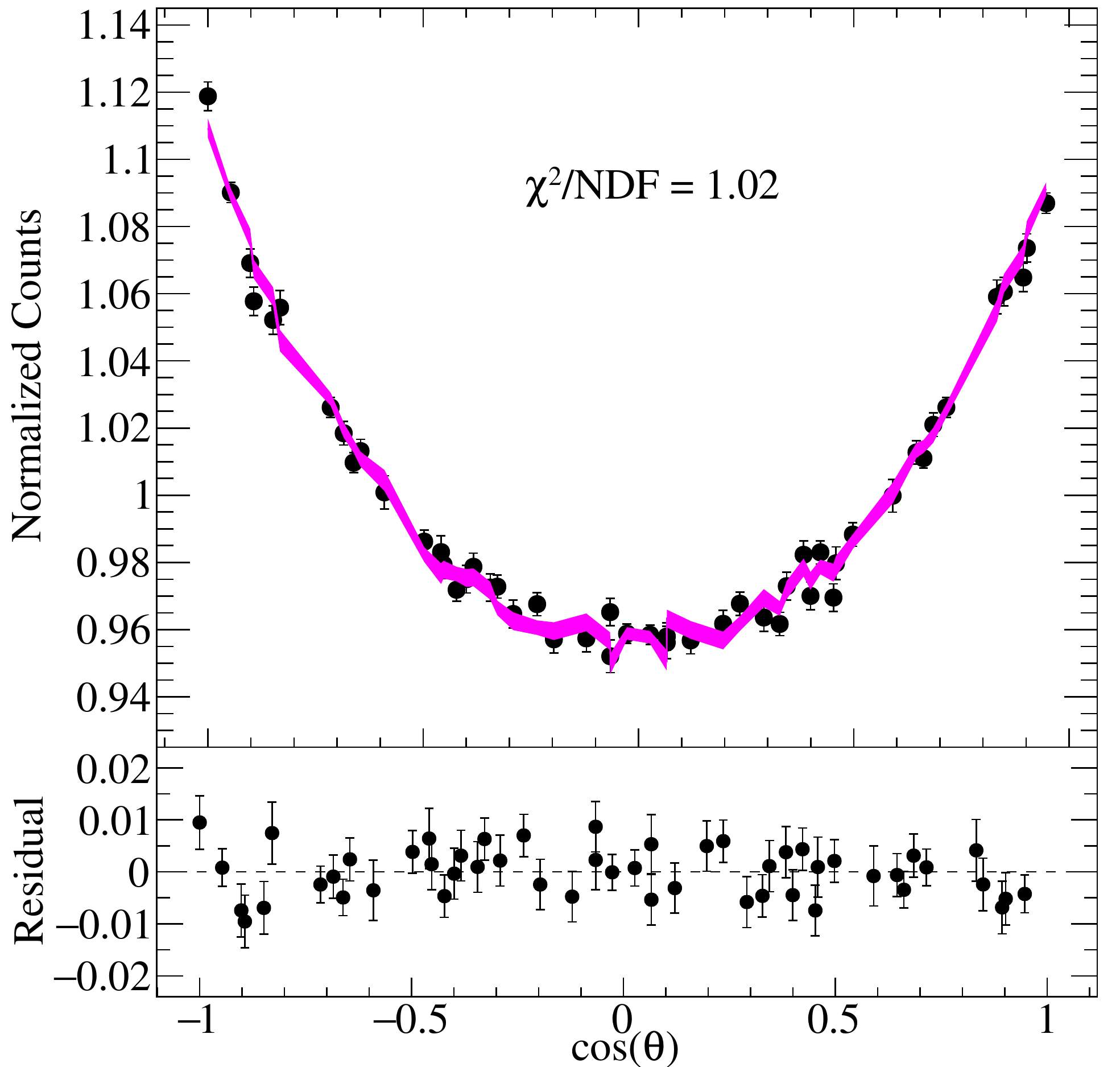}}
      \put(50,180){(a)}
     }
     \end{picture}
     \hfill
     \begin{picture}(224,200)
     \subfloat{%
       \put(0,0){\includegraphics[width=0.45\textwidth]{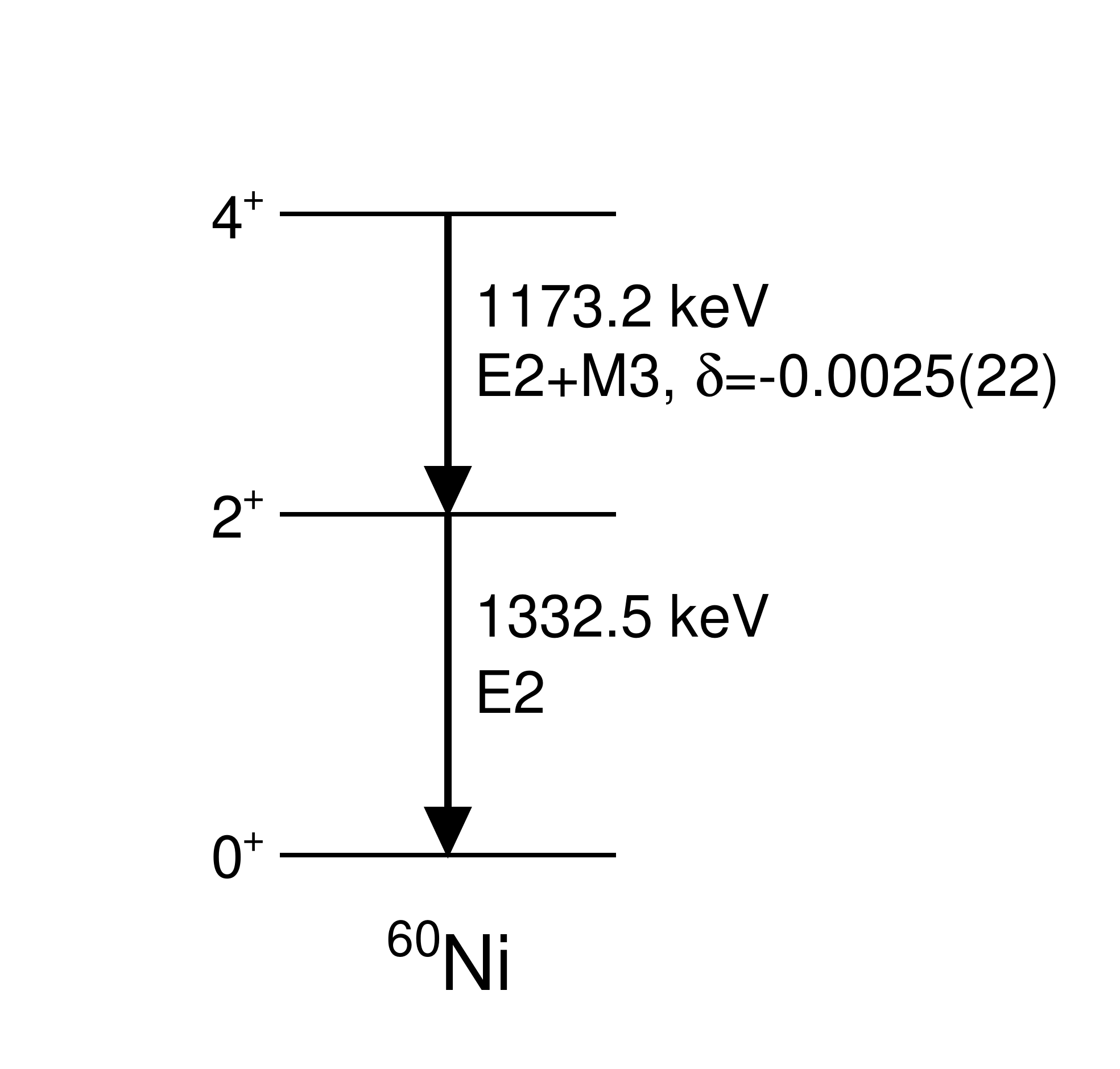}}
       \put(50,180){(b)}
     }
     \end{picture}
     \\
     \begin{picture}(100,200)
     \subfloat{%
      \put(0,0){\includegraphics[width=0.45\textwidth,height=0.42\textwidth]{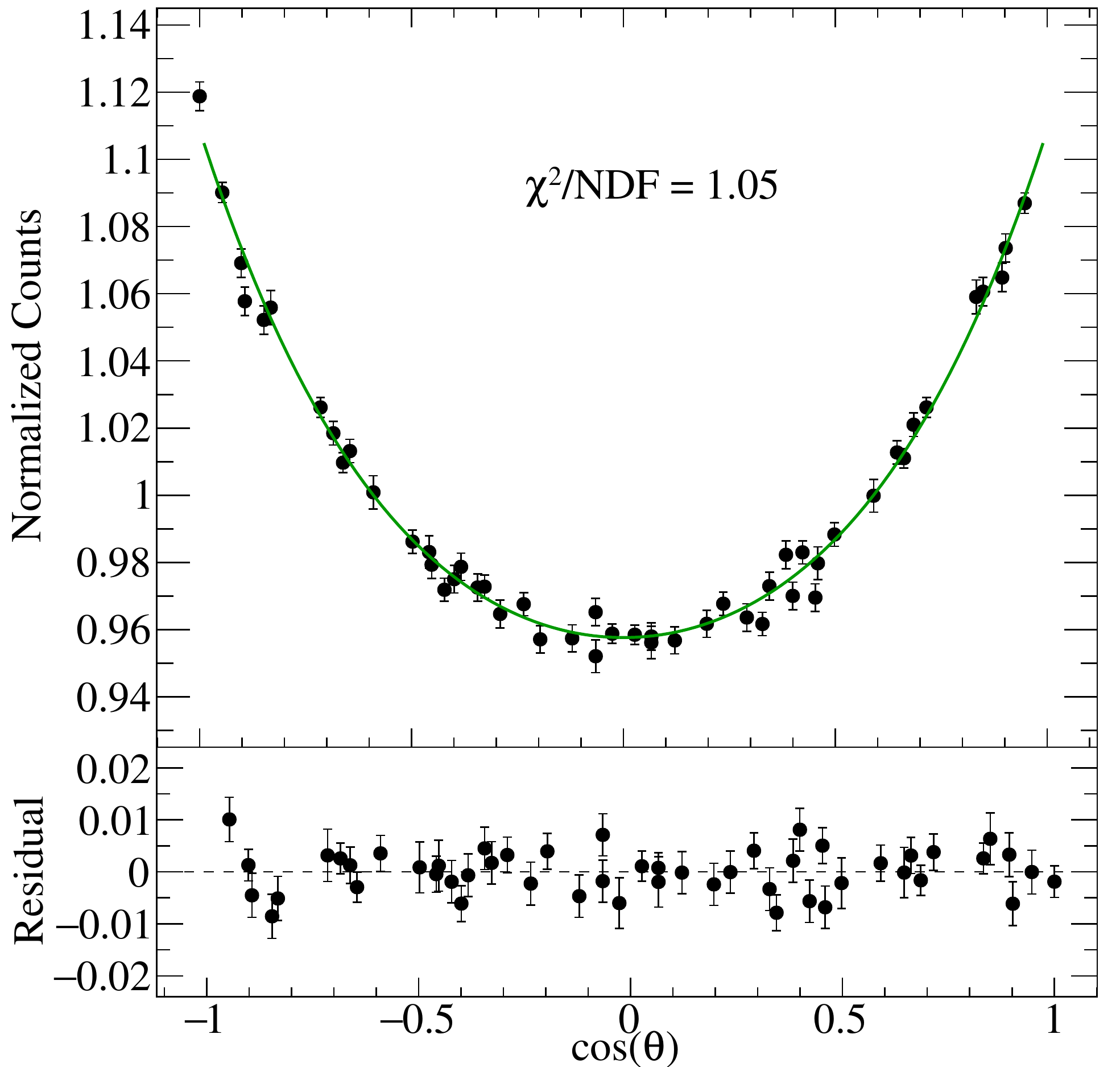}}
      \put(45,170){(c)}
     }
     \end{picture}
     \hfill
     \begin{picture}(224,200)
     \subfloat{%
       \put(0,0){\includegraphics[width=0.45\textwidth]{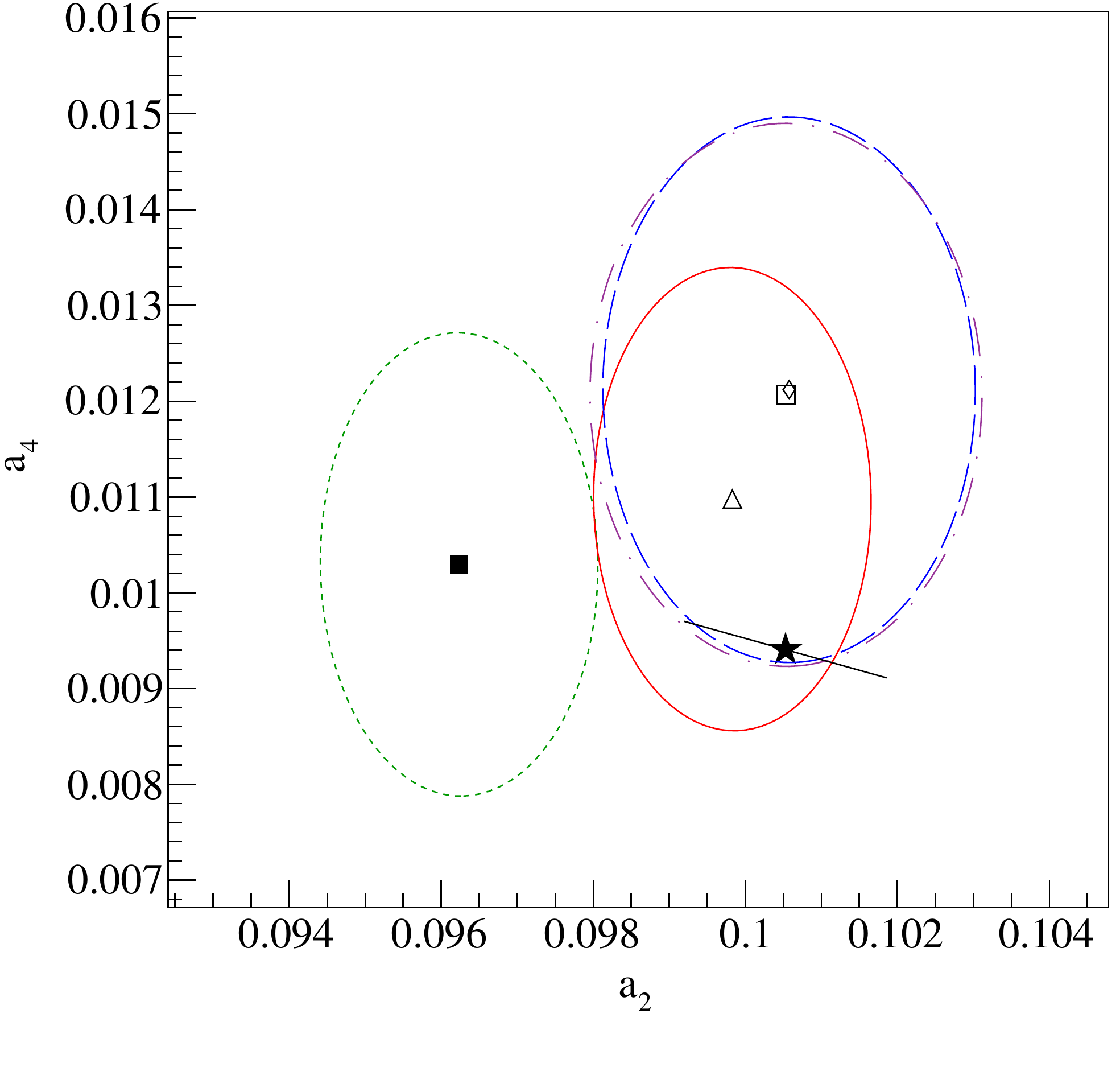}}
       \put(40,185){(d)}
     }
     \end{picture}
     \caption{Plots for the $^{60}$Ni $4^+ - 2^+ - 0^+$ 1332 keV - 1173 keV cascade. (a) The best Method 1 fit (magenta filled line) to the data (black points) has a $\chi^2$/NDF=1.02. The residual of the fit is shown in the lower panel. (b) A partial level scheme showing the experimental details of this cascade. (c) A bare fit of Equation \ref{eq:ang-corr} to the data (green line) minimizes with $a_2$=0.096(2) and $a_4$=0.010(3). (d) A comparison of minimized $a_2$ and $a_4$ values and 1$\sigma$ error fitted to the data with methods described in the paper. The expected $a_2$ and $a_4$ values are indicated by the star, with the black line representing values within the $\delta$ uncertainty. Minimized $a_2$, $a_4$ values and $1\sigma$ confidence intervals are shown for a bare fit of Equation \ref{eq:ang-corr} (filled square, green dotted ellipse), Method 2 (open triangle, red solid ellipse), Method 3 (open diamond, blue dashed ellipse), and Method 4 (open square, purple dot-dashed ellipse).}
     \label{fig:Co60-4Panel-Fig2}
\end{figure*}

\begin{table}
\caption{\label{table:coeffs-pureE2E2} Results from the fitting of simulated $\mathcal{Z}$ distributions covering a wide range of $\gamma-\gamma$ cascade energies. The uncertainties in the $\beta$ and $\gamma$ coefficients are given at the $1\sigma$ level and have been inflated during the fitting procedure as described in Section \ref{sec:alp-bet-gam}.}
\begin{center}
\begin{tabular}{c|c|c} 
\hline
\hline
$E_\gamma - E_\gamma$ & $\beta$ & $\gamma$ \\ \hline
68-20 & 0.87521(96) & 0.7114(12) \\
68-30 & 0.88712(33) & 0.70514(43) \\
68-40 & 0.90844(26) & 0.73539(34) \\
68-50 & 0.92580(25) & 0.78237(32) \\
75-68 & 0.93205(24) & 0.78644(31) \\
90-68 & 0.93431(24) & 0.78880(31) \\
100-20 & 0.87911(95) & 0.7197(12) \\
100-68 & 0.93733(24) & 0.79610(32) \\
120-68 & 0.93678(24) & 0.79468(31) \\
135-68 & 0.93742(24) & 0.79568(32) \\
150-68 & 0.93805(25) & 0.79775(32) \\
200-68 & 0.94029(26) & 0.80128(34) \\
300-68 & 0.94340(29) & 0.81066(37) \\
500-68 & 0.94196(33) & 0.80930(43) \\
779-68 & 0.94486(38) & 0.81714(50) \\
779-344 & 0.95396(47) & 0.84243(61) \\
1039-833 & 0.95627(65) & 0.84723(85) \\
1332-1172 & 0.95684(76) & 0.84949(99) \\
1333-1039 & 0.95571(74) & 0.84976(96) \\
2752-68 & 0.94535(57) & 0.82352(74) \\
2752-1039 & 0.95666(95) & 0.8542(12) \\
2752-2013 & 0.9584(12) & 0.8557(15) \\
5000-68 & 0.94742(76) & 0.82441(99) \\
5000-1039 & 0.9578(13) & 0.8567(17) \\
\hline
\hline
\end{tabular}
\end{center}
\end{table}

\bibliographystyle{apsrev}
\bibliography{ACtechniques.bib}







\end{document}